\documentclass[journal,twocolumn,10pt,twoside]{IEEEtran}
\normalsize

\ifCLASSINFOpdf
\else
\fi

\usepackage{amsmath}
\usepackage{amssymb}
\usepackage{cite}
\usepackage{graphicx}
\usepackage{algorithm}
\usepackage{algpseudocode}
\usepackage{subfigure}
\usepackage{multirow}
\usepackage{longtable}
\usepackage{threeparttable}
\usepackage{caption3}
\usepackage{array}
\usepackage{cases}
\usepackage{color}
\usepackage[T1]{fontenc}
\usepackage[utf8]{inputenc}
\usepackage{authblk}
\usepackage{subfigmat}
\usepackage{fixltx2e}

\newtheorem{remark}{\bf Remark}

\def\E{\mathsf{E}}

\def\phi{\varphi}

\def\l{\left}
\def\r{\right}
\def\({\left(}
\def\){\right)}

\def\b0{{\mathbf{0}}}

\newcommand{\tx}[1]{\texttt{#1}}

\newcommand{\sinc}{\tx{sinc}}

\newcommand{\nn}{\nonumber}

\newtheorem{observation}{Observation}

\newcommand{\Tr}{\mathrm{Tr}}

\newcommand{\rf}{\mathrm{rf}}
\newcommand{\dc}{\mathrm{dc}}
\newcommand{\In}{\mathrm{in}}
\newcommand{\out}{\mathrm{out}}
\newcommand{\ant}{\mathrm{ant}}
\newcommand{\R}{\textnormal{R}}
\newcommand\mycom[2]{\genfrac{}{}{0pt}{}{#1}{#2}}

\begin{document}

%\title{Wireless Power Transfer for Future Networks: \\ Overview of Challenges and Opportunities for Signal Processing and Machine Learning}
%\title{Wireless Power Transfer for Future Networks: \\ Techniques, Challenges and Opportunities for Signal Processing and Machine Learning}
%\title{Wireless Power Transfer for Future Networks: \\ Techniques, Challenges and Opportunities for Signal Processing and Deep Learning}
%\title{An Overview of Signal Processing and Deep Learning Techniques for Wireless Powered Networks}
%\title{Wireless Powered Networks: \\ Techniques, Challenges and Opportunities for Signal Processing and Machine Learning}
%\title{Signal Processing and Deep Learning Techniques for Wireless Powered Networks}
%\title{Signal Processing and Deep Learning Techniques for Wireless Powered Networks: \\ An Overview}
%\title{Wireless Power Transfer for Future Networks: \\ An Overview of Signal Processing and Machine Learning Techniques}
%\title{Signal Processing and Machine Learning Techniques for Future Wireless Powered Networks}
%\title{Signal Processing, Machine Learning, and Computing for Future Wireless Powered Networks}
%\title{Future Wireless Powered Systems and Networks: Signal Processing, Machine Learning, \\ Sensing and Computing}
%\title{Wireless Power Transfer for Next-Generation Wireless Networks: Signal Processing, Machine Learning, and Computing}
%\title{Future Networks with Wireless Power Transfer: Signal Processing, Machine Learning, \\ Sensing and Computing}
\title{Wireless Power Transfer for Future Networks: Signal Processing, Machine Learning, \\ Computing, and Sensing}

\author{Bruno Clerckx, \textit{Senior Member, IEEE}, Kaibin Huang, \textit{Fellow, IEEE}, Lav R.\ Varshney, \textit{Senior Member, IEEE}, Sennur Ulukus, \textit{Fellow, IEEE}, and Mohamed-Slim Alouini, \textit{Fellow, IEEE}% <-this % stops a space
\thanks{B. Clerckx is with the Electrical and Electronic Engineering Department at Imperial College London, London SW7 2AZ, UK (email: b.clerckx@imperial.ac.uk).
\par K. Huang is with the Department of Electrical and
Electronic Engineering, The University of Hong Kong, Hong Kong (e-mail:
huangkb@eee.hku.hk).
\par L.~R.\ Varshney is with the Coordinated Science Laboratory and the Department of Electrical and Computer Engineering, University of Illinois at Urbana–Champaign, Urbana, IL 61801 USA (e-mail: varshney@illinois.edu).
\par S. Ulukus is with the Department of Electrical and Computer Engineering, University of Maryland at College Park, College Park, MD 20742 USA (e-mail: ulukus@umd.edu).

\par M.-S. Alouini is with the CEMSE Division, King Abdullah University of Science, and Technology (KAUST), Thuwal 23955-6900, Saudi Arabia
(e-mail: slim.alouini@kaust.edu.sa).

\par This work has been partially supported by the EPSRC of UK under grant EP/P003885/1 and EP/R511547/1.}}

\IEEEspecialpapernotice{(Overview Paper)}


\maketitle

\begin{abstract}

Wireless power transfer (WPT) is an emerging paradigm that will enable using wireless to its full potential in future networks, not only to convey information but also to deliver energy. Such networks will enable trillions of future low-power devices to sense, compute, connect, and energize anywhere, anytime, and on the move. The design of such future networks brings new challenges and opportunities for signal processing, machine learning, sensing, and computing so as to make the best use of the RF radiations, spectrum, and network infrastructure in providing cost-effective and real-time power supplies to wireless devices and enable wireless-powered applications. In this paper, we first review recent signal processing techniques to make WPT and wireless information and power transfer (WIPT) as efficient as possible. Topics include high-power amplifier and energy harvester nonlinearities, active and passive beamforming, intelligent reflecting surfaces, receive combining with multi-antenna harvester, modulation, coding, waveform, large-scale (massive) multiple-input multiple-output (MIMO), channel acquisition, transmit diversity, multi-user power region characterization, coordinated multipoint, and distributed antenna systems. Then, we overview two different design methodologies: the \emph{model and optimize} approach relying on analytical system models, modern convex optimization, and communication/information theory, and the \emph{learning} approach based on data-driven end-to-end learning and physics-based learning. We discuss the pros and cons of each approach, especially when accounting for various nonlinearities in wireless-powered networks, and identify interesting emerging opportunities for the approaches to complement each other. Finally, we identify new emerging wireless technologies where WPT may play a key role---wireless-powered mobile edge computing and wireless-powered sensing---arguing WPT, communication, computation, and sensing must be jointly designed.  
\end{abstract}


\IEEEpeerreviewmaketitle

\section{Introduction}\label{Intro_section}

%%%%%%%%%%%%%%%%%%%%%%%%%
\IEEEPARstart{T}{wenty} years from now, according to Koomey's law \cite{Koomey:2011}, devices will require 10000 less energy to compute a given task, due to the reduction in power requirements of their electronics. Moreover, trillions of Internet-of-Things (IoT) devices will emerge. This explosion of low-power devices demands a re-thinking of future network design, where wireless will be used to its full potential, not only to convey information but also to deliver energy. Wireless power will bring new opportunities, namely proactive and controllable energy supply with genuine mobility---no wires, no contact, no or reduced batteries---and therefore small, light, and compact devices. This will not only yield environmental benefits by eliminating the need to produce, maintain, or dispose of trillions of batteries, but also enable myriad new wireless applications such as autonomous low-power sensing and computing due to prolonged lifetime and a long-term, predictable, and reliable energy supply unlike ambient energy-harvesting technologies including solar, thermal, or vibration.
\par Wireless power and wireless communications have, however, evolved as two separate fields in academia and industry \cite{Varshney:2012}. This separation has consequences: first, current wireless networks broadcast RF energy into air (for communication purposes) but do not use it for charge devices; second, providing ubiquitous wireless power would require deploying a separate network of dedicated energy transmitters. Imagine instead future network where information and energy flow together through the wireless medium. Wireless Information Transfer (WIT) and Wireless Power Transfer (WPT) would refer to two extreme strategies respectively targeting communication-only and power-only. A unified Wireless Information and Power Transfer (WIPT) design would be able to softly evolve between the two extremes to best use the RF spectrum/radiation and network infrastructure to communicate and energize, thereby outperforming traditional systems that separate communications and power.
\par
\begin{figure}%[!hhh]
   \centerline{\includegraphics[width=8cm]{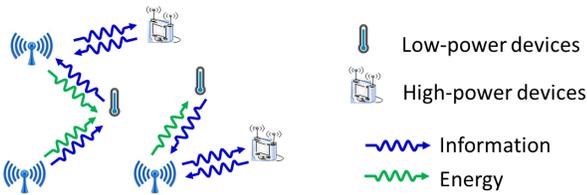}}
  \caption{Future wireless powered networks {\color{blue} to update to integrate sensing, computing, cloud}}
  \label{network}
\end{figure}

 Such a network will enable the creation of highly efficient wireless power resources, such that low-power devices (e.g. sensors) with or without a communication capability can be wirelessly powered anywhere, anytime and on the move and such that low-power devices with communication capabilities can experience a true ubiquitous wireless connectivity. It will also enable low-power and high-power devices to co-exist in such a way that transmit signals simultaneously charge remote low-power devices and carry information to high-power devices (e.g.\ smartphones, tablets), as illustrated in Fig.~\ref{network}. Wireless power will also enable  emerging wireless applications, such as wireless-powered edge intelligence, wireless-powered computing, wireless-powered sensing, and wireless-powered autonomous systems.
\par The design of efficient wireless power resources, the integration of wireless power and communications, sensing and computing, and the positioning of wireless power as a key enabler of new wireless applications brings new challenges, ideas and opportunities, and calls for a paradigm shift in wireless system and network design. Numerous research problems must be addressed that cover a wide range of disciplines, including circuit and systems, sensors, antenna and propagation, microwave theory and techniques, communication, signal processing, machine learning, sensing, computing, and information theory.

\subsection{Wireless Power for Future Networks: Overview of Challenges and Technologies}

\par Wireless power, especially in its most promising form of WPT, will be a fundamental building block of future wireless networks. WPT research over the past decades has largely focused on RF theories and techniques regarding the energy receptor with the design of efficient RF solutions, circuits, antennas, rectifiers and power management units \cite {Hemour:2014,OptBehaviour,Valenta:2014,Costanzo:2016}. Nevertheless, more recently, a new complementary line of research on communications and signal design for WPT has attracted significant attention in the communication and signal processing literature \cite{Zeng:2017}. Additionally, there has been growing interest in bridging RF, signal, and system designs to bring these two communities closer together and better understand the fundamentals of an effective wireless powered network architecture \cite{Clerckx:2018}. This has resulted in new understanding of signal and system design for WPT and WIPT \cite{Clerckx:2019}.

\par There are numerous design challenges of the envisioned future network : 1) \textit{Range}: Deliver wireless power at distances of 5-100s meters (m) for energizing low-power devices in indoor/outdoor settings; 2) \textit{Efficiency}: Boost the end-to-end power transfer efficiency (up to a fraction of a percent/a few percent), or equivalently the DC power level at the energy harvester for a given transmit power; 3) \textit{Non-line of sight (NLoS)}: Support Line of sight (LoS) and NLoS to widen real-world applications of future WIPT networks; 4) \textit{Mobility support}: Support mobile devices, at least for those at pedestrian speed; 5) \textit{Ubiquitous accessibility}: Provide power ubiquitously within the network coverage area;  6) \textit{Safety and health}: Make RF system safe and comply with the regulations; 7) \textit{Energy consumption}: Limit the energy consumption of wireless powered devices; 8) \textit{Seamless integration of wireless communication and wireless power}: Unify wireless communication and wireless power into WIPT; 9) \textit{Integrated WPT, sensing, computing, and communication}: Integrate WPT with sensing/computing and communication in 5G-and-beyond systems with virtualization and network slicing. 

\begin{figure*}%[!hhh]
   \centerline{\includegraphics[width=14cm]{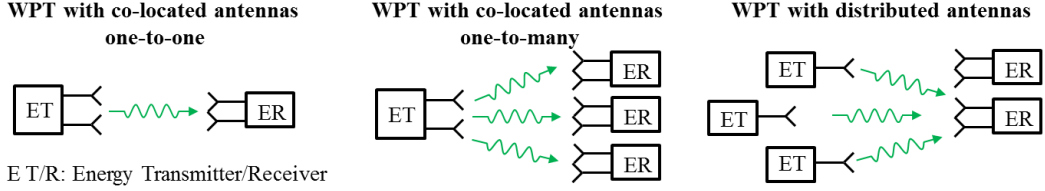}}
  \caption{WPT deployment scenarios.}
  \label{WPT_figure}
\end{figure*}

\par Challenges (1)--(7) are being studied in various communities \cite{Shinohara:2014,Carvalho:2014,Costanzo:2016,Zeng:2017,Clerckx:2018}. Solutions cover a wide range of areas spanning sensors, devices, RF, communication, signal and system designs for WPT. Typical WPT scenarios under study are illustrated in Fig.\ \ref{WPT_figure} and include:
\begin{itemize}
\item \textit{Single-user (point-to-point) WPT}: The focus here is on a single energy transmitter (ET) and a single energy receiver (ER). Both ET and ER may be equipped with multiple co-located antennas. This scenario is the fundamental building block of future wireless networks, since most of the challenges (1)--(7) must be tackled for this setup before considering multi-user scenarios. 
\item \textit{Multi-user WPT}: The focus  here is on transmit antennas being either co-located or distributed and delivering energy to multiple ERs equipped with one or multiple antennas. 
\end{itemize}

\par Challenge (8) has recently been reviewed in \cite{Clerckx:2019} in an attempt to lay the fundamentals of WIPT from energy harvester modeling to signal and system designs. In contrast to WPT and WIT, where the emphasis of the system design is to exclusively deliver energy and information, respectively, in WIPT, both energy and information are to be delivered. The challenge is therefore to understand how to make the best use of the RF radiation and the RF spectrum to provide both information and energy, and requires the characterization of the fundamental trade-off between the amount of information and the amount of energy that can be delivered in a wireless network and how signals should be designed to achieve this trade-off.

\begin{figure*}%[!hhh]
   \centerline{\includegraphics[width=14cm]{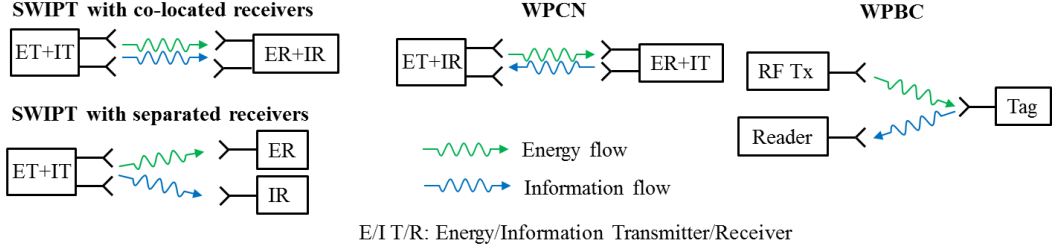}}
  \caption{Different WIPT scenarios and architectures.}
  \label{SWIPT_figure}
\end{figure*}

\par As illustrated in Fig. \ref{SWIPT_figure}, WIPT can be categorized into three different types.
\begin{itemize}
\item \textit{Simultaneous Wireless Information and Power Transfer (SWIPT)}: Energy and information are simultaneously transmitted from one or multiple transmitter(s) to one or multiple receiver(s)\cite{Varshney:2008,Grover:2010,Zhang:2013,Xu:2014,Son:2014,Zhou:2013,Liu:2013,Park:2013,Park:2014,Park:2015,Huang:2013,Zhou:2014,Ng:2013,VarshneyS:2019,Nasir:2013,Huang:2015,Xiang:2012,Timotheou:2014,Khandaker:2014,Fouladgar:2012,Ding:2014}. The information receiver(s) (IR) and ER can be co-located or separated. With co-located receivers, each receiver is a single (typically low-power) device that is simultaneously being charged and receiving data. With separate receivers, ER and IR are different devices, the former being a low-power device being energized, the latter being a device receiving data.
\item \textit{Wirelessly Powered Communication Networks (WPCNs)}: Energy is transmitted in the downlink from an access point to a receiver and information is transmitted in the uplink \cite{Ju:2014,Huang_Lau:2014,Lee:2016}. The receiver is a device that harvests energy in the downlink and uses the harvested energy to transmit data in the uplink.
\item \textit{Wirelessly Powered Backscatter Communication (WPBC)}: Energy is transmitted in the downlink and information is transmitted in the uplink using backscatter modulation at a tag to reflect and modulate the incoming RF signal for communication with a reader \cite{Han:2017,Clerckx:2017b,Zawawi:2018}. Backscatter communications benefit from several orders-of-magnitude lower power consumption than conventional wireless communications because tags do not require oscillators to generate carrier signals \cite{Boyer:2014}.
\end{itemize}
Moreover, a network could have a mixture of all of these types of transmissions with multiple co-located and/or distributed ETs and information transmitter(s) (IT).

\par Challenge (9) is new and arises since next-generation Internet-of-Things (IoT) that build on the 5G/6G platform are seeing an increasing level of integration between storage, compute, and communication so as to efficiently enable a wide range of new applications ranging from distributed sensing to edge computing and artificial intelligence (AI). Thus, wirelessly powering next-generation IoT calls for the joint control of WPT, sensing, computing, and communication so as to optimize efficiency of a system supporting specific applications. In particular, there exist trade-offs between transferred energy and energy consumption of sensing/computing (e.g., on-device AI model training) and communication (e.g., mobile computation offloading). Quantifying and exploiting such trade-offs can substantially improve system performance.   

%\begin{figure*}%[!hhh]
%   \centerline{\includegraphics[width=14cm]{}}
%  \caption{Emerging wireless-powered scenarios and applications. {\color{green} For Kaibin}}
%  \label{WPT_enabler_figure}
%\end{figure*}

\subsection{Objectives and Organization}

\par Various review papers have appeared in past years on WPT, emphasizing separately RF, circuit and antenna solutions \cite{OptBehaviour,Valenta:2014,Shinohara:2014,Carvalho:2014,Costanzo:2016}, and communications, signal and system design solutions \cite{Zeng:2017}. More recently attempts have been made to bridge RF, signal and system designs to get a better understanding of the fundamental building blocks of an efficient WPT network architecture \cite{Clerckx:2018}. This synthesis of work in different areas of WPT has yielded critical observations and given a fresh new look to promising avenues for WPT signal and system design. As an example, \cite{Clerckx:2018} shows that the \textit{nonlinear} nature of the WPT design problem, both for the ET and the ER, must be accounted for at the signal and the circuit-level design. 

\par Similarly, review papers on WIPT have also appeared \cite{Bi:2015,Tabassum:2015,KHuang:2015,Bi:2016,KHuang:2016,Krikidis:2014,Ding:2015,Chen:2015,Lu:2015,Ulukus:2015,Niyato:2017}. Emphasis was put at that time on characterizing the fundamental tradeoff between conveying information and energy, so-called rate-energy (R-E) tradeoff, under the assumption of a very simple \textit{linear} model of the ET and ER. In recent years, the validity of this linear model has been questioned and there has been an increasing departure from simple linear assumptions in the WIPT literature. It turns out that the linear model is inaccurate and leads to inefficient WIPT designs, and that WIPT design radically changes once we adopt more realistic \textit{nonlinear} models of the energy harvester (EH) \cite{Clerckx:2019}. Recently, \cite{Clerckx:2019} showed how crucial the EH model is to WIPT signal and system designs and how WIPT signal and system designs revolve around the underlying EH model. It highlighted different linear and nonlinear EH models, and showed in a systematic way how WIPT designs and R-E tradeoff differ for each of them. In particular, the paper showed how the modeling of the EH can have tremendous influence on the design of the physical and higher layers of WIPT networks. 

\par This paper overviews recent advances and emerging opportunities for signal processing, machine learning, computing, and sensing in the broad area of future wireless powered networks (including WPT, WIPT, and other emerging wireless powered applications). The objectives are threefold. 

\par \textit{First}, this paper aims to provide a review of recent signal processing techniques to tackle the challenges of WPT and WIPT and make them a reality. Topics discussed include high power amplifier (HPA) and EH nonlinearities, transmit active and passive beamforming and intelligent reflecting surfaces, receive combining with multi-antenna harvester, modulation, coding, waveform, joint beamforming, combining and waveform, large-scale (massive) multiple-input multiple-output (MIMO), channel acquisition, transmit diversity, power region characterization in multi-user WPT, coordinated multipoint and distributed antenna systems for wireless powered networks. A particular emphasis is on how the design of those techniques is deeply rooted in the EH nonlinearity and contrasts with a previous tutorial \cite{Zeng:2017} where nonlinearity was highlighted only as part of the waveform design.  

\par \textit{Second}, this paper aims to provide an overview of various design methodologies. Instead of relying exclusively on the traditional model-and-optimize approach used in all past tutorial and review papers that derive an analytical system model (under some assumptions) and then use modern convex optimization tools to optimize it, here we also discuss the role machine learning, in the form of model-based  and data-based end-to-end learning and physics-based learning, can play to design future wireless-powered networks. This is particularly relevant due to the importance of accounting for various sources of nonlinearity in wireless power. We identify the pros and cons of the \emph{model and optimize} approach and the \emph{learning} approach and identify interesting emerging opportunities for machine learning to complement human expertise.

\par \textit{Third}, this paper aims to identify emerging wireless technologies where WPT will play a key role. In particular, we discuss and study how WPT can enable wireless-powered computing, wireless-powered sensing, and wireless-powered edge/federated learning.
 
\emph{Organization:} In Section \ref{WPT_section}, we introduce the system model of WPT, discuss the HPA and EH nonlinearity and EH architecture, and review various signal processing techniques used to increase the end-to-end power transfer efficiency of single-user and multi-user WPT. Section \ref{WIPT_section} builds upon previous section and introduces the system model of WIPT before reviewing various signal processing techniques to achieve the best R-E tradeoff of WIPT. Section \ref{methodology_section} discusses and contrasts the pros and cons of two major design methodologies to design WPT and WIPT, namely the model-and-optimize approach and the learning approach. Section \ref{WPT_enabler_section} discusses how wireless power will enable new and emerging scenarios and applications in future wireless powered networks such wireless-powered computing and sensing. Section \ref{conclusions} concludes the paper and discusses future works.

\emph{Notation:} In this paper, scalars are denoted by italic letters. Boldface lower- and upper-case letters denote vectors and matrices, respectively. $\mathbb{C}^{M\times N}$ denotes the space of $M\times N$ complex matrices. $j$ denotes the imaginary unit, i.e., $j^2=-1$. $\mathbb{E}[\cdot]$ denotes statistical expectation and $\Re\{\cdot\}$ represents the real part of a complex number. $\mathbf{I}_M$ denotes an $M\times M$ identity matrix and $\mathbf{0}$ denotes an all-zero vector/matrix. $|.|$ and $\left\|.\right\|$ refer to the absolute value of a scalar and the 2-norm of a vector. For an arbitrary-size matrix $\mathbf{A}$,  its complex conjugate, transpose, Hermitian transpose, and Frobenius  norm are respectively denoted as $\mathbf A^*$, $\mathbf{A}^{T}$, $\mathbf{A}^{H}$, and $\|\mathbf{A}\|_F$. $[\mathbf A]_{im}$ denotes the $(i,m)$th element of matrix $\mathbf A$. For a square Hermitian matrix $\mathbf{S}$, $\mathrm{Tr}(\mathbf{S})$ denotes its trace, while $\lambda_{\max}(\mathbf S)$ and $\mathbf v_{\max}(\mathbf S)$ denote its largest eigenvalue and the corresponding eigenvector, respectively. In the context of random variables, i.i.d.\ stands for independent and identically distributed. The distribution of a Circularly Symmetric Complex Gaussian (CSCG) random variable with zero-mean and variance $\sigma^2$ is denoted by $\mathcal{CN}(0,\sigma^2)$; hence with the real/imaginary part distributed as $\mathcal{N}(0,\sigma^2/2)$. $\sim$ stands for ``distributed as''. We use the notation $\textnormal{sinc}\left(t\right)=\frac{\sin\left(\pi t\right)}{\pi t}$. $\mathrm{diag}(\mathbf{A}_1, \ldots, \mathbf{A}_N)$ refers to a block diagonal matrix with blocks being $\mathbf{A}_1$, \dots, $\mathbf{A}_N$.

\section{Wireless Power Transfer: Key Technologies to Increase Efficiency}\label{WPT_section}

\par In the past decade, there has been a significant interest in WPT and ambient wireless energy harvesting (WEH) for low-power (e.g., from $\mu$W to a few W) delivery over distances of a few m to hundreds of m \cite{Falkenstein:2012,Popovic:2012}, due to the increasing need to build reliable and convenient wireless power systems for remotely energizing low-power devices, such as sensors, RFID tags, and consumer electronics \cite{Visser:2013,Popovic:2013,Clerckx:2018}. 

\begin{figure}%[!hhh]
   \centerline{\includegraphics[width=0.9\columnwidth]{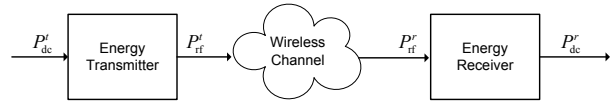}}
	\caption{The block diagram of a generic WPT system \cite{Zeng:2017}.}\label{F:BasicArchitecture}
\end{figure}

\par Fig. \ref{F:BasicArchitecture} shows a generic WPT system, which consists of an RF ET and an ER. A DC power source is used to generate a signal which is upconverted to the RF domain at the ET, then transmitted over the air, and collected at an ER in the RF domain before being converted to DC. The ER is made of an antenna combined with a rectifier (rectenna) and a power management unit (PMU). Since the majority of the electronics requires a DC power source, a rectifier is required to convert RF to DC. The recovered DC power then either supplies a low power device directly, or is stored in a battery or a super capacitor for high power low duty-cycle operations. The recovered DC power can also be managed by a DC-to-DC converter before being stored. In WPT, the entire link, including ET and ER, of Fig. \ref{F:BasicArchitecture} can be fully optimized. Therefore, in contrast to ambient WEH, WPT offers full control of the design and room to enhance the end-to-end power transfer efficiency $e$
\begin{align}\label{e_equation}
e=\frac{P_{\dc}^r}{P_{\dc}^t}=\underbrace{\frac{P_{\rf}^t}{P_{\dc}^t}}_{e_1}\underbrace{\frac{P_{\rf}^r}{P_{\rf}^t}}_{e_2}\underbrace{\frac{P_{\dc}^r}{P_{\rf}^r}}_{e_3},
\end{align}
where $e_1$, $e_2$, and $e_3$ denote the DC-to-RF, RF-to-RF, and RF-to-DC power conversion/transmission efficiency, respectively. 

\subsection{Signal and System Model}\label{WPT_system_model}

\par We consider a single-user point-to-point MIMO WPT system in a general multipath environment. This setup is referred to as ``WPT with co-located antennas one-to-one'' in Fig. \ref{WPT_figure}. The ET is equipped with $M$ antennas that transmit power to a ER equipped with $Q$ receive antennas. We consider the general setup of a multi-subband transmission (with a single subband being a special case) employing $N$ orthogonal subbands where the $n^{\textnormal{th}}$ subband has carrier frequency $f_n$ and all subbands employ equal bandwidth $f_{\mathrm{w}}$, $n=0,...,N-1$. The carrier frequencies (also called tones) are evenly spaced such that $f_n=f_0+n \Delta_f$ with the inter-carrier frequency spacing $\Delta_f$ (with $f_{\mathrm{w}}\leq \Delta_f$).

\par The WPT signal transmitted on antenna $m$, $x_{\rf,m}(t)$, is a multi-carrier modulated waveform with frequencies $f_n$, $n=0,...,N-1$, carrying independent symbols on subband $n=0,...,N-1$. The input WPT signal at time $t$ to the HPA of antenna $m=1,...,M$ is given by
\begin{equation}
x_{\In,m}(t)=\sqrt{2}\Re\left\{\sum_{n=0}^{N-1} x_{m,n}(t) e^{j 2\pi f_n t}\right\}\label{SWIPT_WF}
\end{equation}
with the baseband equivalent signal $x_{m,n}(t)$ given by
\begin{equation}
x_{m,n}(t)=\sum_{k=-\infty}^{\infty} x_{m,n,k}\: \sinc(f_{\mathrm{w}} t-k)
\end{equation}
where $x_{m,n,k}$ denotes the complex-valued power carrying symbol at time index $k$, modeled as a random variable generated in an i.i.d. fashion. $x_{m,n}(t)$ has bandwidth $[-f_{\mathrm{w}}/2,f_{\mathrm{w}}/2]$. For the special case of unmodulated WPT, $x_{m,n}(t)$ is constant across $t$, i.e., $x_{m,n}(t)=x_{m,n}=s_{m,n}e^{j \phi_{m,n}}$, $\forall t$. In this case, $x_{\In,m}(t)$ is a summation of $N$ sinewaves inter-separated by $\Delta_f$ Hz, and hence essentially occupies zero bandwidth.

\par The power at the transmitter before HPA is written as  
\begin{align}
P_{\dc}^t=\sum_{n=0}^{N-1} \Tr(\mathbf Q_n) =\Tr(\mathbf Q),\label{eq:prft}
\end{align}
with $\mathbf{Q}=\textnormal{diag}\left\{\mathbf{Q}_0,...,\mathbf{Q}_{N-1}\right\}$ where the positive semidefinite input covariance matrix $\mathbf Q_n$ at subband $n$ is defined as $\mathbf Q_n \triangleq \mathbb {E} \left[\mathbf x_n(t) \mathbf x_n^H(t) \right]\in \mathbb{C}^{M\times M}$ and $\mathbf{x}_{n}(t)\triangleq\big[x_{1,n}(t),...,x_{M,n}(t)\big]^T$ denotes the signal vector across the $M$ antennas in subband $n$. For convenience, we also define $P_n=\Tr(\mathbf Q_n)$ as the transmit power in subband $n$, such that $P_{\dc}^t=\sum_{n=0}^{N-1}P_n$.

\par The input signal $x_{\In,m}(t)$ on each antenna $m$ is then amplified by a HPA and filtered using a band-pass filter (BPF) into the transmit WPT signal $x_{\rf,m}(t)$ 
\begin{equation}
x_{\rf,m}(t)=\sqrt{2}\Re\left\{\sum_{n=0}^{N-1} x_{\rf,m,n}(t) e^{j 2\pi f_n t}\right\},\label{SWIPT__HPA}
\end{equation}
with
\begin{equation}
x_{\rf,m,n}(t)=\sum_{k=-\infty}^{\infty} x_{\rf,m,n,k}\: \sinc(f_{\mathrm{w}} t-k).
\end{equation}
Realistically, the relationship between $x_{\In,m}(t)$ and $x_{\rf,m}(t)$ is nonlinear and accounts for coupling across frequencies as well as magnitude and phase distortions induced by the HPA and BPF. The transmit WPT signal $x_{\rf,m}(t)$ is then transmitted over the air by antenna $m$. The total average transmit power is expressed as $P_{\rf}^t=\sum_{m=1}^{M} \mathbb{E}[x_{\rf,m}(t)^2]$ and is subject to the constraint $P_{\rf}^t\leq P$. 

\par The transmit WPT signal propagates through a multipath channel, characterized by $L$ paths. Let $\tau_l$ and $\alpha_l$ be the delay and amplitude gain of the $l^{\textnormal{th}}$ path, respectively. Further, denote by $\zeta_{q,m,n,l}$ the phase shift of the $l^{\textnormal{th}}$ path between transmit antenna $m$ and receive antenna $q$ for subband $n$. The signal received at antenna $q$ ($q=1,...,Q$) from transmit antenna $m$ can be expressed as
\begin{align}
\!y_{\rf,q,m}(t)\!&=\!\sqrt{2}\Re\Bigg\{\sum_{l=0}^{L-1}\sum_{n=0}^{N-1} \alpha_l x_{\rf,m,n}(t-\tau_l)\Bigg. \nonumber\\
&\hspace{2.3cm}\Bigg.e^{j 2\pi f_n (t-\tau_l)+\zeta_{q,m,n,l}}\Bigg\}\!,\nonumber\\
&\approx\!\sqrt{2}\Re\left\{\sum_{n=0}^{N-1} h_{q,m,n}x_{\rf,m,n}(t) e^{j 2\pi f_n t}\right\}.\label{received_signal_ant_m}
\end{align}
We have assumed $\max_{l\neq l'}\left|\tau_l-\tau_{l'}\right|<1/f_{\mathrm{w}}$ so that, for each subband, $x_{\rf,n,m}(t)$ are narrowband signals, thus $x_{\rf,m,n}(t-\tau_l)=x_{\rf,m,n}(t)$, $\forall l$. Variable $h_{q,m,n}=\!\sum_{l=0}^{L-1}\alpha_l e^{j(-2\pi f_n\tau_l+\zeta_{q,m,n,l})}$ is the baseband channel frequency response between transmit antenna $m$ and receive antenna $q$ at frequency $f_n$.

\par The total signal and noise received at antenna $q$ is the superposition of the signals received from all $M$ transmit antennas, i.e.,
\begin{equation}
y_{\rf,q}(t) =  \sqrt{2}  \Re \left\{  \sum_{n=0}^{N-1} \mathbf{h}_{q,n} \mathbf x_{\rf,n}(t) e^{j 2\pi f_n t}  \right\}+w_{\textnormal{A},q}(t), \label{eq:yit}
\end{equation}
where $w_{\textnormal{A},q}(t)$ is the antenna noise, $\mathbf{h}_{q,n}\!\triangleq\!\big[h_{q,1,n},...,h_{q,M,n}\big]$ denotes the channel vector from the $M$ transmit antennas to receive antenna $i$, and $\mathbf{x}_{\rf,n}(t)\triangleq\big[x_{\rf,1,n}(t),...,x_{\rf,M,n}(t)\big]^T$. 

\par Ignoring the noise power, the total RF power received by all $Q$ antennas of the receiver can be expressed as
\begin{align}
P_{\rf}^r&=\sum_{q=1}^{Q} \mathbb{E} \left[y_{\rf,q}(t)^2\right]=\sum_{q=1}^{Q} \sum_{n=0}^{N-1} \mathbb{E} \left[|\mathbf h_{q,n} \mathbf x_{\rf,n}(t)|^2 \right].
\end{align}

\par Finally, unless stated explicitly, we assume perfect Channel State Information at the Transmitter (CSIT).
\par Next, the output DC power $P_{\dc}^r$ depends on the exact ER architecture to be discussed in the next sections. 

\begin{remark}
The system model is written using a general form assuming that complex-valued symbols are random variables and occupy a non-zero bandwidth. This is used to ease and harmonize the system model with WIPT discussed in Section \ref{WIPT_section}. It is nevertheless to be noted that if the aim is to design WPT without any consideration for communications, one would strictly speaking not need complex-valued symbols to be random, and one could assume them deterministic (with zero bandwidth), therefore transforming the above system model into unmodulated WPT with multisine waveforms (with $N$ sinewaves) transmitted from each antenna.
\end{remark} 

\subsection{Transmitter (HPA) Nonlinearity}

We here discuss the modeling of the HPA. The HPA input-output relationship is realistically nonlinear, though this source of nonlinearity is commonly ignored.  

\subsubsection{Linear HPA}: If we ignore the HPA nonlinearity and assume the relationship between $x_{\In,m}(t)$ and $x_{\rf,m}(t)$ is linear such that $x_{\rf,m}(t)= G x_{\In,m}(t)$ with $G$ the amplification gain, and taking $G=1$ for simplicity of exposure, we have $P_{\rf}^t=\sum_{m=1}^{M} \mathbb{E}[x_{\rf,m}(t)^2]= P_{\dc}^t$. In other words, referring to Fig. \ref{F:BasicArchitecture}, the DC-to-RF conversion efficiency $e_1$ is equal to 1.
More realistically, under the linear regime of the HPA, $P_{\rf}^t=e_1 P_{\dc}^t$ with $e_1$ a constant strictly smaller than 1 and independent of the input signal (but whose exact value depends on the HPA technology). 
\par The total RF power received by all $Q$ antennas can then be expressed more easily as
\begin{align}
P_{\rf}^r&=e_1 \sum_{n=0}^{N-1} \mathrm{Tr}\left(\mathbf H_n^H \mathbf H_n \mathbf Q_n\right),\label{eq:prfr}
\end{align}
where $\mathbf H_n \triangleq \left [\mathbf h_{1,n}^H,\cdots, \mathbf h_{Q,n}^H\right]^H\in \mathbb{C}^{Q\times M}$ denotes the MIMO channel matrix from the $M$ transmit antennas to the $Q$ receive antennas at subband $n$.

\subsubsection{Nonlinear HPA}
\begin{figure}%[t!]
\centering
\includegraphics[width=0.9\columnwidth]{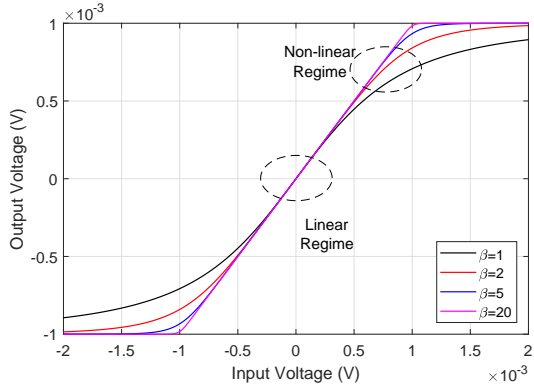}
\caption{Input-output voltage characteristics for solid state HPA with $A_{\mathrm{s}}=10^{-3}$V and $G=1$.}
\label{SSPA}
\end{figure}

: The HPA has a nonlinear characteristics that distorts its input signal and makes it challenging to analyze. Indeed, real HPAs do not exhibit a pure linear behavior and $x_{\rf,m}(t)= f_{\mathrm{HPA}}\left(x_{\In,m}(t)\right)$ where $f_{\mathrm{HPA}}$ is a nonlinear function, which leads to $P_{\rf}^t=e_1(x_{\In,m}(t)) P_{\dc}^t$, i.e.\ $e_1(x_{\In,m}(t))$ is itself a nonlinear function of $x_{\In,m}(t)$. A common model for solid state HPA \cite{Rapp:1991,Dehos:2007} is written as 
\begin{equation}\label{HPA_NL_model}
f_{\mathrm{HPA}}(x_{\In,m}(t))=\frac{G }{\left(1+\left(\frac{G\left|x_{\In,m}(t)\right|}{A_\mathrm{s}}\right)^{2\beta}\right)^{\frac{1}{2\beta}}}x_{\In,m}(t)
\end{equation}
where $A_\mathrm{s}$ is the output saturation voltage, $G$ is the amplification gain, and $\beta$ represents the smoothness of the transition from the linear regime to the saturation. In Fig. \ref{SSPA}, \eqref{HPA_NL_model} is illustrated for $A_{\mathrm{s}}=10^{-3}$V and $G=1$. The HPA would operate in the linear regime if the input voltage is significantly smaller than $A_s$, and would operate in the nonlinear regime (leading to saturation) otherwise. 

%{\color{blue} For multi-carrier transmission, aside the effect of phase and amplitude distortions, we need to consider the coupling across frequencies.
%The input-output relationship of the HPA-BPF is such that 
%\begin{multline}
%\left[x_{\rf,m,0,k},\ldots,x_{\rf,m,N-1,k}\right]\\=f_{\mathrm{HPA-BPF}}\left([x_{m,0,k},\ldots,x_{m,N-1,k}]\right)
%\end{multline}
%where $f_{\mathrm{HPA-BPF}}$ is a nonlinear function accounting for coupling across frequencies as well as magnitude and phase distortions.}

\subsection{Energy Receiver Nonlinearity and Architecture}\label{EH_section}

We here discuss the architecture and related nonlinearity of single-antenna and multi-antenna ER. 

\subsubsection{Single-Antenna Energy Receiver} 
\begin{figure}%[t!]
\centering
\includegraphics[width=0.8\columnwidth]{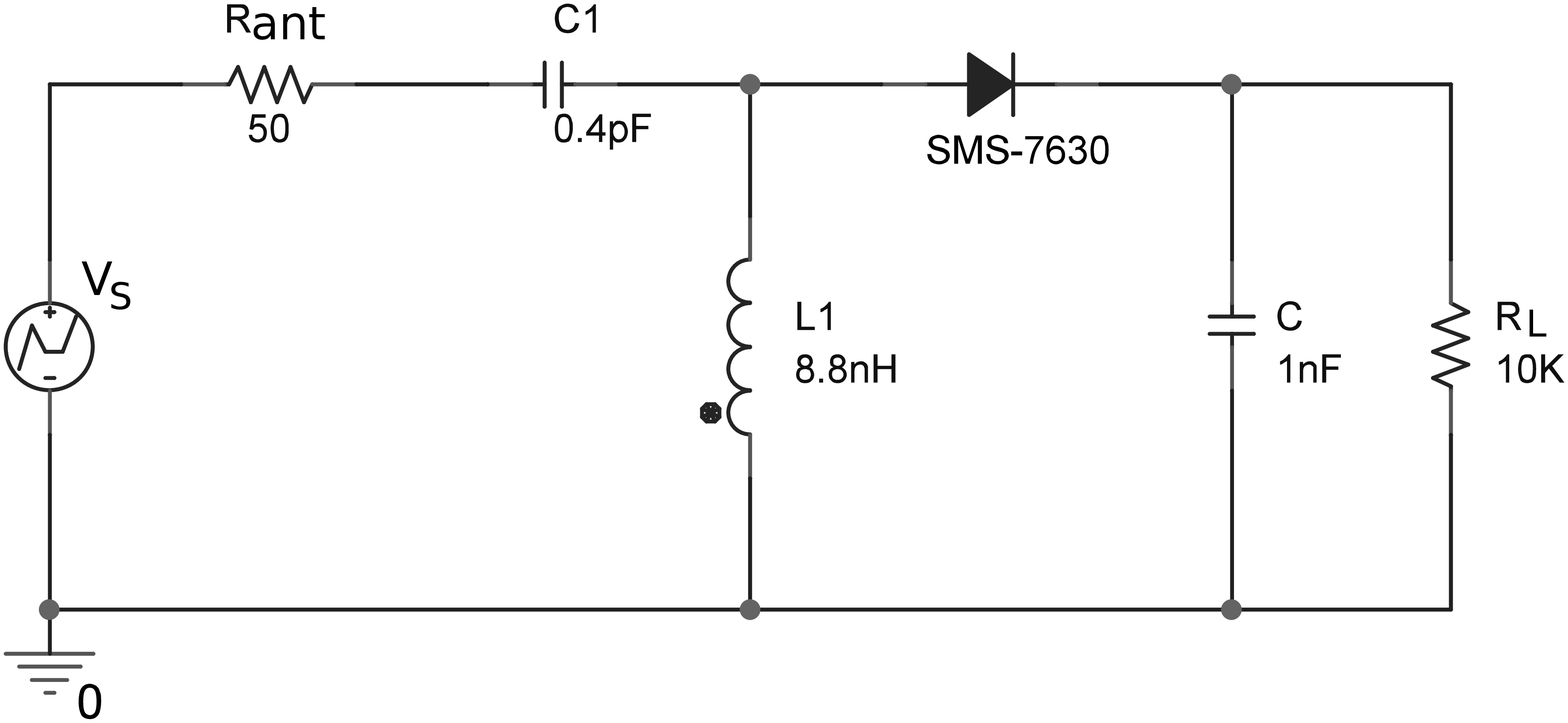}
\caption{Single series rectifier designed for an average RF input power of -20dBm (10$\mu$W) at 2.45GHz \cite{Clerckx:2018c}. $v_{\mathrm{s}}$ is the voltage source of the antenna. R1 models the antenna impedance. C1 and L1 form the matching network. SMS-7630 refers to the type of Schottky diode. C and R\textsubscript{L} form the low-pass filter with R\textsubscript{L} being the output load.}
\label{TD_schematic}
\end{figure}

The key building block of the ER is the rectenna. A rectenna harvests electromagnetic energy, then rectifies and filters it using a low pass filter. The rectenna can be optimized for the specific operating frequencies, input power level and input waveforms. Various rectifier technologies (including the popular Schottky diodes) and topologies (with single and multiple diode rectifier) have been studied \cite{OptBehaviour,Valenta:2014,Costanzo:2016}. The simplest form of rectifier, so-called single series rectifier, is illustrated by the circuit in Fig. \ref{TD_schematic} \cite{Clerckx:2018c}. It is made of a matching network (to match the antenna impedance to the rectifier input impedance) followed by a single diode and a low-pass filter. This circuit was designed for 10$\mu$W input power at 2.45GHz. 

\par
\begin{figure}%[!hhh]
\centerline{\includegraphics[width=0.9\columnwidth]{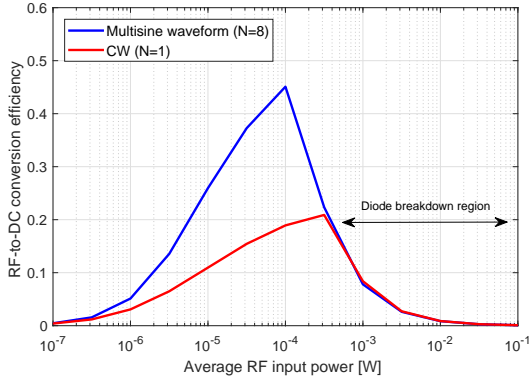}}
  \caption{RF-to-DC conversion efficiency $e_3$ vs average RF input power $P_{\rf}^{r}$ with rectifier from Fig. \ref{TD_schematic} obtained from circuit simulations \cite{Clerckx:2018c}. The input signal is a CW at 5.18 GHz and rectifier is designed for -20dBm input power.}
  \label{Pdc_Pin}
\end{figure}

 Using circuit simulations and the single-series rectifier from Fig. \ref{TD_schematic}, Fig. \ref{Pdc_Pin} illustrates the dependency of the RF-to-DC conversion efficiency $e_3$ to the average signal power and shape at the input of the rectifier, when continuous wave (CW), i.e.\ a single sinewave , and a multisine waveform (with $N=8$ equispaced frequencies) are used for excitation \cite{Clerckx:2018c}. Those two excitations have the same average RF input power $P_{\rf}^r$, but their shape is different.  
\par We note that $e_3$ is particularly low at low input power for both types of excitations. This is due to the rectifier sensitivity with the diode not being easily turned on at low input power. Nevertheless, the multisine waveform manages to boost $e_3$ in the low power regime much better than CW. Importantly, for a given waveform, be it CW or multisine, $e_3$ increases with $P_{\rf}^r$ in the normal region of operation of the rectifier, namely whenever the diode is not in the breakdown region. Beyond a few hundreds of $\mu$W input power, irrespectively of the input signal shape, the output DC power saturates and $e_3$ suddenly significantly drops when the rectifier enters the diode breakdown region\footnote{The diode SMS-7630 becomes reverse biased at $P_{\rf}^r \approx 500\mu$W to 1mW for CW. To operate beyond such input power, multiple diode rectifier is preferred to avoid the saturation problem \cite{OptBehaviour,Costanzo:2016,Sun:2013}.}, which is not the intended region of operation of the rectifier.
\par The key observation of Fig. \ref{Pdc_Pin} is that due to the EH nonlinearity, $e_3$ is clearly not a constant, but depends on 1) the input power level and 2) the shape of the input signal $y_{\rf}$ \cite{Trotter:2009,Boaventura:2011,Collado:2014,Valenta:2015}. Mathematically, this is reflected by the fact that the output DC voltage $v_{\out}=f_{\mathrm{EH}}\left(y_{\rf}(t)\right)$ where $f_{\mathrm{EH}}\left(y_{\rf}(t)\right)$ is a nonlinear function of $y_{\rf}(t)$, which has as consequence that $P_{\dc}^r=e_3\left(y_{\rf}(t)\right)P_{\rf}^r$, i.e. $e_3$ is not a constant but rather a nonlinear function of the input signal to the rectenna. Note the importance of writing $f_{\mathrm{EH}}\left(y_{\rf}(t)\right)$ and $e_3\left(y_{\rf}(t)\right)$ instead of simply $f_{\mathrm{EH}}\left(P_{\rf}^r\right)$ and $e_3\left(P_{\rf}^r\right)$. $f_{\mathrm{EH}}$ and $e_3$ are not simply a nonlinear function of the average RF input power $P_{\rf}^r$ of the input waveform $y_{\rf}(t)$, but also of the shape of this input waveform! 

\begin{figure}%[!hhh]
   \centerline{\includegraphics[width=0.9\columnwidth]{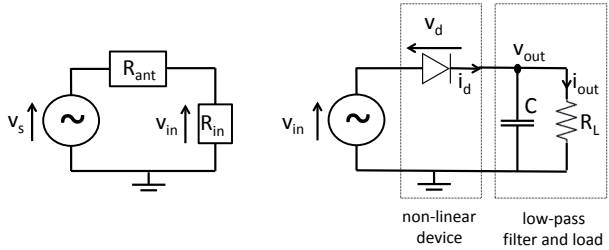}}
  \caption{Antenna equivalent circuit (left) and a single diode rectifier (right) \cite{Clerckx:2016b}. The rectifier comprises a non-linear device (diode) and a low-pass filter (consisting of a capacitor C and a load R\textsubscript{L}).}
  \label{antenna_model}
\end{figure}

\par The first and only model available in the WPT signal design literature that captures power and shape dependency on output DC power was derived in \cite{Clerckx:2015,Clerckx:2016b,Clerckx:2018b}, and is briefly summarized in the sequel. Let us abstract the rectifier in Fig. \ref{TD_schematic} into the simplified representation in Fig. \ref{antenna_model}. We consider for simplicity a rectifier with input impedance $R_{\In}$ composed of a single series diode followed by a low-pass filter with a load. We consider this setup as it is the simplest rectifier configuration\footnote{The model is not limited to a single series diode but also holds for more general rectifiers with many diodes as per \cite{Clerckx:2017}.}. As per the system model, the RF signal $y_{\rf}(t)$ impinging on the receive antenna has an average power $P_{\rf}^r$. The receive antenna is assumed lossless and modeled as an equivalent voltage source $v_{\mathrm{s}}(t)$ in series with an impedance $R_{\ant}$ as shown in Fig. \ref{antenna_model}. With perfect matching ($R_{\In} = R_{\ant}$), the input voltage of the rectifier $v_{\In}(t)$ can be related to the received signal $y_{\rf}(t)$ by $v_{\In}(t)\!=\!y_{\rf}(t)\sqrt{ R_{\ant}}$. A rectifier is always made of a nonlinear rectifying component such as diode followed by a low pass filter with load as shown in Fig. \ref{antenna_model}.
\par The current $i_d(t)$ flowing through an ideal diode (neglecting its series resistance) relates
to the voltage drop across the diode $v_{\mathrm{d}}(t)=v_{\In}(t)-v_{\out}(t)$ as
\begin{equation}
i_d(t)=i_{\mathrm{s}} \big(e^{\frac{v_{\mathrm{d}}(t)}{n v_{{\mathrm{t}}}}}-1 \big),\label{id}
\end{equation}
where $i_{{\mathrm{s}}}$ is the reverse bias saturation current, $v_{{\mathrm{t}}}$ is the thermal voltage, $n$ is the ideality factor (assumed equal to 1.05). 
\par Taking the polynomial (Taylor) expansion of the diode I-V characteristics $i_d(t)$, truncating it at the $n_o^{th}$ order, making use of some physical assumptions on an ideal low-pass filter that removes the non-DC components in $i_d(t)$ and the rectenna output voltage $v_{\out}(t)$, the output DC voltage of the rectifier $v_{\out}$ can be approximated as the following nonlinear function of $y_{\rf}(t)$
\begin{equation}\label{vout_def}
v_{\out}=f_{\mathrm{EH}}\left(y_{\rf}(t)\right)=\sum_{i \hspace{0.1cm}\textnormal{even}, i\geq 2}^{n_o} \beta_i \mathbb{E}\left[y_{\rf}(t)^i\right]
\end{equation}
where $\beta_i=\frac{R_{\ant}^{i/2}}{i!\left(n v_{\mathrm{t}}\right)^{(i-1)}}$ \cite{Clerckx:2016b,Huang:2017}. The operator $\mathbb{E}[\cdot]$ in \eqref{vout_def} has the effect of taking the DC component of the diode current $i_{\mathrm{d}}(t)$ but also averaging over the potential randomness carried by the input signal $y_{\rf}(t)$. Consequently, the harvested DC power $P_{\dc}^r$ of the single-antenna receiver is then given by
\begin{equation}\label{diode_current_power}
P_{\dc}^r=\frac{v_{\out}^2}{R_{\mathrm{L}}}.
\end{equation}

\par We clearly see that $f_{\mathrm{EH}}\left(y_{\rf}(t)\right)$ is a nonlinear function of $y_{\rf}(t)$. Specifically, it is a function of the input signal average power $P_{\rf}^r=\mathbb{E}\left[y_{\rf}(t)^2\right]$ (i.e. the second moment of $y_{\rf}(t)$) but also of its higher order moments $\mathbb{E}\left[y_{\rf}(t)^i\right]$ for $i$ even and $i>2$. This dependency on the second and higher order moments of $y_{\rf}(t)$ explains why multisine outperforms CW in Fig.\ \ref{Pdc_Pin} \cite{Clerckx:2016b}, but also explains why $e_3$ is an increasing function of $P_{\rf}^r$. Indeed, due to the convexity of the I-V characteristics and the polynomial expansion, using Jensen's inequality, we have  
\begin{equation}\label{Jensen}
\mathbb{E}\left[y_{\rf}(t)^i\right]\geq (\mathbb{E}\left[y_{\rf}(t)^{2}\right])^{\frac{i}{2}} = \left(P_{\rf}^r\right)^{\frac{i}{2}}
\end{equation}
for $i$ even and $i\geq 2$, so that
\begin{equation}\label{effect_Pin}
v_{\out}\geq \sum_{i \hspace{0.1cm}\textnormal{even}, i\geq 2}^{n_o} \beta_i \left(P_{\rf}^r\right)^{\frac{i}{2}}.
\end{equation}
Taking for instance as $y_{\rf}(t)$ a multisine waveform with average power $P_{\rf}^r$ uniformly distributed across the $N$ sinewaves, we can easily show that $\mathbb{E}\left[y_{\rf}(t)^4\right]$ scales proportionally to $N \left(P_{\rf}^r\right)^{2}$, therefore demonstrating that $\mathbb{E}\left[y_{\rf}(t)^4\right]>\left(P_{\rf}^r\right)^{2}$ for sufficiently large $N$ and explaining mathematically why multisine (and other types of signals) can outperform CW ($N=1$) \cite{Clerckx:2016b}. We can draw two crucial observations from relationships \eqref{Jensen} and \eqref{effect_Pin}, respectively. 
\begin{observation}\label{higher_order} Relationship \eqref{Jensen} highlights the key role of choosing input signals with large $\mathbb{E}\left[y_{\rf}(t)^i\right]$. Two input signals may indeed have the same $\mathbb{E}\left[y_{\rf}(t)^{2}\right]=P_{\rf}^r$ but very different $\mathbb{E}\left[y_{\rf}(t)^4\right]$. This explains mathematically the dependence of $e_3$ (and $P_{\dc}^r$) on the shape of the input signal in Fig.\ \ref{Pdc_Pin}.  
\end{observation}
\begin{observation}\label{higher_input_power} The lower bound \eqref{effect_Pin} highlights that $e_3$ increases with $P_{\rf}^r$. This explains mathematically the dependence of $e_3$ on the input power level in Fig.\ \ref{Pdc_Pin} for the practical operation regime of the rectifier (not in breakdown), and highlights that the strategy that maximizes $P_{\rf}^r$ \textit{does not} maximize $P_{\dc}^r$, but only maximizes a lower bound on $P_{\dc}^r$.
\end{observation}

\par Those two observations highlight that a signal theory, design and processing of basic building blocks of wireless powered networks such as modulation, waveform, and input distribution, are influenced by the EH nonlinearity, and motivates efficient signal and system designs that leverage the EH nonlinearity. The crucial role played by this EH nonlinearity in the signal designs and evaluations of WPT, SWIPT, and WPBC was first highlighted in \cite{Clerckx:2016b}, \cite{Clerckx:2018b}, and \cite{Clerckx:2017b}, respectively.

\begin{remark} Slightly different formulations of the above EH model are available in \cite{Clerckx:2016b,Huang:2017,Moghadam:2017,Varasteh:2019_ICASSP,Morsi:2017}, where the output is expressed in terms of DC current instead of voltage, or where operator $\mathbb{E}\left[.\right]$ is applied without performing the polynomial expansion.
\end{remark}

\begin{remark} Other models for $P_{\dc}^r$ are available in the literature as discussed in greater details in \cite{Clerckx:2019}. Those models either assume $e_3$ constant (so-called linear model \cite{Zeng:2017}) or only capture the dependency of $e_3$ on $\mathbb{E}\left[y_{\rf}(t)^{2}\right]=P_{\rf}^r$ (e.g. so-called saturation nonlinear model \cite{Boshkovska:2015}). The linear model is very inaccurate \cite{Clerckx:2016b,Kim:2018}. The saturation model is more accurate since it is based on curve fitting, but does not capture the dependency of the rectification process on the shape of the input signal and arguably over-emphasizes the importance of saturation in the EH. Saturation is unlikely a major problem in wireless powered networks since the typical input RF power levels (below $100\mu$W) are smaller than the saturation level, as demonstrated by over the air measurements with various types of signals in \cite{Kim:2018,Kim:2020} (and also in Fig. \ref{cdf_8ant} and \ref{fit_all} below). Moreover, if saturation happens to lead to a significant performance loss, it implies that the rectifier was not designed carefully enough for the expected range of input power levels. Saturation can indeed be avoided by a proper design of the rectifier \cite{OptBehaviour,Costanzo:2016,Sun:2013,Clerckx:2018b,Clerckx:2019}. The interested reader is referred to \cite{Clerckx:2019,Clerckx:2018b} and references therein for more discussions on EH models.
\end{remark}

\subsubsection{Multi-Antenna Energy Receiver} 
\begin{figure}%[!hhh]
	\centerline{\includegraphics[width=0.9\columnwidth]{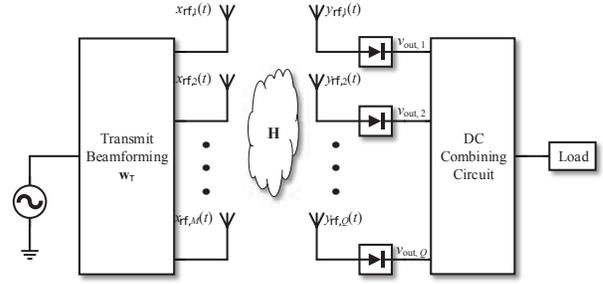}}
	\caption{Schematic of the MIMO WPT system with DC combining at the receiver \cite{Shen:2020a}.}
	\label{DC_Combining}
\end{figure}

\begin{figure}%[!hhh]
	\centerline{\includegraphics[width=0.9\columnwidth]{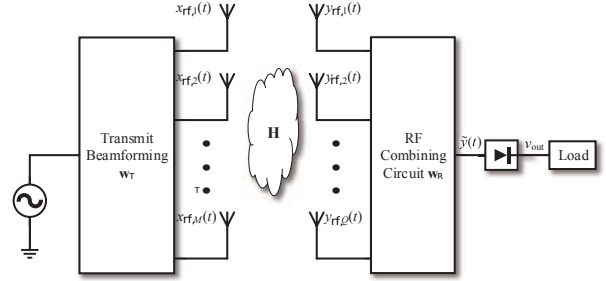}}
	\caption{Schematic of the MIMO WPT system with RF combining at the receiver. \cite{Shen:2020a}.}
	\label{RF_Combining}
\end{figure}

Two main combining strategies exist, namely DC combining and RF combining, as illustrated in Fig.\ \ref{DC_Combining} and Fig.\ \ref{RF_Combining}, respectively \cite{Shen:2020a}. In DC combining, each receive antenna is connected to a rectifier and the number of rectifiers increases with the number of receive antennas. However, in RF combining, the RF signals from all receive antennas are first combined in the RF domain before being fed to a single rectifier used to rectify the combined RF signal. A combination of those two architectures is also possible, as well as other variants based on the use of power splitters and power combiners \cite{Ma:2019}.

\par In the DC combiner architecture, 
\begin{equation}\label{diode_current_power_DC}
P_{\dc}^r=\sum_{q=1}^Q\frac{v_{\out,q}^2}{R_{\mathrm{L}}},
\end{equation}
where $v_{\out,q}=f_{\mathrm{EH}}\left(y_{\rf,q}(t)\right)$ is the output DC voltage of the rectifier connected to receive antenna $q$. 
\par In the RF combiner architecture, a frequency-dependent analogue combiner $\mathbf{w}_{\mathrm{R},n}$ is applied to the received signals \eqref{eq:yit} such that the received signal after combining fed to the single rectifier is given by
\begin{equation}
\tilde{y}(t) =  \sqrt{2}  \Re \left\{  \sum_{n=0}^{N-1} \mathbf{w}_{\mathrm{R},n}^H \mathbf H_n \mathbf x_{\rf,n}(t) e^{j 2\pi f_n t}  \right\}+\tilde{w}_{\textnormal{A}}(t),
\end{equation}
where $\tilde{w}_{\textnormal{A}}$ is the effective combined noise. Note that in practice, it may be difficult to design frequency-dependent combiner, in which case $\mathbf{w}_{\mathrm{R}}=\mathbf{w}_{\mathrm{R},n}$ is constant across frequency. The combiner is subject to the constraint $\left\|\mathbf{w}_{\mathrm{R}}\right\|^2\leq 1$ originating from the fact that since the RF combining circuit is passive, the output power of the RF combining circuit should be no larger than its input power.
Additionally, the combiner may be subject to constant modulus constraint so as to be implemented by $Q$ phase shifters of the form 
\begin{equation}\label{phase_shifters}
\mathbf{w}_{\mathrm{R}}=\frac{1}{\sqrt{Q}}\left[e^{-j\theta_1},e^{-j\theta_2},\ldots,e^{-j\theta_Q}\right]^T,
\end{equation}
where $\theta_q$ denotes the $q$th phase shift for $q=1,\ldots,Q$. Finally, the output DC power is given by $P_{\dc}^r=\frac{v_{\out}^2}{R_{\mathrm{L}}}$ where $v_{\out}=f_{\mathrm{EH}}\left(\tilde{y}(t)\right)$.


\subsection{End-to-End Efficiency, Energy Maximization and Problem Formulation}

\par A major and interesting technical challenge in WPT system design is that the maximization of $e$ is not achieved by maximizing  $e_1$, $e_2$, $e_3$ independently from each other. This is because $e_1$, $e_2$, $e_3$ are coupled due to the aforementioned nonlinearities, especially at practical input RF power range 1 $\mu$W -1 mW. Indeed, since $e_3$ is a function of the input signal shape and power to the rectifier and therefore a function of the transmit signal and the wireless channel state. Similarly, $e_2$ depends on the transmit signal and the channel state and so is $e_1$, since it is a function of the HPA nonlinearity. 

\par One possible problem formulation is therefore to find the signaling strategies that maximizes $e$, which writes as 
\begin{align}%\label{SWIPT_opt_problem}
\max_{p(\mathbf{x}_0,...,\mathbf{x}_{N-1})} \hspace{0.3cm}& e(\mathbf{x}_0,...,\mathbf{x}_{N-1}) \label{e_max}\\
\mathrm{subject\,\,to} \hspace{0.3cm} & P_{\rf}^t\leq P.
\end{align}
where maximization is here performed over the input distributions $p(\mathbf{x}_0,...,\mathbf{x}_{N-1})$ that satisfies the average transmit power constraint $P_{\rf}^t\leq P$. An alternative formulation that is more common consists in maximizing the harvested DC output power
\begin{align}%\label{SWIPT_opt_problem}
\max_{p(\mathbf{x}_0,...,\mathbf{x}_{N-1})} \hspace{0.3cm}& P_{\dc}^r(\mathbf{x}_0,...,\mathbf{x}_{N-1}) \label{Pdc}\\
\mathrm{subject\,\,to} \hspace{0.3cm} &P_{\rf}^t\leq P,
\end{align}
In those two formulations, if the ER is equipped with an RF combiner, the optimization would have to be additionally performed over $\mathbf{w}_{\mathrm{R}}$ subject to the constraint $\left\|\mathbf{w}_{\mathrm{R}}\right\|^2\leq 1$ or structure as in \eqref{phase_shifters}. Note that in the event power bearing symbols are deterministic, $\max_{p(\mathbf{x}_0,...,\mathbf{x}_{N-1})}$ can be replaced by $\max_{\mathbf{x}_0,...,\mathbf{x}_{N-1}}$.
\par Note that those two formulations are not equivalent. The main difference is that \eqref{e_max} specifically accounts for $e_1$, while \eqref{Pdc} does not. To account for $e_1$ and HPA efficiency, additional constraints can be added to problem \eqref{Pdc} for instance in the form of peak-to-average power (PAPR) constraints \cite{Clerckx:2016b}.

\subsection{Signal Processing Techniques for Single-User WPT}\label{subsection_SU_WPT}

\par In this section, we review recent signal processing techniques developed to tackle the challenges of WPT, increase its efficiency and its range in a single-user setting. Techniques discussed include transmit active beamforming, transmit passive beamforming and intelligent reflecting surfaces, receive combining with multi-antenna harvester, waveform, joint beamforming, combining and waveform, large-scale (massive) multiple-input multiple-output (MIMO), channel acquisition, transmit diversity, time-reversal, and retrodirective arrays. Importantly, while some of those techniques focus on enhancing $P_{\rf}^r$ and $e_2$ (and therefore a lower bound on $P_{\dc}^r$), others such as waveform, transmit diversity, receive combiner, joint waveform and beamforming are deeply rooted in the EH nonlinearity (and therefore the maximization of $P_{\dc}^r$ itself) and only appeared to light once the nonlinearity is accounted for in the signal design.

%{\color{blue} we need to have a table to explain how each technique contribute to e, i.e. max $e_2$, e2 times e3, and what can be combined together, etc..., highlight the different gains and thick how strategies fit, transmit beamforming gain, receive combining gain, frequency diversity gain, rectenna nonlinearity gain. }

\subsubsection{Transmit Active Beamforming}

\par
\begin{figure}%[!hhh]
	\centerline{\includegraphics[width=0.8\columnwidth]{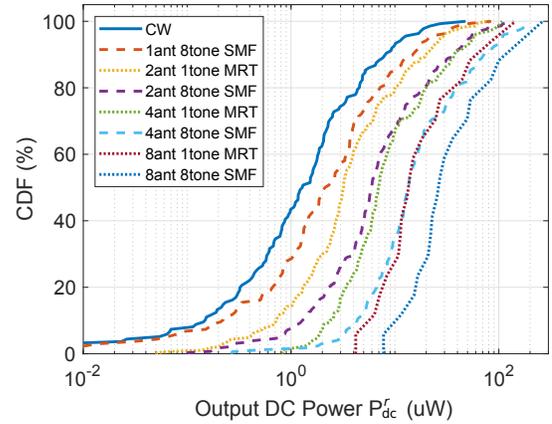}}
	\caption{CDF of output DC power ($P_{\dc}^r$) measurement results at
different distances from 0.6 to 5.4 m with the number of transmit antennas $M$=1, 2, 4, 8, and the number of carriers (tones) $N$=1 and 8 \cite{Kim:2020}.}
	\label{cdf_8ant}
\end{figure}

\begin{figure}%[!hhh]
	\centerline{\includegraphics[width=0.8\columnwidth]{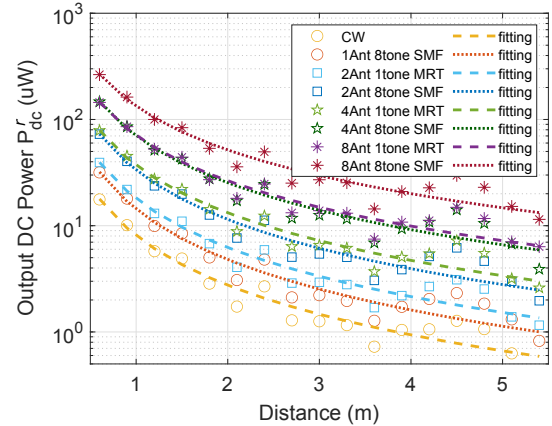}}
	\caption{Curve fitting and measurement results of output DC power ($P_{\dc}^r$) versus distance with $N$=1, 8 and $M$=1,2,4,8 \cite{Kim:2020}.}
	\label{fit_all}
\end{figure}

 Leveraging the presence of multiple antennas at the transmitter, each equipped with an RF chain, the simplest strategy is transmit active beamforming to increase $P_{\rf}^r$. Considering a MISO setup ($Q=1$) with $N=1$ and a linear HPA, \eqref{eq:yit} boils down to $y_{\rf}(t) =  \sqrt{2}  \Re \left\{ \mathbf{h} \mathbf{w}_{\mathrm{T}} x(t) e^{j 2\pi f t}  \right\}+w_{\textnormal{A}}(t)$, with $\mathbf{w}_{\mathrm{T}}$ the transmit beamformer. The transmitter simply performs conventional Maximum Ratio Transmission (MRT) $\mathbf{w}_{\mathrm{T}}=\sqrt{P}\bar{\mathbf{h}}^H$, with $\bar{\mathbf{h}}=\mathbf{h}/\left\|\mathbf{h}\right\|$, and $x(t)$ being any chosen random input with unit power (with $x(t)=1$ corresponding to a CW). Fig.\ \ref{cdf_8ant} and \ref{fit_all} illustrate the benefits in terms of output DC power $P_{\dc}^r$ and range of WPT by adopting MRT beamforming with 1, 2, 4, 8 transmit antennas and continuous wave ($N=1$, 1 tone), based on experimental data gathered in a typical indoor environment at 2.4GHz with a rectenna similar to Fig. \ref{TD_schematic} under an Effective Isotropic Radiated Power (EIRP) of 36dBm \cite{Kim:2020}. Other experimental results of such beamforming technique can be found in \cite{ChoiKim2017a,ChoiKim2018a}. 

%\begin{figure}%[!hhh]
	%\centerline{\includegraphics[width=0.8\columnwidth]{Figures/fit_antennas.eps}}
	%\caption{Experimental results and curve fitting of output DC power ($P_{\dc}^r$) versus distance for MRT beamforming with continuous-wave (1-tone signal) and multiple transmit antennas ($M$=1,2,4,8) \cite{Kim:2020}.}
	%\label{fit_antennas}
%\end{figure} 

\subsubsection{Transmit Passive Beamforming}

\par Transmit Passive Beamforming through \textit{intelligent reflecting surface} (IRS), also known as \textsl{reconfigurable intelligent surface}, has gained popularity as an emerging technology for wireless networks \cite{Basar:2019,Gong:2019,Wu:2020}. IRS consists of a large number of $L$ reconfigurable passive elements (without any need for an RF chain) integrated into the propagation environment. By collaboratively adjusting the impedance of all passive elements at the IRS, the reflected signals add coherently with the signals from other paths at the desired receiver to increase the received RF signal power, therefore enabling a passive beamforming gain. Owing to the passive structure, IRS has several advantages including low cost, low profile, light weight, conformal geometry, low power consumption and no additive thermal noise during the reflection.

\par Considering a SISO setup ($M=1$, $Q=1$) with $N=1$ and a linear HPA, $y_{\rf}(t) =  \sqrt{2}  \Re \left\{ h x(t) e^{j 2\pi f t}  \right\}+w_{\textnormal{A}}(t)$ with $h=g_{\mathrm{d}}+\mathbf{g}_{\mathrm{r}}\mathbf{\Theta}\mathbf{g}_{\mathrm{i}}$ where $g_{\mathrm{d}}$ refers to the direct channel between the ET and ER, $\mathbf{g}_{\mathrm{r}}$ is a $1\times L$ vector channel between the IRS (equipped with $L$ elements) and the ER, and $\mathbf{g}_{\mathrm{i}}$ is the $L\times 1$ vector channel between the ET and the IRS. $\mathbf{\Theta}$ is the scattering matrix of the $L$-port reconfigurable impedance network and is subject to the constraints $\mathbf{\Theta}=\mathbf{\Theta}^T$ and $\mathbf{\Theta}^H\mathbf{\Theta}=\mathbf{I}_L$ \cite{Shen:2020d}. The $L$-port reconfigurable impedance network is constructed with reconfigurable and passive elements so that it can reflect the incident signal with a reconfiguration that can be adapted to the channel. Three different architecture are possible, namely single connected reconfigurable impedance network characterized by a diagonal $\mathbf{\Theta}$, group connected reconfigurable impedance network characterized by a block diagonal $\mathbf{\Theta}=\mathrm{diag}(\mathbf{\Theta}_1, \mathbf{\Theta}_2, \ldots, \mathbf{\Theta}_G)$ where the $L$ elements have been divided into $G$ groups with each group having $L_G=L/G$ elements, and fully connected reconfigurable impedance network characterized by a full $\mathbf{\Theta}$ \cite{Shen:2020d}. %Single and fully connected network architectures are illustrated in Fig. \ref{IRS} for $L=4$. 
%\begin{figure}%[!hhh]
	%\centerline{\includegraphics[width=0.8\columnwidth]{Figures/4_port_IRS3.pdf}}
	%\caption{(a) 4-element IRS with single connected reconfigurable impedance network and (b) 4-element IRS with fully connected reconfigurable impedance network \cite{Shen:2020d}.}
	%\label{IRS}
%\end{figure}

\par
\begin{figure}%[!hhh]
	\centerline{\includegraphics[width=0.9\columnwidth]{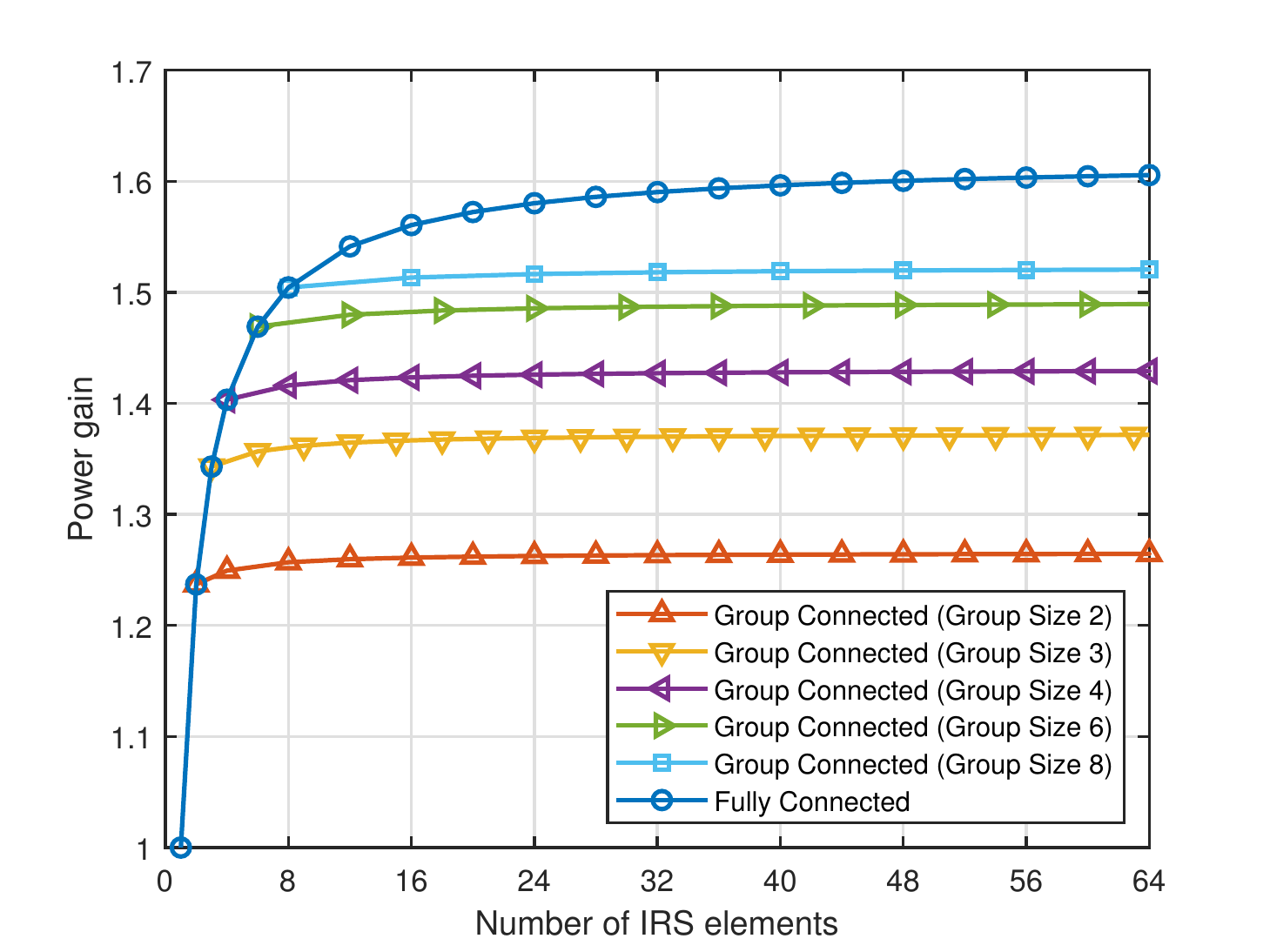}}
	\caption{Power gain of the group connected and fully connected reconfigurable impedance networks over the single connected
reconfigurable impedance network. \cite{Shen:2020d}.}
	\label{IRS_powergain}
\end{figure}

 Considering a group connected reconfigurable impedance network, the design of $\mathbf{\Theta}$ that maximizes $e_2$ is the solution of the optimization problem
\begin{align}
\max_{\mathbf{\Theta}} \hspace{0.3cm}& \left|h_{\mathrm{d}}+\mathbf{h}_{\mathrm{r}}\mathbf{\Theta}\mathbf{h}_{\mathrm{i}}\right|^2 \label{opt_IRS_1} \\
\mathrm{subject\,\,to} \hspace{0.3cm}&\mathbf{\Theta}=\mathrm{diag}(\mathbf{\Theta}_1, \mathbf{\Theta}_2, \ldots, \mathbf{\Theta}_G),\\
&\mathbf{\Theta}_g^H\mathbf{\Theta}_g=\mathbf{I}_{L_G}, \forall g,\\
&\mathbf{\Theta}_g=\mathbf{\Theta}_g^T, \forall g.\label{opt_IRS_4}
\end{align}
The single and fully connected reconfigurable impedance networks can be designed similarly by noting that they are two special cases of the group connected reconfigurable impedance network, i.e. with $G=L$ ($L_G=1$) and $G=1$ ($L_G=L$), respectively. One way to solve \eqref{opt_IRS_1}-\eqref{opt_IRS_4} is by reformulating it as an unconstrained optimization problem \cite{Shen:2020d}. It has been shown in \cite{Shen:2020d}, that for a given $L$, the larger $L_G$ (and the smaller $G$) the higher the received RF power (and therefore the higher $e_2$). In other words, the received RF power $P_{\rf,\mathrm{full}}^r$ of fully connected networks is larger than that of the group connected network ($P_{\rf,\mathrm{group}}^r$) and single connected network ($P_{\rf,\mathrm{single}}^r$), at the cost of a higher implementation complexity. Group connected network exhibit a nice tradeoff between complexity and performance. This is illustrated in Fig. \ref{IRS_powergain} where the power gains $P_{\rf,\mathrm{group}}^r/P_{\rf,\mathrm{single}}^r$ and $P_{\rf,\mathrm{full}}^r/P_{\rf,\mathrm{single}}^r$ of the group connected and fully connected reconfigurable impedance networks over the single connected reconfigurable impedance network are displayed as a function of $L$ for several values of group size $L_G$ \cite{Shen:2020d}.  Compared with the single connected reconfigurable impedance network, fully connected reconfigurable impedance network can increase the received signal power by up to 62\%.  For group connected with $L_G = 2, 3, 4, 6, 8$, gains of 26\%, 37\%, 43\%, 49\%, 52\% are achieved over the single connected network, respectively.

\par In contrast to active antenna arrays where the amplitudes and phases can be adjusted freely at each antenna and at each frequency, the elements in IRS are subject to less flexibility due to the passive nature of the IRS and the hardware constraints. Specifically, taking a single-connected network, due to constraints $\mathbf{\Theta}^H\mathbf{\Theta}=\mathbf{I}$, the amplitude of the diagonal entries is fixed to unity and only the phases of those entries are optimized. Moreover the IRS is commonly assumed frequency flat in the sense that the phases of the passive elements are kept constant across frequency. Despite those constraints, the passive beamforming gain can be significant and IRS brings some natural benefits to WPT since IRS can help increasing the RF power level $P_{\rf}^r$ at the input of the rectenna \cite{Pan:2020,Wu:2019,Tang:2019,Wu:2019b,Feng:2020,Zhao:2020}. The presence of active antennas and passive IRS leads to a joint design and optimization of active and passive beamforming.

\subsubsection{Receive Combining}

\par Beamforming is not limited to the ET and can also be used at the ER subject to a proper design of the DC and RF combiner schemes of Fig. \ref{DC_Combining} and \ref{RF_Combining} \cite{Shen:2020a}.
Assuming $N=1$ and a linear HPA, $y_{\rf,q}(t) =  \sqrt{2}  \Re \left\{ \mathbf{h}_q \mathbf{w}_{\mathrm{T}} x(t) e^{j 2\pi f t}  \right\}+w_{\textnormal{A},q}(t)$. In MIMO WPT with DC combiner, only the transmit beamformer $\mathbf{w}_{\mathrm{T}}$ is optimized and problem \eqref{Pdc} is equivalent to 
\begin{align}\label{Pdc_combiner}
\max_{\mathbf{w}_{\mathrm{T}}} &\frac{1}{R_{\mathrm{L}}}\sum_{q=1}^Q \left(\sum_{i \hspace{0.1cm}\textnormal{even}, i\geq 2}^{n_o} \beta_i \zeta_i  \left|\mathbf{h}_q \mathbf{w}_{\mathrm{T}}\right|^i\right)^2\\
\mathrm{subject\,\,to} \hspace{0.3cm} &\left\|\mathbf{w}_{\mathrm{T}}\right\|^2\leq P,
\end{align}
where $\zeta_2=1/2$, $\zeta_4=3/8$, $\zeta_6=5/16$ \cite{Shen:2020a}.
With RF combiner, $\tilde{y}_{\rf}(t) =  \sqrt{2}  \Re \left\{ \mathbf{w}_{\mathrm{R}}^H\mathbf{H} \mathbf{w}_{\mathrm{T}} x(t) e^{j 2\pi f t}  \right\}+\tilde{w}_{\textnormal{A}}(t)$, and $\mathbf{w}_{\mathrm{T}}$ and $\mathbf{w}_{\mathrm{R}}$ need to be jointly optimized. Subject to combiner structure \eqref{phase_shifters}, Problem \eqref{Pdc} is equivalent to  
\begin{align}
\max_{\mathbf{w}_{\mathrm{T}},\left\{\theta_q\right\}}&\left|\mathbf{w}_{\mathrm{R}}^H\mathbf{H} \mathbf{w}_{\mathrm{T}}\right|^2\\
\mathrm{subject\,\,to} \hspace{0.3cm} &\left\|\mathbf{w}_{\mathrm{T}}\right\|^2\leq P,\\
&\mathbf{w}_{\mathrm{R}}=\frac{1}{\sqrt{Q}}\left[e^{-j\theta_1},e^{-j\theta_2},\ldots,e^{-j\theta_Q}\right]^T,\\
&-\pi \leq \theta_q \leq \pi, 1\leq q \leq Q.
\end{align}
Those non-convex optimization problems can be solved by involving geometric program (GP) and semi-definite relaxation (SDR) \cite{Shen:2020a}.

\par
\begin{figure}%[!hhh]
	\centerline{\includegraphics[width=0.9\columnwidth]{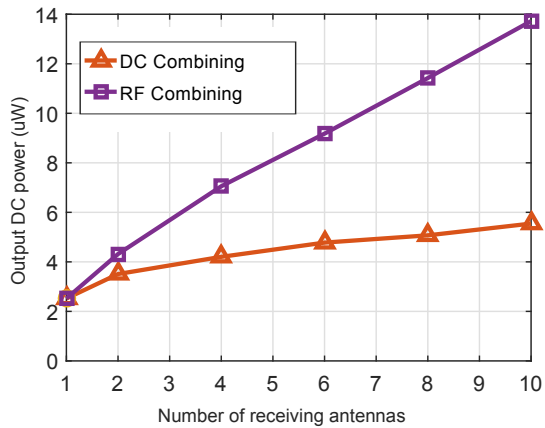}}
	\caption{Average output DC power $P_{dc}^{r}$ versus the number of receive antennas $Q$
for $M=10$ transmit antennas based on circuit simulations \cite{Shen:2020a}.}
	\label{PDC_SPICE_M10}
\end{figure}

 Interestingly, due to the rectenna nonlinearity that induces a higher $e_3$ for higher input power level (recall Fig.\ \ref{Pdc_Pin} for power levels lower than saturation and Observation \ref{higher_input_power}), it turns out that RF combining outperforms DC combining since the rectifier in RF combining operates on a higher RF power input signal. In other words, RF combining can leverage the nonlinearity more efficiently than DC combining. The performance gains of RF combining methods over DC combining can be quite significant as shown in the circuit simulations of Fig. \ref{PDC_SPICE_M10}. We can see that $P_{\dc}^{r}$ increases with the number of receive antennas $Q$ (this is in a way reminiscent of increasing $M$ in transmit beamforming), but the increase is faster with RF combining than DC combining. Hence, in MIMO WPT, while increasing the number of transmit antennas $M$ and receive antennas $Q$ helps to increase $P_{\rf}^{r}$ and therefore $e_2$, a suitable choice of the combiner at the receiver further helps by increasing $e_3$.  

\par The challenge with RF combining is that it not only needs CSIT but also Channel State Information at the Receiver (CSIR) for the joint transmit beamforming and receive combiner optimization. In contrast, DC combining only needs CSIT for transmit beamforming optimization. 

\subsubsection{Waveform}\label{waveform_WPT_section}

\par Another promising strategy is the design of transmit multi-carrier ($N>1$) waveform to utilize the nonlinear characteristic of the rectenna so as to boost $e_2 \times e_{3}$ \cite{Clerckx:2016b}. Such design originates from the fact that the output of the EH and $e_3$ are a nonlinear function of the rectenna input signal shape, as shown in Fig.\ \ref{Pdc_Pin} and discussed in Observation \ref{higher_order}. The transmit waveform design has a significant influence on $P_{\dc}^r$, namely it not only affects $e_2$ and $P_{\rf}^r$, but also $e_3$. 

%\par A multisine is characterized by a high PAPR, and the envelope of the transmitted RF signal is designed so that there are large peaks, while the average power is kept the same as in the continuous wave case. Consider indeed multiple in-phase sinewaves (with equal magnitudes) at frequencies $f_n=f_0+n\Delta_f$, $n=0,\ldots,N-1$, as the voltage source of the rectenna. As the number of tones $N$ increases, the time domain waveform appears as a sequence of pulses with a period equal to $1/\Delta_f$. The signal power is therefore concentrated into a series of high energy pulses, each of which triggers the diode that then conducts and helps charging the output capacitor. Once a pulse has passed, the diode stops conducting and the capacitor is discharging. The larger the number of tones $N$, the larger is the magnitude of the pulses and therefore the larger is the output voltage at the time of discharge. Since peaks of high power drive the rectenna with a much higher efficiency than the average low level input, they contribute more to the output DC voltage, and the rectifier sensitivity, range and RF-to-DC conversion efficiency $e_3$ increase. Such behavior was first highlighted in \cite{Trotter:2009,Trotter:2010}, wherein the authors proposed the use of a multisine waveform instead of a continuous wave (single sinewave) to provide a higher charge pump efficiency and thus to increase the range of RFID readers. 

\par
\begin{figure}%[!hhh]
	\centerline{\includegraphics[width=0.9\columnwidth]{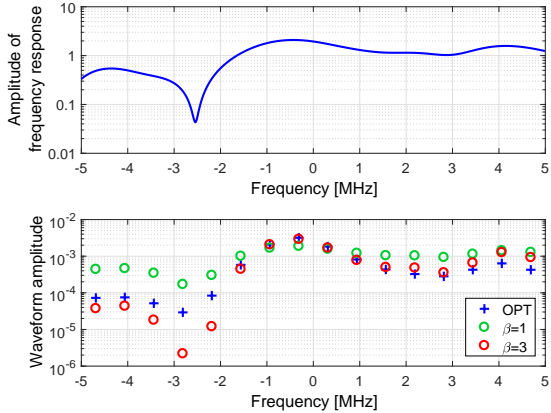}}
	\caption{Frequency response of the wireless channel and WPT waveform
magnitudes (N = 16) for 10 MHz bandwidth \cite{Clerckx:2017}.}
	\label{lowcomplexityWF}
\end{figure}

 In \cite{Clerckx:2016b}, a systematic methodology was derived to design and optimize waveforms for WPT. The optimal waveform design in \cite{Clerckx:2016b} is adaptive to the frequency selective channel (with frequency flat channel being a special case) and is rooted in the tradeoff between allocating the power to the strongest carrier so as to leverage the frequency diversity/selectivity and maximize $e_{2}$ and allocate power across $N$ carriers so as to leverage the rectifier nonlinearity and maximize $e_{3}$. As a result, the optimal waveform allocates power non-uniformly across the $N$ carriers, with the carriers corresponding to stronger channel gain allocated more power. Due to the EH nonlinearity, the waveform design results from a non-convex and computationally involved optimization problem. Assuming $M=1$, $Q=1$, linear HPA and deterministic multisine waveform, $y_{\rf}(t) =  \sqrt{2}  \Re \left\{  \sum_{n=0}^{N-1} h_{n} x_{n} e^{j 2\pi f_n t}  \right\}+w_{\textnormal{A}}(t)$. Denoting $x_{n}=s_n e^{j \phi_n}$ and $h_n=A_n e^{j \varphi_n}$, the optimal set of phases $\left\{\phi_n\right\}$ and magnitudes $\left\{s_n\right\}$ that are solutions of problem \eqref{Pdc} are given by $\phi_n^{\star}=-\varphi_n$ and by the solutions of the optimization problem (for $n_o=4$) 
\begin{align}\label{waveform_opt}
\max_{\left\{s_n\right\}} \hspace{0.2cm}&\alpha\left[\sum_{n=0}^{N-1}s_n^2A_n^2\right]+\sum_{\mycom{n_0,n_1,n_2,n_3}{n_0+n_1=n_2+n_3}}\prod_{j=0}^3s_{n_j}A_{n_j}\\
\mathrm{subject\,\,to} \hspace{0.2cm}& \sum_{n=0}^{N-1}s_n^2 \leq P
\end{align}
where $\alpha=\beta_2\zeta_2/(\beta_4\zeta_4)$. The first term in \eqref{waveform_opt} relates to $P_{\rf}^r=\mathbb{E}\left[y_{\rf}(t)^2\right]$ and the second term to $\mathbb{E}\left[y_{\rf}(t)^4\right]$. The challenge is due to the nonlinear coupling across frequency captured by the second term. The first term will favor a single-sinewave power allocation strategy, i.e. allocating all the power to sinewave corresponding to $\max_n A_n$. However due to the presence of the second term, such a single-sinewave strategy
is in general sub-optimal. Indeed, the optimal solution results from a tradeoff between maximizing the first term (and therefore maximize $P_{\rf}^r$) by allocating power to a single sinewave and leveraging the nonlinearity of the second term (and therefore maximize $e_3$) by allocating power across multiple sinewaves. Consequently, the optimal solution, obtained using reverse GP, reveals that the power is allocated across all sinewaves but more power is allocated to frequencies corresponding to larger channel gains \cite{Clerckx:2016b}. This is illustrated in Fig. \ref{lowcomplexityWF} where the upper graph is the magnitude of the channel frequency response, and the lower figure illustrates the solution of problem \eqref{waveform_opt} (``opt'') at $N=16$ uniformly spaced frequencies. Doing so, the waveform exploits a channel frequency diversity gain and the EH nonlinearity.

\par GP does not lend itself easily to implementation due to high complexity. Other optimization frameworks to design waveforms have therefore been proposed in \cite{Huang:2017,Moghadam:2017}. Suboptimal low complexity methods, called SMF, have also been proposed in \cite{Clerckx:2017}. A simple way to allocate power across frequencies is as follows $s_n^2=c A_n^{2\beta}$ where $c$ is a constant satisfying the average transmit power constraint. By scaling the channel gain using an exponent proportional to $\beta>1$, the waveform allocates more (resp.\ less) power to the frequency components corresponding to large (resp.\ weak) channel gains and replicates the main behavior of the ``opt'' solution. This is illustrated in Fig. \ref{lowcomplexityWF} with $\beta=1,3$ \cite{Clerckx:2017}. By adjusting $\beta>1$, we amplify the strong frequency components and attenuate the weak ones, so as to come close to the optimal power allocation. Though suboptimal, the SMF design was shown to perform close to the ``opt'' GP design.  

\par Such optimized and low complexity waveforms were shown using circuit simulations to provide significant benefits of 100\%-200\% over conventional continuous-wave signal and non-optimized waveforms in a wide range of rectifier topologies by leveraging the channel frequency diversity gain and a gain originating from the rectifier nonlinearity \cite{Clerckx:2016b,Clerckx:2017}. They have been successfully experimentally validated, demonstrating gains of 105\%-170\% in real-time over-the-air experimentation, in \cite{Kim:2018}. 
In Fig. \ref{cdf_8ant} and \ref{fit_all}, the benefit in terms of output DC power and range with using $N=8$ over conventional CW ($N=1$) is illustrated. 

\par In the presence of HPA nonlinearity, PAPR constraints could be added to problem \eqref{waveform_opt} as in \cite{Clerckx:2016b}, though a more involved and interesting problem would be to revise problem \eqref{waveform_opt} accounting for HPA nonlinearity \eqref{HPA_NL_model}.

\subsubsection{Joint Beamforming, Combining and Waveform}

\par Remarkably, such waveforms can also be designed for a multi-antenna transmitter so as to additionally exploit a beamforming gain \cite{Clerckx:2016b,Huang:2017}. Joint waveform and beamforming enables to simultaneously harvest three different gains, namely a beamforming gain, a frequency diversity gain and a gain related to the rectifier nonlinearity, and therefore offers additional opportunities over spatial domain processing/beamforming-only or over frequency domain waveform-only to boost $e_2 \times e_3$. 

\par Though the optimal design of joint waveform and beamforming results from the solution of an optimization problem \cite{Clerckx:2016b,Huang:2017}, a simple combination of the MRT beamforming and the low-complexity SMF waveform was demonstrated experimentally in \cite{Kim:2020} to significantly boost the output DC power and the range of WPT, as illustrated in Fig.\ \ref{cdf_8ant} and Fig.\ \ref{fit_all}. We observe that WPT performance gains can be obtained by exploiting either the frequency domain, the spatial domain, or both domains jointly. Besides, the 8-antenna 1-tone waveform shows a similar performance to that of the 4-antenna 8-tone waveform. In the same manner, 4-antenna single-tone and 2-antenna 8-tone, and 2-antenna single-tone and 1-antenna 8-tone show similar performance. Such behavior demonstrates that one can trade the spatial domain (number of antennas) processing with the frequency domain (number of tones) processing and inversely, and the gains in terms of output DC power and range can be accumulated using a joint beamforming and waveform strategy.

\par The accumulated gains of beamforming and waveform also applies to MIMO WPT where a joint waveform, transmit beamforming and receive combining was shown to provide significant gains over individual techniques \cite{Shen:2020c}. It was shown that the joint waveform and beamforming design provides a higher output DC power than the beamforming only design with a relative gain exceeding 180\% when $M = 2$, $N = 16$, and $Q = 2$. Moreover, RF combining was shown to provide a higher output DC power than DC combining with a relative gain which can be up to 550\% when $M = 2$, $N = 8$, and $Q = 10$.

\par Similarly, the transmit waveform and active beamforming can be jointly designed together with the passive beamforming at the IRS so as to efficiently exploit frequency and spatial domain gains \cite{Feng:2020,Zhao:2020}. Note those gains were demonstrated despite the frequency flat constraints of the passive elements of the IRS (the scattering matrix $\Theta$ of the IRS is constant across frequency).

\subsubsection{Large-Scale (Massive) MIMO}

\par The results in \cite{Clerckx:2016b} also highlight the potential of a \textit{large-scale multisine multiantenna} ($M>>1$, $N>>1$) closed-loop WPT architecture, reminiscent of Massive MIMO with OFDM in communications. In \cite{Huang:2017}, such a promising architecture was studied and shown to enhance $e$ and increase the range of WPT. It enables highly efficient WPT by jointly optimizing transmit signals over a large number of frequency components and transmit antennas, thereby combining the benefits of pencil beams (as in Massive MIMO) and waveform design to exploit the large beamforming gain of the transmit antenna array and the nonlinearity of the rectifier at long distances. The challenge is the large number of dimensions $N$ and $M$, which requires a reformulation of the optimization problem. The new design offers significantly lower complexity in signal design compared to the GP approach \cite{Huang:2017}.  
\par Interestingly in the limit of large $M$, the design of the joint multiantenna multisine waveform is simplified thanks to the channel hardening. Indeed, with $y_{\rf}(t) =  \sqrt{2}  \Re \left\{  \sum_{n=0}^{N-1} \mathbf{h}_{n} \mathbf{x}_{n} e^{j 2\pi f_n t}  \right\}+w_{\textnormal{A}}(t)$, and writing $\mathbf{x}_{n}=s_n\bar{\mathbf{h}}_{n}^H$ (with $\bar{\mathbf{h}}_{n}=\mathbf{h}_{n}/\left\|\mathbf{h}_{n}\right\|$ and $\sum_{n=0}^{N-1}s_n^2 \leq P$), $\lim_{M\rightarrow\infty}\left\|\mathbf{h}_{n}\right\|/\sqrt{M}=1$ and the channel after beamforming becomes effectively frequency flat due to channel hardening on all frequencies. In the limit of large $M$, $\left\{s_n\right\}$ is therefore simply obtained as the solution \eqref{waveform_opt} over an effective frequency-flat channel ($A_n=A$). A good (close to optimum) strategy is to allocate power uniformly across frequencies, i.e. $s_n=\sqrt{P/N}$.

\subsubsection{Channel Acquisition}
\par The aforementioned techniques have been designed assuming perfect CSIT. In practice, the CSI should be acquired by the ET and several strategies have been proposed, including forward-link training with CSI feedback, reverse-link training via channel reciprocity, power probing with limited feedback, and channel estimation based on backscatter communications \cite{Zeng:2017,Xu_Zhang:2014,Zeng_Zhang:2015,Xu_Zhang:2015a,Xu_Zhang:2016,Abeywickrama:2018,ChoiKim2017b,Huang:2018,Yang:2015}. The first two are similar to strategies used in modern communication systems, but incur too high energy consumption and/or too complex processing for low power nodes. The third is more promising and tailored to WPT because it is implementable with very low communication and signal processing requirements at the ER. The fourth one is also promising and is based on the idea that the ET exploits its observed backscatter signals to estimate the backscatter-channel (i.e., ET-to-ER-to-ET) state information (BS-CSI) directly instead of estimating the forward channel ET-ER as in previous three techniques. The BS-CSI is then used in the transmit signal design.

\par
\begin{figure}%[!hhh]
   \centerline{\includegraphics[width=3.3 in]{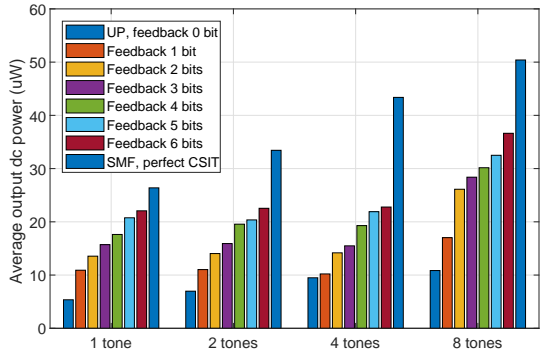}}
  \caption{Measurement results of output DC power $P_{\dc}^r$ of joint waveform and beamforming with limited feedback for $M=4$, $N=1,2,4,8$ and 1 to 6 bits of feedback \cite{Shen:2021}.}
  \label{LF}
\end{figure}

 The framework for power probing with limited feedback of \cite{Huang:2018} focuses on general setup of multi-antenna multi-carrier WPT over frequency-selective channels. It demonstrates that one can jointly exploit a beamforming gain, the channel frequency selectivity, and the EH nonlinearity through a joint waveform and beamforming based on limited feedback. To that end, it relies on the output DC power measurement and a limited number of feedback bits for the selection or the refinement of the joint waveform and beamforming. In the selection strategy, the ET transmits over multiple time slots with a different (joint waveform and beamforming) precoder within a codebook at each time slot, and the ER reports the index of the precoder in the codebook that offers the largest $P_{\dc}^r$. In the refinement strategy, the ET sequentially transmits using two precoders in each stage, and the ER reports one feedback bit, indicating an increase or a decrease in $P_{\dc}^r$ during this stage. Based on multiple one-bit feedback, the ET successively refines precoders (across space and frequency) in a tree-structured codebook over multiple stages. The optimization of the codebook of joint waveform and beamformers is pretty challenging and employs the framework of the generalized Lloyd’s algorithm. Fig. \ref{LF} illustrates the experimental results obtained with such a strategy for 1 to 6 bits of feedback with $M=4$ and $N=1,2,4,8$ \cite{Shen:2021}. We note the significant increase in $P_{\dc}^r$ as $N$ increases with perfect CSIT, and the need for larger codebook sizes to come closer to the perfect CSIT performance. 

\subsubsection{Transmit Diversity}
Another WPT signal strategy, denoted as transmit diversity \cite{Clerckx:2018c}, relies on $M$ dumb transmit antennas to induce fast fluctuations of the wireless channel through a simple phase sweeping method consisting of the transmission of a signal $x(t)$ on each antenna with an antenna dependent time-varying phase $\psi_m(t)$, namely $x_{\In,m}(t)=\sqrt{2}\Re\left\{x(t) e^{j (2\pi f t+\psi_m(t))}\right\}$. Those fluctuations are shown to boost $P_{\dc}^r$. This is another consequence of EH nonlinearity in Observation \ref{higher_order}, namely that fading and fast fluctuations of the wireless channel do not increase $P_{\rf}^r$ but increase $\mathbb{E}\left[y_{\rf}(t)^4\right]$ and therefore $e_3$. In contrast to the beamforming strategies, transmit diversity does not rely on any form of CSIT. Interestingly, this highlights that multiple transmit antennas are useful in WPT even in the absence of CSIT. In \cite{Clerckx:2018c,Kim:2018} real-time over-the-air measured gains of 50\% to 100\% were demonstrated with a two-antenna transmit diversity strategy over single-antenna setup, without any need for CSIT.

%\begin{figure}%[!hhh]
   %\centerline{\includegraphics[width=3.1 in]{Figures/TD_drawing.eps}}
  %\caption{General architecture of transmit diversity for WPT with $M$ dumb antennas. Signal $x(t)$ can be a continuous wave, modulated carrier or waveform \cite{Clerckx:2018c}.{\color{blue} change s(t) to x(t) and change color green to blue}}
  %\label{TD_drawing}
%\end{figure} 

\par Transmit diversity has a number of practical benefits leading to low cost deployments, namely the use of dumb antennas fed with a low PAPR CW (hence making a better use of $e_1$), no need for synchronization among transmit antennas, applicable to co-located and distributed antenna deployments, transparent to the ERs (which eases the system implementation), applicable to deployments with a massive number of devices (massive IoT deployments) for which CSIT acquisition is unpractical. Another related CSI-free multiantenna techniques for WPT has been proposed in \cite{Lopez:2019}.

\subsubsection{Other Techniques}
\par Another technique that can be seen as an alternative to multi-antenna beamforming to enable directional/energy focusing transmission for WPT, is \textit{time-reversal} \cite{Ku:2016,Ku:2017}. With time-reversal, the multipaths in the wireless channel are used as virtual antennas to enable spatial-temporal focusing effect and enhance $e_2$. Upon acquiring the channel impulse response, the transmitter sends a time-reversed conjugate waveform, using the principle of match filtering, in order to leverage the multipath channel and focus the signal power at the receiver input. Time-reversal can be applied to a single-antenna or multi-antenna transmitters and requires large bandwidth in order to distinguish as many paths in the channel as possible. Note that time reversal waveform design is different from aforementioned waveform design and is not rooted in the EH nonlinearity. It may be promising to investigate how those two types of waveforms can be designed in a unified manner.
\par Another low complexity (without the need for sophisticated digital signal processing) alternative to enable beamforming gain in multi-antenna settings is by using \textit{retrodirective arrays}. Upon receiving a signal from any direction, retrodirective arrays exploit channel reciprocity to transmit a signal response, in the form of a phase-conjugated version of the received signal, back to the same direction without the need of knowing the source direction or performing explicit channel estimation/feedback \cite{Miyamoto:2002,Pon:1964}. Two well known retrodirective array structures are
Van Atta arrays and the heterodyne retrodirective arrays with phase-conjugating circuits. WPT using retrodirective techniques have been studied in \cite{Lee:2017} and experimentally demonstrated in different setups \cite{Ren:2006,Li:2012,Wang:2014}. It would be interesting to explore how waveform and retrodirective array could be jointly designed so as to exploit the beamforming and waveform gains.

\subsection{Signal Processing Techniques for Multi-User WPT}\label{subsection_MU_WPT} 

\par WPT is not limited to a single ET and ER. In a multiuser WPT setting with one ET and $K$ ERs, with each ER having one rectenna (Fig. \ref{WPT_figure}), the output DC power $P_{\dc,k}^r$ at a given rectenna $k$ depends on $P_{\dc,j}^r$ at another rectenna $j\neq k$; i.e., a given signal, e.g. waveform or beamformer, may be suitable for one rectenna but inefficient for another. Therefore, a tradeoff exists between the output DC power of the different rectennas. The energy region formulates this tradeoff by expressing the set of output DC power at all rectennas that can be achieved simultaneously, which is written mathematically as a weighted sum of output DC power as 
\begin{align}%\label{SWIPT_opt_problem}
\max_{p(\mathbf{x}_0,...,\mathbf{x}_{N-1})} \hspace{0.3cm}& \sum_{k=1}^K v_k P_{\dc,k}^r(\mathbf{x}_0,...,\mathbf{x}_{N-1}) \label{Pdc_mu}\\
\mathrm{subject\,\,to} \hspace{0.3cm} &P_{\rf}^t\leq P,
\end{align}
where, by changing the weights $v_k$, we can operate on a different point of the energy region boundary and therefore favor one user over another one. An alternative problem formulation can be written as a maximization of minimum energy among all $K$ users
\begin{align}%\label{SWIPT_opt_problem}
\max_{p(\mathbf{x}_0,...,\mathbf{x}_{N-1})} \min_{k=1,\ldots,K} \hspace{0.3cm}& P_{\dc,k}^r(\mathbf{x}_0,...,\mathbf{x}_{N-1}) \label{Pdc_mu_mmf}\\
\mathrm{subject\,\,to} \hspace{0.3cm} &P_{\rf}^t\leq P.
\end{align}
Those two problems have been studied in \cite{Huang:2017}.
\par
\begin{figure}%[!hhh]
   \centerline{\includegraphics[width=3.1 in]{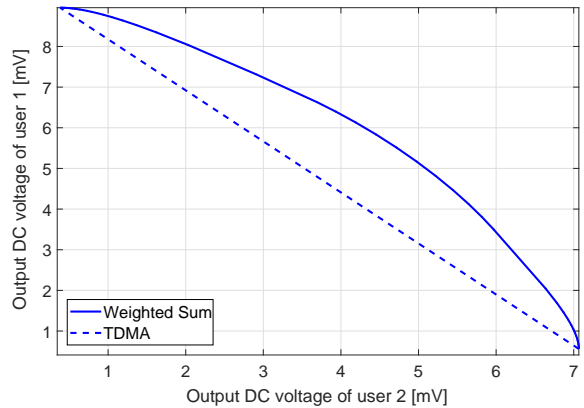}}
  \caption{The two-user energy region with joint beamforming and waveform ($M$ = 20 and $N$ = 10) \cite{Huang:2017}.}
  \label{twouser}
	\vspace{-0.2cm}
\end{figure}

 All aforementioned single-user techniques can be designed for multiuser WPT. Among the most advanced ones leveraging combined frequency-domain and spatial-domain gains, we note the joint beamforming and waveform for multiuser WPT of \cite{Clerckx:2016b,Huang:2017}, and the joint passive beamforming and waveform design for multiuser IRS-aided WPT of \cite{Feng:2020}. Fig.\ \ref{twouser} illustrates the energy region for a two user MISO WPT scenario with a joint beamforming and waveform spanning twenty transmit antennas ($M=20$) and ten frequencies ($N=10$), obtained by solving problem \eqref{Pdc_mu} \cite{Huang:2017}. The challenge is that solving \eqref{Pdc_mu} in this two user setup results in a coupled optimization of the frequency and the spatial domains. Indeed, while decoupling the spatial and frequency domains by first designing the spatial beamformer in each frequency and then designing the power allocation across subbands is optimal in the single-user/rectenna case, it is clearly suboptimal in the multi-user/rectenna case \cite{Huang:2017}. The key takeaway here is that, by optimizing the waveform to jointly transfer power to multiple users simultaneously, we obtain an energy region (``weighted sum'') that is larger than that achieved by a time-sharing approach, such as time-division multiple access (TDMA), where the transmit waveform and beamforming is optimized for a single user at a time and each user is scheduled to receive energy during a fraction of the time. Other multi-user waveform designs have appeared in \cite{Kim:2019_mu,Kim:2020_mu}.

\par Considering an entire network consisting of many ETs and ERs (Fig. \ref{WPT_figure}), \cite{Zeng:2017} defines various network architectures. All ETs can cooperate jointly to design the transmit signals for multiple ERs (in the form of a coordinated multipoint (CoMP)-based WPT) or locally coordinate their efforts such that a given ER is served by a subset of ETs (or, in the simplest scenario, where each ER is served by a single ET). As a consequence, different resource allocation strategies (centralized versus distributed) in terms of CSI sharing and acquisition at the different ETs can be considered. 
\par In \cite{Zeng:2017}, it was shown that distributing antennas (DAS) across a coverage area (as in Fig. \ref{WPT_figure}) and enabling cooperation among them distributes energy more evenly in space and, therefore, enhances the ubiquitous accessibility of wireless power as compared to a co-located deployment. Strong energy beams in the direction of users are also avoided, which is desirable from a safety perspective.
\par Recent experimental results of WPT architecture based on DAS in \cite{Shen:2020b} show that that WPT DAS can boost $P_{\dc}^r$ by up to 30 dB in a single-user setting and the sum of output DC power by up to 21.8 dB in a two-user setting and broaden the service coverage area in a low cost, low complexity, and flexible manner by suitably and dynamically selecting transmit antenna and frequency. Other DAS WPT studies have been reported in \cite{ChoiKim2018b,Zhou:2015,Lee:2017b,Zhang:2018,Huang:2019,Salem:2019}.

\section{Wireless Information and Power Transfer: Achieving the Best Rate-Energy Tradeoff}\label{WIPT_section}

Building upon WPT signal design and processing techniques, we can study how to integrate communications and power into WIPT. The objective is here to achieve the best R-E tradeoff.

\subsection{Signal and System Model}\label{WIPT_system_model}
We focus on a single-user (point-to-point) multi-subband MIMO SWIPT system (referred to as ``SWIPT with co-located receivers'' in Fig. \ref{SWIPT_figure}). The system model is the same as in Section \ref{WPT_system_model} though $x_{m,n,k}$ now denotes the complex-valued information and power carrying symbol, instead of just a power carrying symbol, since both information and power are transmitted simultaneously. The information is captured in $S$ possible messages $s\in \mathcal{M} = \{1,2,...,S\}$, where $\mathcal{M}$ denotes the set of messages. The mapping from $\mathcal{M}$ to the sequence of complex-valued transmitted information and power carrying symbols $x_{m,n,k}$ is denoted by $g_{\mathrm{\theta_T}}$, where $\mathrm{\theta_T}$ refers to the set of transmitter design parameters.

\par The processing then depends on the exact SWIPT receiver architecture. One commonality nevertheless exists among all considered types of receivers. Namely, from an energy perspective, $y_{\rf,q}(t)$ (or a fraction of it) is conveyed to an ER, where energy is harvested directly from the RF-domain signal. From an information perspective, an IR downconverts $y_{\rf,q}(t)$ (or a fraction of it) and filters it to produce the baseband signal for subband $n$
\begin{equation}
y_{q,n}(t) =  \mathbf{h}_{q,n} \mathbf x_n(t) +w_{q,n}(t),
\end{equation}
where $w_{q,n}(t)$ is the downconverted received filtered noise and accounts for both the antenna and the RF-to-baseband processing noise. After sampling at a frequency $f_{\mathrm{w}}$, the sampled outputs at time instants $k$ (multiples of the sampling period) can be expressed as
\begin{equation}\label{BB_model_with_k}
y_{q,n,k} =  \mathbf{h}_{q,n} \mathbf x_{n,k} +w_{q,n,k}
\end{equation}
with $\mathbf x_{n,k}\triangleq\big[x_{1,n,k},...,x_{M,n,k}\big]^T$.
Following the i.i.d. channel inputs and the discrete memoryless channel assumptions, we drop the time index $k$ such that
\begin{equation}\label{BB_model}
y_{q,n} =  \mathbf{h}_{q,n} \mathbf x_{n} +w_{q,n}.
\end{equation}
We model $w_{q,n}$ as an i.i.d.\ and CSCG random variable with
variance $\sigma^2$, i.e., $w_{q,n}\sim\mathcal{CN}(0,\sigma^2)$, where $\sigma^2=\sigma_A^2+\sigma_{P}^2$ is the total Additive White Gaussian Noise (AWGN) power originating from the antenna ($\sigma_A^2$) and the RF-to-baseband processing ($\sigma_{P}^2$).

\par The observations from all receive antennas can then be stacked to obtain
\begin{equation}\label{BB_model_vector}
\mathbf y_{n} =  \mathbf{H}_{n} \mathbf x_{n} +\mathbf w_{n},
\end{equation}
where $\mathbf y_{n}\!\triangleq\!\big[y_{1,n},...,y_{Q,n}\big]^T$, $\mathbf w_{n}\!\triangleq\!\big[w_{1,n},...,w_{Q,n}\big]^T$.

\par The estimated message $\hat{s}$ is then produced by the information decoder (ID) which  maps the received noisy sequence $\mathbf y_{n}$ to $\hat{s}\in\mathcal{M}$ through a parametric function denoted by $h_{\mathrm{\theta_R}}$, where $\mathrm{\theta_R}$ refers to the set of receiver design parameters at the ID.

\par Finally, we assume perfect Channel State Information at the Transmitter (CSIT) and perfect Channel State Information at the Receiver (CSIR).

\subsection{Receiver Architectures}

\par Several receiver architectures for SWIPT have been proposed in Fig. \ref{SWIPT_figure}. 

\begin{figure}
\begin{center}
\subfigure[Ideal Receiver]{\scalebox{0.5}{\includegraphics*{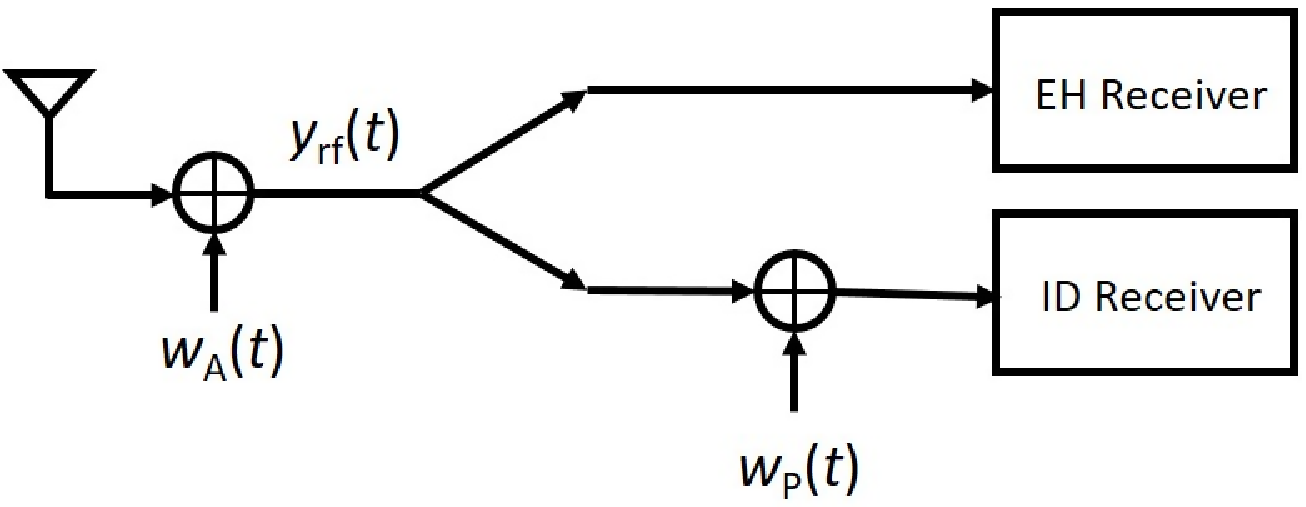}}}
\subfigure[TS Receiver]{\scalebox{0.5}{\includegraphics*{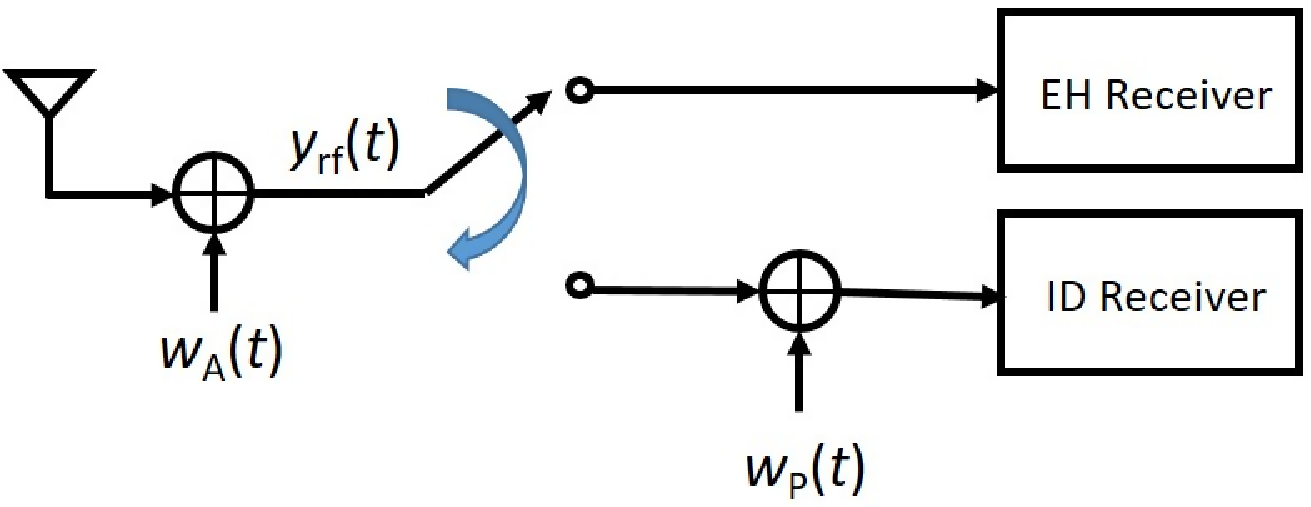}}}
\subfigure[PS Receiver]{\scalebox{0.5}{\includegraphics*{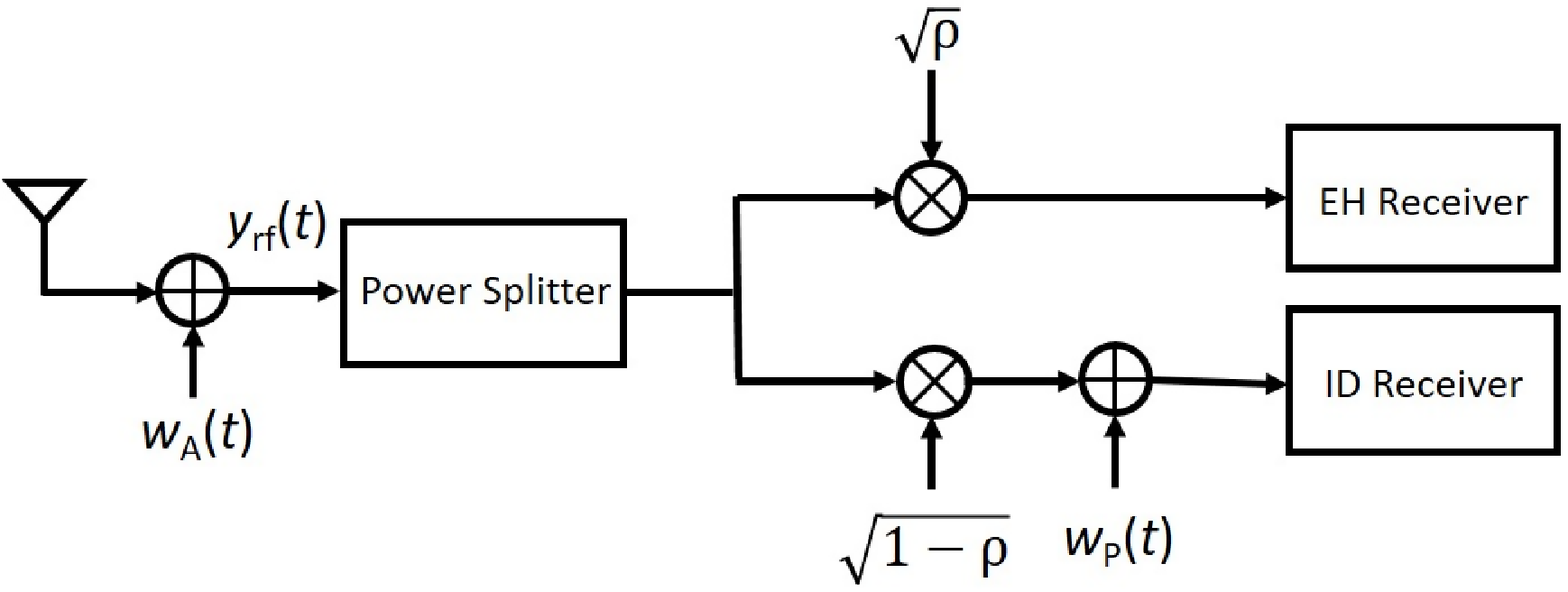}}}
\caption{Three single-antenna receiver architectures for SWIPT: (a) Ideal receiver (using the same signal for both the ID and EH receivers); (b) TS receiver (switching the signal to either ID or EH receiver); and (c) PS receiver (splitting a portion of the signal to ID receiver and the rest to EH receiver) \cite{Clerckx:2019}.}\label{three receiver architectures}
\end{center}
\end{figure}

\par An \textit{Ideal Receiver} (Fig. \ref{three receiver architectures}(a)) assumes the same signal $y_{\rf,q}(t)$ is conveyed to the EH and also simultaneously RF-to-baseband downconverted and conveyed to the ID \cite{Varshney:2008,Grover:2010}; however, no practical circuits can currently realize this operation. Different R-E tradeoffs could be realized by varying the design of the transmit signals to favor rate or energy.

\par A \textit{Time Switching (TS) Receiver} (Fig. \ref{three receiver architectures}(b)) consists of a conventional ID and an EH (following the structure in Section \ref{EH_section}) that are co-located \cite{Zhang:2013,Zhou:2013,Park:2013}. The transmission block is divided into two orthogonal time slots, one for power transfer and the other for data transfer. In each time slot, the transmit waveforms are optimized for either WPT or WIT. Accordingly, the receiver switches its operation periodically between harvesting energy and decoding information in the two time slots. Then, different R-E tradeoffs are realized by varying the length of the WPT time slot, jointly with the transmit signals \cite{Varshney:2010}.

\par In a \textit{Power Splitting (PS) Receiver} (Fig. \ref{three receiver architectures}(c)), the EH and ID receiver components are the same as those of a TS receiver. The transmitted signals are jointly optimized for information and energy transfer and the received signal is split into two streams, where one stream with PS ratio $0 \leq \rho \leq 1$ is exploited for EH, and the other with power ratio $1-\rho$ is utilized for ID \cite{Zhang:2013,Zhou:2013,Liu:2013}. Hence, assuming perfect matching (as in Section \ref{EH_section}), the input voltage signals $\sqrt{\rho R_{\ant}}y_{\rf}(t)$ and $\sqrt{(1-\rho)R_{\ant}}y_{\rf}(t)$ are respectively conveyed to the EH and the ID.  Different R-E tradeoffs are realized by adjusting the value of $\rho$ jointly with the transmit signals.

\par It is common to assume that the power of the processing noise is much larger than that of the antenna noise, i.e., $\sigma_P^2\gg \sigma_A^2$, such that $\sigma^2=\sigma_A^2+\sigma_P^2\approx \sigma_P^2$. As explained in \cite{Zhang:2013}, the above setting results in the worst-case R-E region for the practical PS receiver. 

\subsection{Rate-Energy Region and Problem Formulation}

\par The R-E region $\mathcal{C}_{\R-\E}$ is defined as the set of all pairs of rate $R$ and energy $E$ such that simultaneously the receiver can communicate at rate $R$ and harvested energy $E$. The R-E region in general is obtained through a collection of input distributions $p(\mathbf{x}_0,...,\mathbf{x}_{N-1})$ that satisfies the average transmit power constraint $P_{\rf}^t\leq P$. Mathematically, we can write
\begin{multline}\label{RE_region_def}
\mathcal{C}_{\R-\E}(P)\!\triangleq\!\bigcup_{\mycom{p(\mathbf{x}_0,...,\mathbf{x}_{N-1}):}{P_{\rf}^t\leq P}}\Bigg\{(R,E):R\leq \sum_{n=0}^{N-1} I\left(\mathbf{x}_n,\mathbf{y}_n\right),
\\ E\leq P_{\dc}^r\left(\mathbf{x}_0,...,\mathbf{x}_{N-1}\right) \Bigg\}
\end{multline}
where $I\left(\mathbf{x}_n,\mathbf{y}_n\right)$ refers to the mutual information between the channel input $\mathbf{x}_n$ and the channel output $\mathbf{y}_n$ on subband $n$ and $P_{\dc}^r$ is a nonlinear function of $\mathbf{x}_0,...,\mathbf{x}_{N-1}$. For multi-antenna harvesters based on RF combining, the receive combiner would also have to be jointly optimized with the input distribution. 

\par One approach to identify the R-E region is to calculate the capacity (supremization of the mutual information over all possible input distributions $p(\mathbf{x}_0,...,\mathbf{x}_{N-1})$) of a complex AWGN channel subject to an average RF power constraint $P_{\rf}^t\leq P$ and a receiver delivered power constraint $P_{\dc}^r(\mathbf{x}_0,...,\mathbf{x}_{N-1})\geq \bar{E}$, for different values of $\bar{E}\geq 0$. Namely,
\begin{align}%\label{SWIPT_opt_problem}
\sup_{p(\mathbf{x}_0,...,\mathbf{x}_{N-1})} \hspace{0.3cm}& \sum_{n=0}^{N-1}I(\mathbf{x}_n;\mathbf{y}_n) \label{MI}\\
\mathrm{subject\,\,to} \hspace{0.3cm}  
& P_{\rf}^t\leq P,\\
& P_{\dc}^r(\mathbf{x}_0,...,\mathbf{x}_{N-1})\geq \bar{E},\label{MI_2}
\end{align}
where $\bar{E}$ is interpreted as the minimum required or target delivered power. 
%Another solution that is also popular to characterize an achievable R-E region is to perform an energy maximization subject to a rate constraint
%\begin{align}%\label{SWIPT_opt_problem}
%\max_{p(\mathbf{x}_0,...,\mathbf{x}_{N-1})} \hspace{0.3cm}& P_{\dc}^r(\mathbf{x}_0,...,\mathbf{x}_{N-1}) \label{MI_3}\\
%\mathrm{subject\,\,to} \hspace{0.3cm} &\Tr\left(\mathbf{Q}\right)\leq P,\\
%& \sum_{n=0}^{N-1}I(\mathbf{x}_n;\mathbf{y}_n) \geq \bar{R},\label{MI_4}
%\end{align}
%where $\bar{R}$ is interpreted as the minimum required or target rate. The two problems are in general not equivalent due to the non-convexity of the objective functions and/or constraints.

\subsection{Signal Processing Techniques for WIPT}

\par Designing SWIPT\footnote{Even though emphasis is put on SWIPT in this section, the analysis and ideas reviewed in the paper also find applications in WPCN and WPBC.} requires the transmit signals to carry information and therefore to be subject to some randomness, and this randomness has an impact on the amount of harvested DC power. This raises interesting questions on how modulated signals perform in comparison to deterministic signals for WPT, and consequently on how to design modulation, waveform, coding and multi-antenna processing for SWIPT. Due to space limitations, emphasis is put on key single-user (point-to-point) techniques, though they can be extended to multi-user settings. Readers are invited to consult \cite{Clerckx:2019} for more discussions on multi-user WIPT.
 
\subsubsection{Modulation and Input Distribution}

\par Let us first assume a SISO ($M = Q = 1$) single-subband ($N = 1$) transmission with a linear HPA (with $e_1=1$) and the ideal receiver. The system model in \eqref{BB_model} simplifies
to $y = hx + w$. Problem \eqref{MI}-\eqref{MI_2} becomes  
\begin{align}
\sup_{p(x)} \hspace{0.3cm}& I(x;y) \label{MI_NL}\\
\mathrm{subject\,\,to} \hspace{0.3cm} &\mathbb{E}\left[|x|^2\right]\leq P,\label{MI_1_NL}\\
& P_{\dc}^r(x)\geq \bar{E}.\label{MI_2_NL}
\end{align}
From \eqref{vout_def} and Observation \ref{higher_order}, $P_{\dc}^r(x)$ is not only a function of $\mathbb{E}\left[|x|^2\right]$ but also a function of $\mathbb{E}\left[|x|^4\right]$ and higher order moments \cite{Clerckx:2018b,Varasteh:2017}.
\par Interestingly, the higher moments of the input distribution have a significant impact on the selection of the input distribution $p(x)$. It was shown in \cite{Clerckx:2018b} that modulation using CSCG inputs leads to a higher $P_{\dc}^r(x)$ compared to an unmodulated input, despite presenting the same average power $P_{\rf}^r$ at the rectenna input. This gain originates from the large higher order ($>2$) moment of CSCG inputs, which is leveraged by the rectifier nonlinearity and modeled by the higher order terms in \eqref{effect_Pin}. Indeed CSCG inputs $x\sim\mathcal{CN}(0,P)$ have $\mathbb{E}\left[|x|^4\right]=2 P^2$ while unmodulated CW inputs with the same average RF power only achieve $\mathbb{E}\left[|x|^4\right]=P^2$ \cite{Clerckx:2018b}.

\par Even larger gains can be obtained using asymmetric Gaussian inputs \cite{Varasteh:2017} and on-off keying (or also called flash signaling) \cite{Varasteh:2017b}. Indeed, real Gaussian modulation outperforms CSCG modulation despite the same average input power $P_{\rf}^r$ to the rectifier. In \cite{Varasteh:2017}, assuming general non-zero mean Gaussian inputs $\Re\left\{x\right\}\!\sim\!\mathcal{N}(\mu_r,P_r)$ and $\Im\left\{x\right\}\!\sim\!\mathcal{N}(\mu_i,P_i)$ with $P_r+P_i \leq P$, it is found that zero mean asymmetric Gaussian inputs with $P_r+P_i = P$ achieve the supremum in Problem \eqref{MI_NL}-\eqref{MI_2_NL}. CSCG input obtained by equally distributing power between the real and the imaginary dimensions, i.e., $\Re\left\{x\right\}\!\sim\!\mathcal{N}(0,P/2)$ and $\Im\left\{x\right\}\!\sim\!\mathcal{N}(0,P/2)$ is optimal for rate maximization. However, as $\bar{E}$ increases, the input distribution becomes asymmetric with $P_r$ increasing and $P_i=P-P_r$ decreasing (or inversely) till the rate is minimized and the energy is maximized by allocating the transmit power to only one dimension, e.g. $\Re\left\{x\right\}\!\sim\!\mathcal{N}(0,P)$. This is because a higher fourth moment is obtained by allocating all power to one dimension. Indeed, $\mathbb{E}\left[x^4\right]=3 P^2$ for $x\sim\mathcal{N}(0,P)$ in contrast to $\mathbb{E}\left[|x|^4\right]=2 P^2$ with $x\sim\mathcal{CN}(0,P)$. 

\par
\begin{figure}%[!hhh]
   \centerline{\includegraphics[width=0.9\columnwidth]{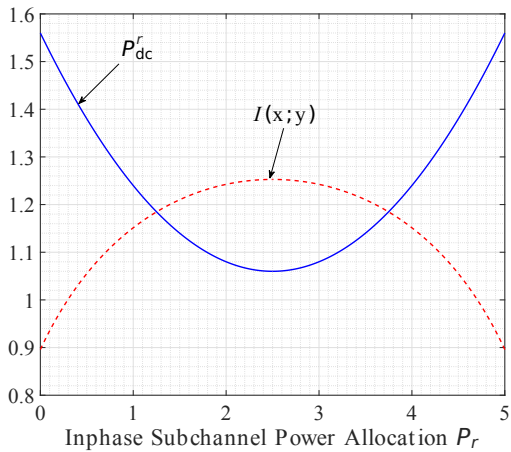}}
  \caption{Mutual information $I(x;y)$ (Red dashed line) and (normalized) output DC power $P_{\dc}^r(x)$ (blue solid line) corresponding to the complex Gaussian inputs with asymmetric power allocation \cite{Varasteh:2017b}. The transmitted information rate is maximized for $P_i = P_r = P/2$ and delivered power is maximized when $P_i = 0$, $P_r = P$ or $P_i = P$, $P_r = 0$.}
  \label{Fig6_TIT}
\end{figure}

 In Fig. \ref{Fig6_TIT}, the information rate $I(x;y)$ and (normalized) output DC power $P_{\dc}^r(x)$ for complex Gaussian inputs is shown versus $P_r$, assuming an ideal receiver. It is observed that the information rate and output DC power are indeed maximized and minimized, respectively, for $P_i = P_r = P/2$. Alternatively the information rate and output DC power are minimized and maximized, respectively when $P_i = 0$, $P_r = P$ or $P_i = P$, $P_r = 0$. This shows that there is a fundamental R-E tradeoff even in the simplest SISO AWGN scenario. It is important to recall that the tradeoff is induced by the presence of the fourth and higher moments of the received signal $y_{\rf}(t)$ in $v_{\out}$. Had we accounted only for the second term in $v_{\out}$, $P_{\dc}^r(x)$ in Fig. \ref{Fig6_TIT} would have been replaced by a flat curve. Ignoring the nonlinearity brought by the higher order terms, there is no tradeoff between R-E and $P_i = P_r = P/2$ simultaneously maximizes the rate and energy \cite{Varshney:2008,Grover:2010,Clerckx:2019}.

%\begin{figure}%[!hhh]
   %\centerline{\includegraphics[width=\columnwidth]{RE_region_gaussian.eps}}
  %\caption{R-E region with asymmetric Gaussian inputs in single-subband transmission ($P=1, \sigma^2=10^{-4}, k_2=0.17, k_4=19.145$) \cite{Varasteh:2017}. By evolving from point D to point A, the input distribution becomes more asymmetric and the harvested energy increases. The dashed line corresponds to the R-E region for the diode linear model. Note that the energy unit is $\mu$A because the energy metric used is $z_{\dc}$, which is a contribution to the output current.}
  %\label{RE_region_gaussian}
%\end{figure}

\par Relaxing the constraints on Gaussian inputs, it is remarkably shown in \cite{Varasteh:2017b} that the capacity of an AWGN channel under transmit average power and target delivered power constraints as characterized by Problem \eqref{MI_NL}-\eqref{MI_2_NL} is obtained by adopting a combination of CSCG and on-off-keying inputs. Let $E_\mathrm{G}$ denote the output DC power with the input $x\sim\mathcal{CN}(0,P)$. For $\bar{E}\leq E_\mathrm{G}$, the capacity is achieved via the unique input $x\sim\mathcal{CN}(0,P)$. For $\bar{E} > E_\mathrm{G}$, the capacity is approached by using time sharing between CSCG distribution and on-off keying, reminiscent of flash signaling, exhibiting a low probability of high amplitude signals. Such a combination of CSCG and on-off-keying inputs achieves a larger R-E region than asymmetric Gaussian inputs. Writing the complex input as $x=r e^{j\theta}$ with its phase $\theta$ uniformly distributed over $\left[0,2\pi\right)$, such a on-off keying distribution is given by the following probability mass function
\begin{align}\label{flash_distr}
p_r(r)=\begin{cases}
1-\frac{1}{l^2}, \ & r=0,\\
\frac{1}{l^2}, \ & r=l\sqrt{P},
\end{cases}
\end{align}
with $l\geq 1$. Such a distribution has indeed a low probability of high amplitude signals since $r=l\sqrt{P}$ increases and $p_r(l\sqrt{P})$ decreases as $l$ increases. We note that $\mathbb{E}\left[|x|^4\right]=l^2 P^2$ and $\mathbb{E}\left[|x|^2\right]=\mathbb{E}\left[r^2\right]=P$, hence achieving large higher moments as $l$ increases while satisfying the average RF power constraint. Choosing $l>\sqrt{3}$ makes the fourth order moment higher than the $2P^2$ and $3P^2$ obtained respectively with real Gaussian and CSCG inputs. Here again, the benefits of departing from Gaussian inputs originate from Observation \ref{higher_order} that favors the use of distributions with large higher moments of the channel input $x$. 

%\begin{figure}%[!hhh]
   %\centerline{\includegraphics[width=\columnwidth]{RE_region_optimal.eps}}
  %\caption{R-E region with optimal inputs in single-subband transmission ($P=5, \sigma^2=2$) \cite{Varasteh:2017b}. Note that the energy unit is $\mu$A because the energy metric used is $z_{\dc}$, which is a contribution to the output current.}
  %\label{RE_region_optimal}
%\end{figure}

\par
\begin{figure}%[!hhh]
	\centerline{\includegraphics[width=0.8\columnwidth]{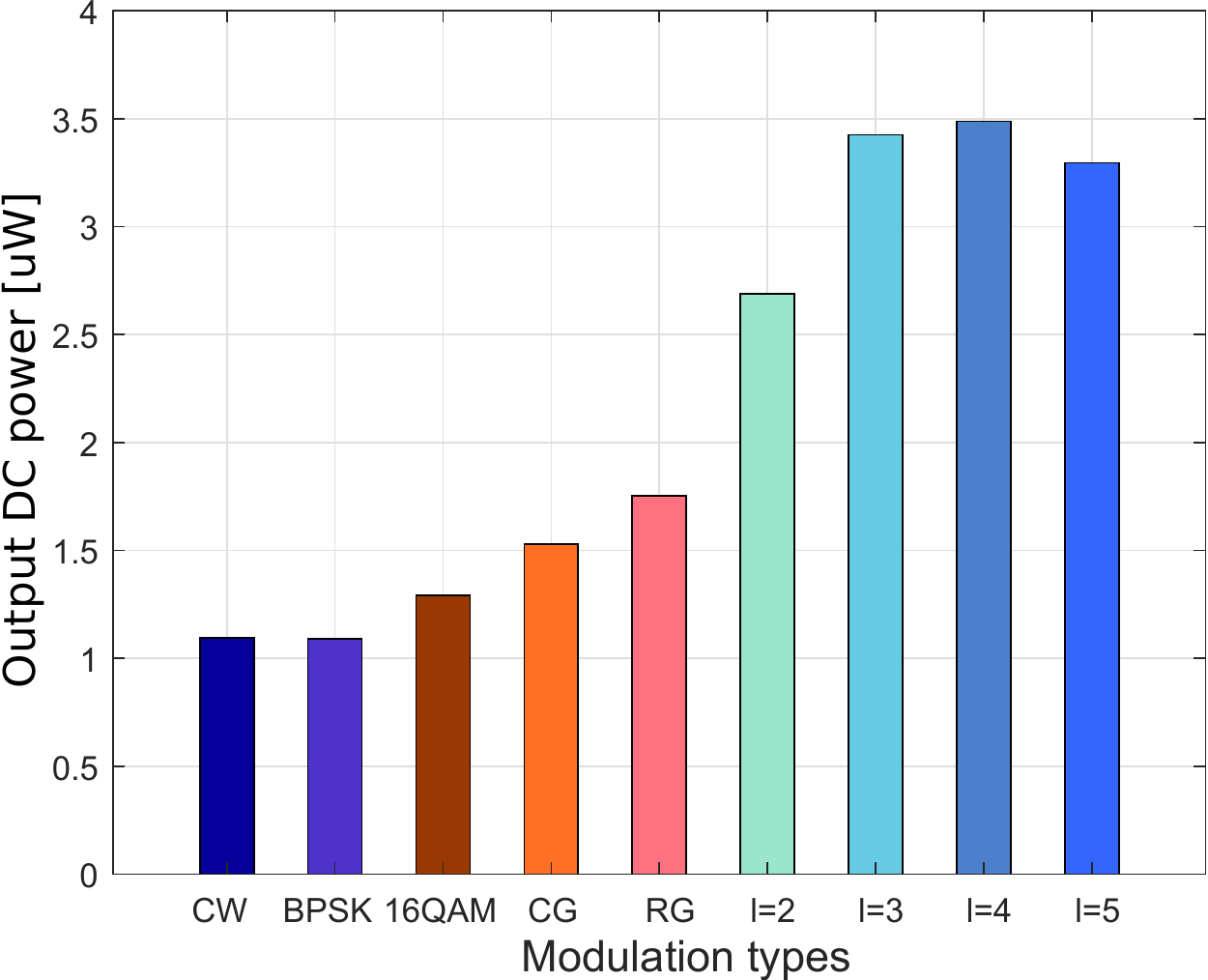}}
	\caption{Output DC power ($P_{\dc}^r$) of conventional and WPT-optimized modulations \cite{Kim:2020}. CG refers to CSCG, and RG to real Gaussian. $l$=1,2,3,4,5 refer to on-off keying modulations.}
	\label{modulation_fig}
\end{figure}

 On-off keying has been shown to significantly boost $e_3$ conversion efficiency over various baselines \cite{Kim:2020,Varasteh:2017b}. In Fig.\ \ref{modulation_fig}, we display circuit simulation results of various modulations using the rectifier of Fig.\ \ref{TD_schematic}. We note that on-off keying ($l$=1,2,3,4,5) provides three times (i.e. gain of over 200\%) more output DC power ($P_{\dc}^r$) than conventional communication modulations such BPSK and 16-QAM despite having the same average RF input power ($P_{\rf}^r$). 

\par The optimal distribution resulting from the use of on-off keying discussed so far assumed a linear HPA. In the presence of nonlinear HPA response as in Fig. \ref{SSPA}, high amplitude signals would be distorted and the optimal distribution would need to be re-assessed. One solution is to introduce an additional amplitude constraint in Problem \eqref{MI_NL}-\eqref{MI_2_NL}. It was shown in \cite{Varasteh:2017b} that under average transmit power, amplitude, and (nonlinear) delivered output DC power constraints, the optimal capacity achieving distributions are discrete with a finite number of mass points for the amplitude and continuous uniform for the phase (see also \cite{Varshney:2012}). Other SWIPT design to account for HPA nonlinearities have been discussed in \cite{Jang:2020,Krikidis:2020} and is further discussed in Section \ref{learning_SWIPT}. 

\par Though the above discussion assumed an ideal receiver, conclusions on the modulation and input distribution are applicable to TS and PS receiver. With a TS receiver, the transmitter would transmit asymmetric Gaussian or preferably on-off keying and CSCG alternatively on the two orthogonal time slots and the receiver would switch accordingly. With a PS receiver, the asymmetry ($P_r$ and $P_i$ values) in the Gaussian input would have to be optimized jointly with the PS ratio $\rho$. Similarly a combination of CSCG and on-off keying jointly optimized with the PS ratio could be used. Given two fixed distributions, one based on CSCG and the other based on on-off keying, an ideal receiver would achieve to a larger R-E region than a PS receiver, which itself has a larger R-E region than a TS receiver. 

\par Other information theoretical studies of SWIPT with nonlinear EH models have appeared in \cite{Varasteh:2018,Morsi:2017,Morsi:2020}.

\subsubsection{Waveform}

\par Let us now consider the SISO multi-subband transmission such that \eqref{BB_model} becomes $y_n = h_n x_n + w_n$ in subband $n$.
The capacity achieving waveform and input distribution remains an open problem. Nevertheless, interesting results are known assuming Gaussian inputs. Assuming a linear HPA (with $e_1=1$) and a SISO multi-carrier waveform with a general non-zero mean Gaussian inputs $\Re\left\{x_n\right\}\!\sim\!\mathcal{N}(\mu_{nr},P_{nr}-\mu_{nr}^2)$ and $\Im\left\{x_n\right\}\!\sim\!\mathcal{N}(\mu_{ni},P_{ni}-\mu_{ni}^2)$ on each carrier/subband $n=0,\ldots,N-1$, with $P_n=P_{nr}+P_{ni}$ and $\sum_{n=0}^{N-1}P_{n} \leq P$, problem \eqref{MI}-\eqref{MI_2} becomes
\begin{align}%\label{SWIPT_opt_problem}
\sup_{\left\{\mu_{nr},\mu_{ni},P_{nr},P_{ni}\right\}_{n=0}^{N-1}} \hspace{0.3cm}& \sum_{n=0}^{N-1}I(x_n;y_n) \label{MI_FS}\\
\mathrm{subject\,\,to} \hspace{0.3cm} &\Tr\left(\mathbf{Q}\right)\leq P,\\
& P_{\dc}^r(\mathbf{x}_0,...,\mathbf{x}_{N-1})\geq \bar{E},\label{MI_2_FS}
\end{align}
In \cite{Clerckx:2018b,Varasteh:2018a}, problem \eqref{MI_FS}-\eqref{MI_2_FS} were investigated for such Gaussian inputs. It was shown that, while single-subband favors asymmetric inputs with a zero mean as described in previous section, multi-subband favors non-zero mean and asymmetric inputs. 

\par
\begin{figure}[!t]
   \centerline{\includegraphics[width=0.9\columnwidth]{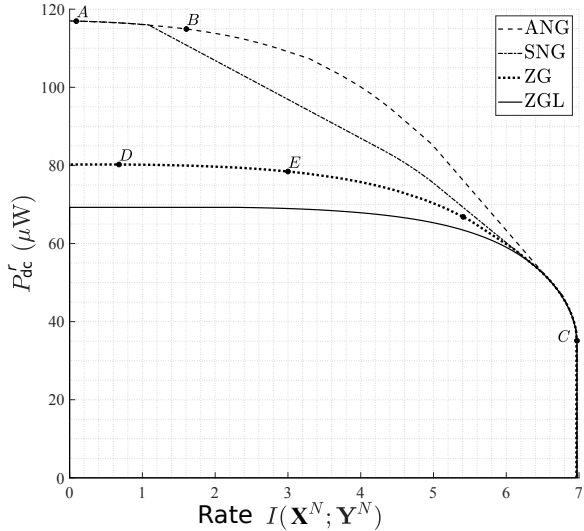}}
  \caption{The optimized R-E regions corresponding to ANG, SNG, ZG and ZGL with an average power constraint $P=100 ~\mu$W and SNR $20$ dB for an ideal receiver \cite{Varasteh:2018a}.}
  \label{Fig3_TCOM_2020_Morteza}
\end{figure}

 In Fig. \ref{Fig3_TCOM_2020_Morteza}, the R-E regions for \textit{Asymmetric Non-zero mean Gaussian} (ANG), \textit{Symmetric Non-zero mean Gaussian} (SNG) and \textit{Zero mean Gaussian} (ZG) are illustrated for $N=9$ over a frequency selective channel for $n_o=4$ in \eqref{vout_def} \cite{Varasteh:2018a}. We also compare with the R-E region corresponding to the optimal power allocations under the linear model assumption with $n_o=2$, denoted by \textit{Zero mean Gaussian for Linear model} (ZGL). As it is observed in Fig. \ref{Fig3_TCOM_2020_Morteza}, due to the asymmetric power allocation in ANG, there is an improvement in the R-E region compared to SNG. This gain is reminiscent of the gain observed for single-carrier modulation. Additionally, it is observed that ANG and SNG achieve larger R-E region compared to optimized ZG and that ANG and SNG perform better than ZGL. This highlights the fact that under EH nonlinearity ($n_o>2$), ZGL is far from optimal and the R-E region enhancement offered by ANG over ZGL in Fig. \ref{Fig3_TCOM_2020_Morteza} illustrates the gain obtained by accounting for the EH nonlinearity in SWIPT signal and system design.

\par The reason why ANG, SNG lead to larger R-E regions is due to the fact that the fourth order term in \eqref{vout_def} (and therefore the output DC power) is boosted by allowing the mean of the channel inputs to be non-zero \cite{Clerckx:2018b,Varasteh:2018a}. Hence, in contrast to ZGL, we note that the EH nonlinearity impacts not only the power allocation strategy across subbands but also the input distribution in each subband.

\par The superiority of non-zero mean inputs over zero mean inputs can be qualitatively explained by the fact that a multi-carrier unmodulated waveform, e.g. multisine, is more efficient in exploiting the EH nonlinearity and therefore boosting $P_{\dc}^r$ compared to a multi-carrier modulated waveform with CSCG inputs. From analysis and circuit simulations in \cite{Clerckx:2016b,Clerckx:2018b}, $P_{\dc}^r$ was shown to scale linearly with $N$ for an unmodulated multisine waveform. This originates from all the carriers being in phase, which turns on the rectifier diode (and therefore boosts its sensitivity) in the low power level in a periodic manner by sending high energy pulses every $1/\Delta_f$. On the other hand, $P_{\dc}^r$ scales at most logarithmically with $N$ for a modulated waveform due to the independent CSCG randomness (and therefore random fluctuations of the amplitudes and phases) of the information-carrying symbols across subbands.

\begin{figure}[!t]
%\centering
\begin{subfigmatrix}{2}
\subfigure[]{\label{a}\includegraphics[width = 4.2cm]{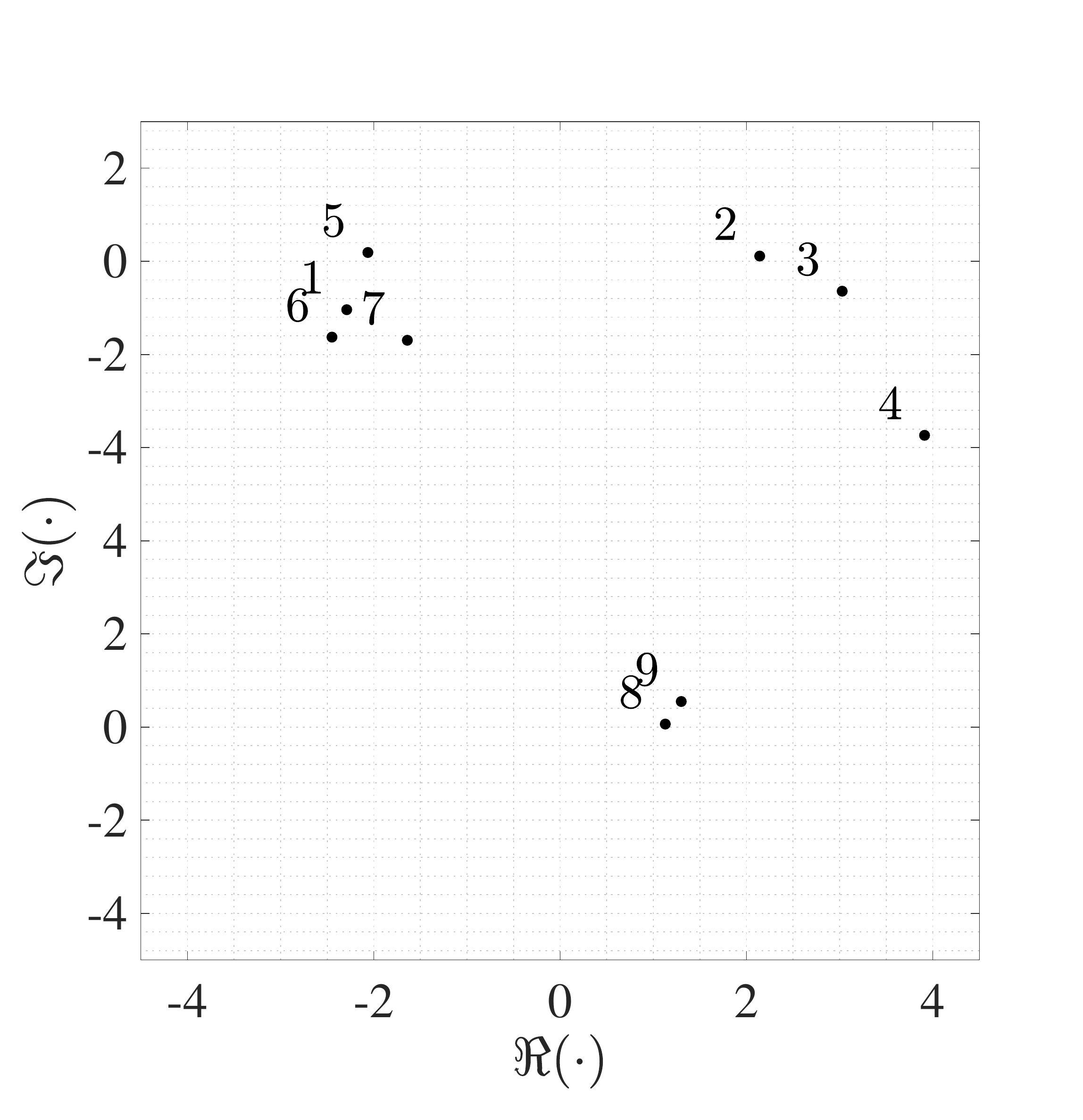}}
\subfigure[]{\label{b}\includegraphics[width = 4.2cm]{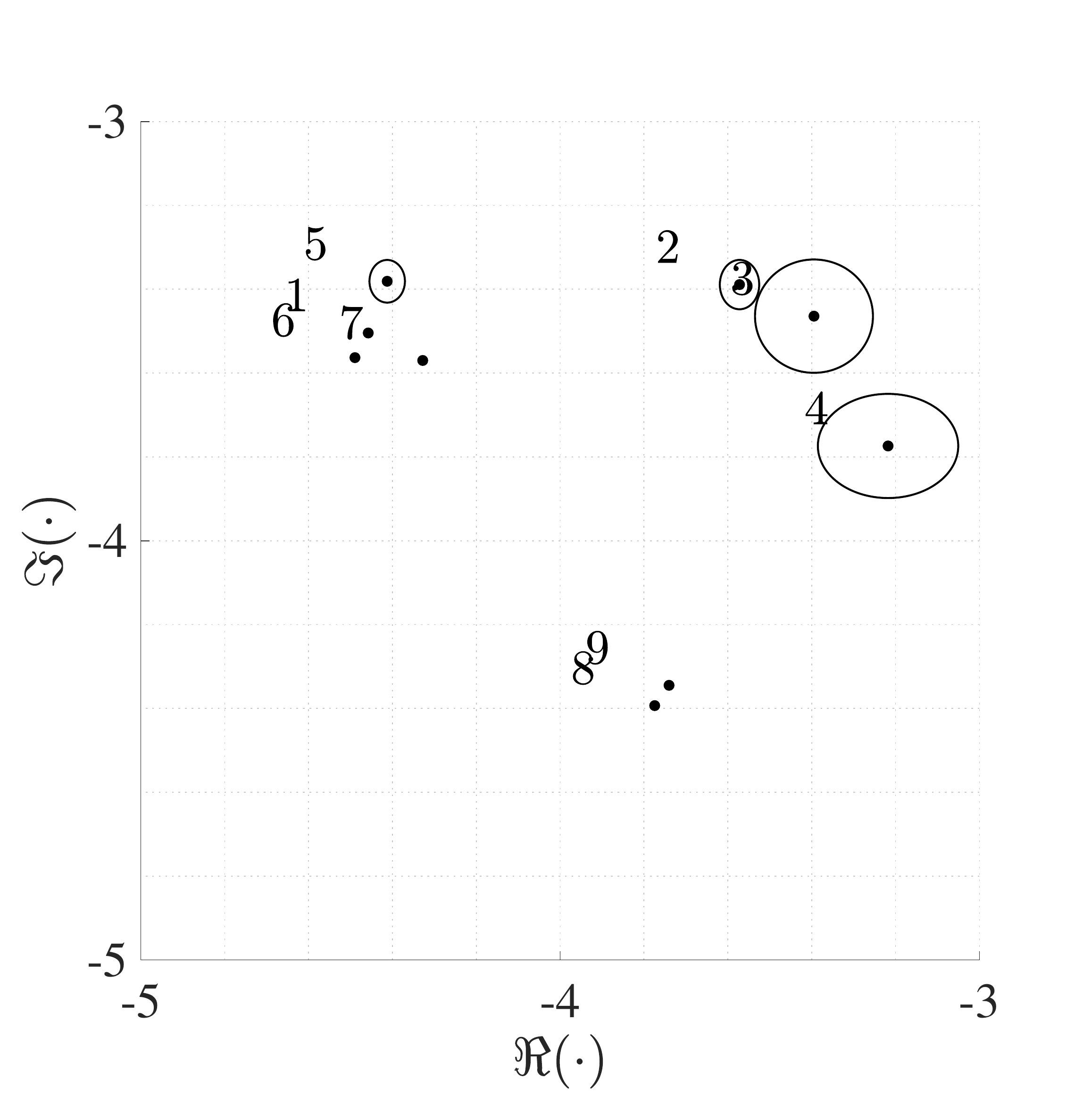}}
\subfigure[]{\label{c}\includegraphics[width = 4.2cm]{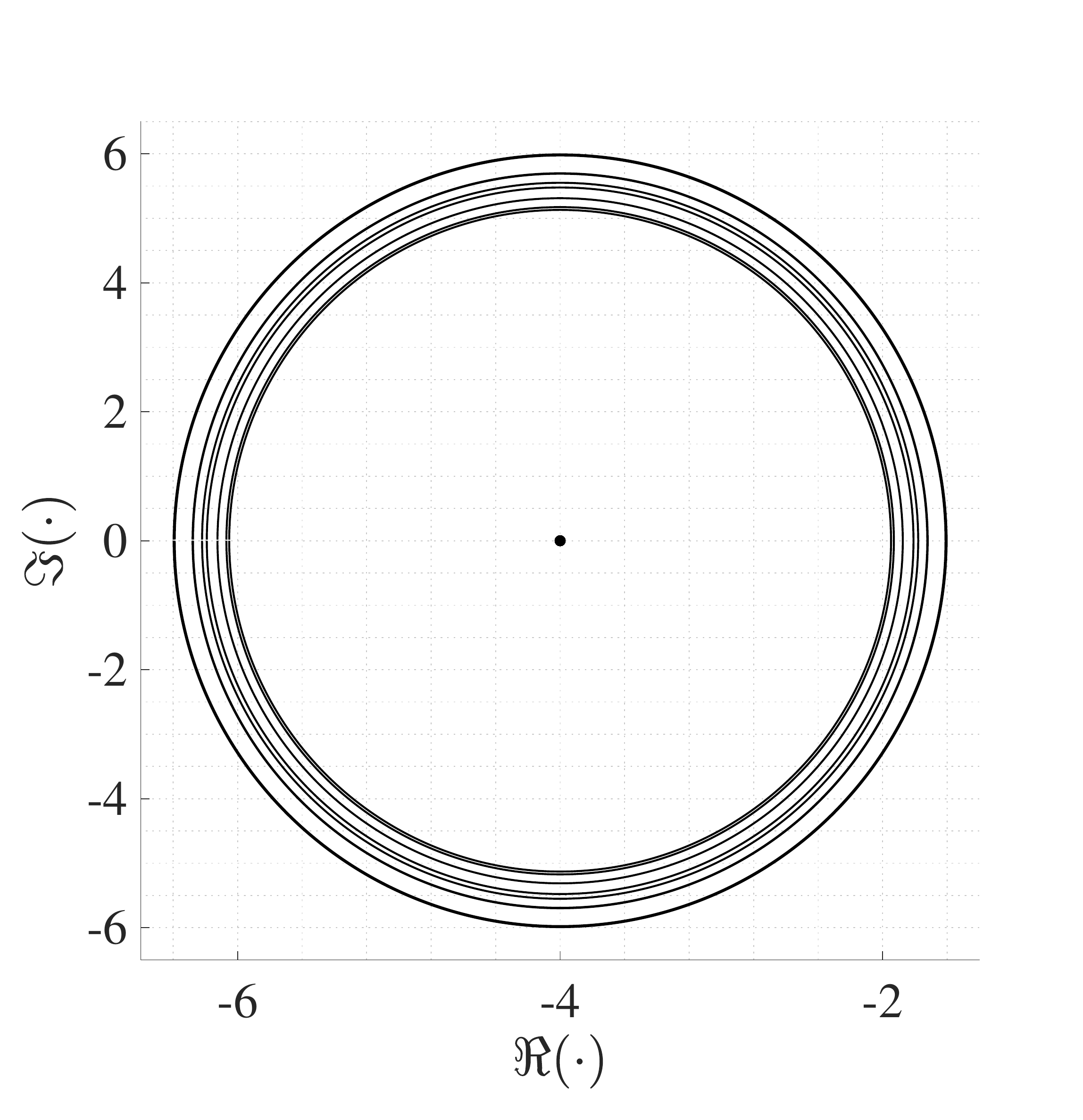}}
\subfigure[]{\label{c}\includegraphics[width = 4.2cm]{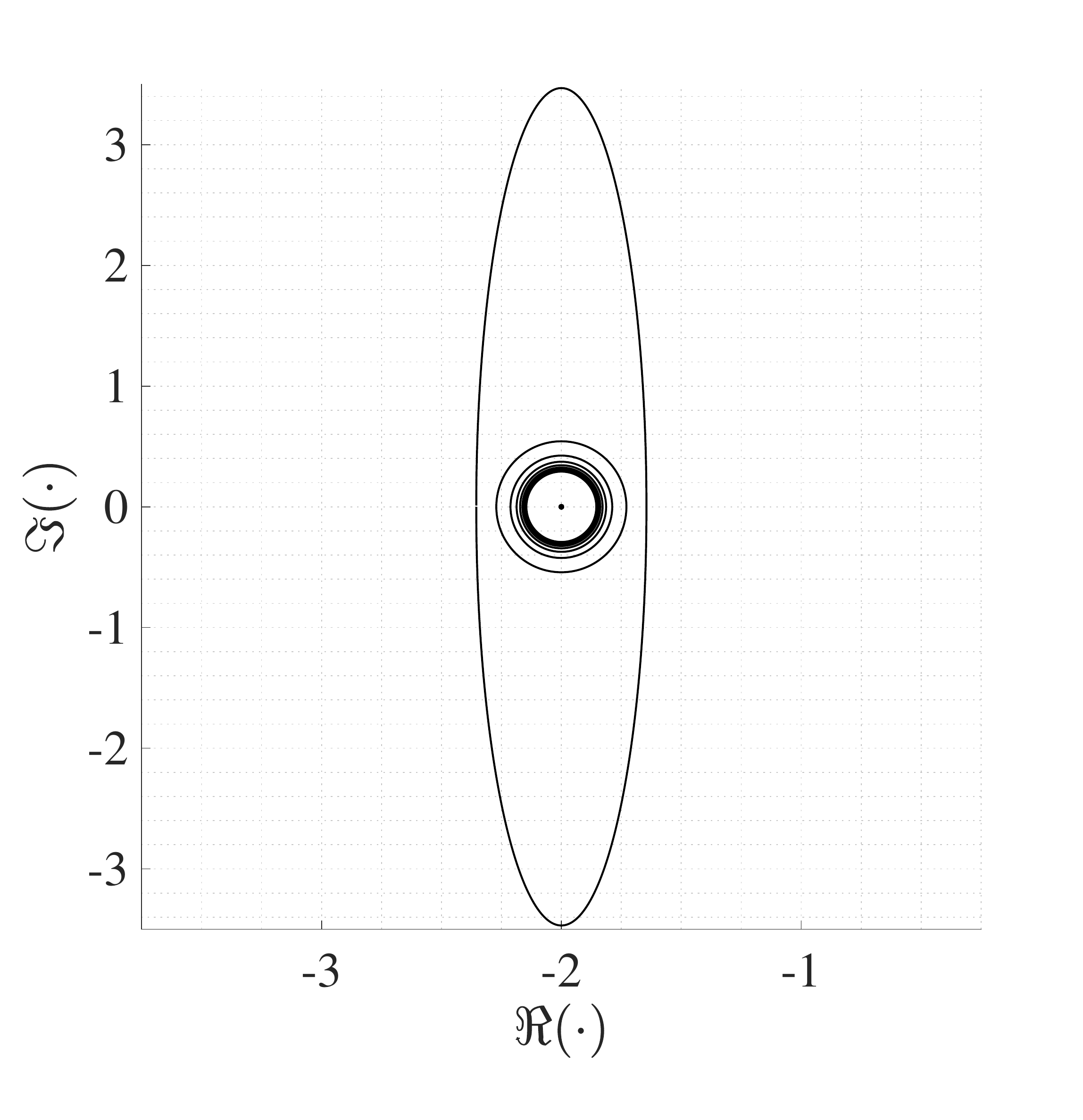}}
\end{subfigmatrix}
\caption{From (a) to (d), the mean and variance of different inputs corresponding to the points $A$, $B$, $C$ and $E$, respectively \cite{Varasteh:2018a}.}
\label{Fig4_TCOM_2020_Morteza}
\end{figure}

\par In Fig. \ref{Fig4_TCOM_2020_Morteza}, from (a) to (d), the optimized inputs in terms of their complex mean $\mu_n,~l=0,\ldots,8$ (represented as dots) and their corresponding variances $\sigma_{nr}^2,\sigma_{ni}^2,~n=0,\ldots,8$ (represented as ellipses) are shown for points $A,~B,~C$ and $E$ in Fig. \ref{Fig3_TCOM_2020_Morteza}, respectively. Point $A$ represents the maximum output DC power. Hence the waveform obtained in Point $A$ corresponds to the optimal deterministic multisine WPT waveform of \eqref{waveform_opt} \cite{Clerckx:2016b}. Point $B$ represents the performance of a typical input used for power and information transfer. Point $C$ represents the performance of an input obtained via the conventional water-filling strategy in multi-subband communications (the delivered power constraint is inactive). These three plots show that as we move from point $A$ to point $C$, the means of the different inputs decrease. Also, as we move to point $C$, the means get to zero with their variances increasing asymmetrically until the power allocation reaches the water-filling solution (where the power allocation between the real and imaginary components are symmetric). 

\par Point $D$ in Fig. \ref{Fig3_TCOM_2020_Morteza} corresponds to the input where all of the subbands other than the strongest one (in terms of the $\max\limits_{n=0,\ldots,N-1}|h_n|^2$) are allocated zero power. All the transmit power is allocated to the strongest subband in an asymmetric manner to either real or imaginary component of the input \cite{Varasteh:2017,Varasteh:2018a}. Note that this is different from the power allocation under the EH linear model assumption (i.e. ZGL), for which all the transmit power is allocated to the strongest subband but equally divided between the real and imaginary components of the input. Fig. \ref{Fig4_TCOM_2020_Morteza} illustrates the variances of the inputs on the different subbands corresponding to the point $E$ in Fig. \ref{Fig3_TCOM_2020_Morteza}. As we move from point $D$ to point $C$ (increasing the information demand at the receiver) in Fig. \ref{Fig3_TCOM_2020_Morteza}, the variance of the strongest subband varies asymmetrically (in its real and imaginary components). 

\par SWIPT with non-zero mean Gaussian inputs translates into an architecture based on the superposition of two waveforms at the transmitter: a power waveform comprising a deterministic multisine (as in section \ref{waveform_WPT_section}) and an information waveform comprising multi-carrier modulated (with CSCG or asymmetric Gaussian inputs) waveforms. The complex-valued information-power symbol on subband $n$ can then be explicitly written as $x_n=x_{\mathrm{P},n}+x_{\mathrm{I},n}$ with $x_{\mathrm{P},n}=\mu_{nr}+j \mu_{ni}$ representing the deterministic power symbol of the multisine waveform on subband $n$ and $x_{\mathrm{I},n}\sim \mathcal{N}(\mu_{nr},P_{nr}-\mu_{nr}^2)+j\mathcal{N}(\mu_{ni},P_{ni}-\mu_{ni}^2)$ representing the zero-mean Gaussian distributed information symbol of the modulated waveform on subband $n$. 

\par Since $x_{\mathrm{P},n}$ is deterministic, the receiver could operate with and without waveform cancellation. In the former case, the contribution of the power waveform is subtracted from the received signal after down-conversion from RF-to-baseband (BB) and Analog-to-Digital Conversion (ADC). In the latter case, a power waveform cancellation operation is not needed and the baseband receiver decodes the translated version of the symbols.

\par The above waveform and input distribution discussion, though introduced for the ideal receiver in Fig. \ref{three receiver architectures}, also holds for TS and PS receivers. Nevertheless, one interesting implication of the non-zero mean Gaussian inputs is that TS can outperform PS in multi-subband transmission. It is indeed shown in \cite{Clerckx:2018b} that for a sufficiently large $N$ (e.g. $N=16$), TS is preferred at low SNR and PS at high SNR, but in general the largest convex hull is obtained by a combination of PS and TS. This contrasts with the single-subband case where PS outperforms TS.

\par The above discussion relies on Gaussian inputs. Leveraging the above observations on input distribution, the design of of multicarrier waveform with finite constellation under EH nonlinearity was studied in \cite{Bayguzina:2019}. The authors adapted PSK modulation to SWIPT requirements and showed the benefits of asymmetric PSK modulation to enable a larger R-E region compared to that obtained with conventional symmetric PSK constellations. Another SWIPT architecture that leverages the waveform design can be found in \cite{Kim:2016}. 

\subsubsection{Coding} 
As we have indicated, there is much insight into information-theoretic, communication-theoretic, and system design aspects of WIPT.  Yet, there has been limited study of practical codes for the WIPT problem: practical codes that approach or achieve R-E limits have only recently been developed \cite{BasuV2019}.  Yet, a coding theory viewpoint on simultaneously transmitting information and energy is important in practice. 

Constrained codes have been developed for the WIPT problem \cite{BarberoRYY2001, FouladgarSE2014, TandonMV2014a}, especially in the noiseless setting. When requiring smooth energy delivery, e.g.\ due to finite-sized battery at the receiver, run-length limited (RLL) codes may not be best for WIPT \cite{TandonMV2018}, but subblock energy-constrained codes (SECCs) \cite{TandonMV2016, TandonKM2018} and skip-sliding window codes \cite{WuTVM2018, WuTMV2020} may be better.  Initial analyses of constrained code performance over noisy channels indicates they do not achieve information-theoretic limits. 

In establishing fundamental limits of WIPT, notice that a certain optimal distribution over the transmitted symbols is required for the receiver to extract a given amount of expected energy. Further, in a number of important cases, this distribution on the signaling constellation is not uniform \cite{Varshney:2008, Varshney:2012, Varasteh:2017b}.  Non-uniform input distribution requirements for WIPT rules out linear codes since linear codes can achieve Shannon capacity only when the optimal input distribution is uniform over the input alphabet. One might consider nonlinear algebraic codes like Kerdock codes and Preparata codes~\cite{VanLint1998}, but there is limited understanding of their performance for asymmetric input distributions and would  likely not achieve capacity.

WIPT was also explored using the concatenation of an inner nonlinear trellis code with an outer LDPC code, showing performance $\sim0.8$ dB away from the capacity of an AWGN channel \cite{DabirniaD2016}, but did not achieve the R-E limit. 

To achieve information-theoretic R-E limits using practical codes, \cite{BasuV2019} built on polar codes for asymmetric channels \cite{HondaY2013, SutterRDR2012}, and proposed two polar coding techniques that achieve R-E limits: one technique involved concatenating nonlinear mappings with linear polar codes, whereas the other involves using randomized rounding \cite{BasuV2019}.  The work focused on  discrete memoryless channels and the AWGN channel with peak output power constraint, but investigating optimal polar codes for realistic nonlinear WPT system models is of interest.

\subsubsection{Multi-Antenna and Intelligent Reflecting Surface}
\par In a MISO setup $y_{n} =  \mathbf{h}_{n} \mathbf x_{n} +w_{n}$, it can be shown for a general multi-band transmission that MRT in each subband is optimal \cite{Clerckx:2018b}. Hence, the optimal input symbol vector can be written as $\mathbf x_{n}=\bar{\mathbf{h}}_{n}^H x_{n}$ with $\bar{\mathbf{h}}_{n}=\mathbf{h}_{n}/\left\|\bar{\mathbf{h}}_{n}\right\|$ and $x_{n}$ designed according to the optimal input distribution/waveform of a SISO transmission. 
\par In a multi-band IRS-aided SWIPT with linear HPA (with $e_1=1$), $M=1$ and $L$ elements, $y_{n} =  \left(h_{\mathrm{d}_n}+\mathbf{h}_{\mathrm{r},n}\mathbf{\Theta}\mathbf{h}_{\mathrm{i},n}\right) x_{n} +w_{n}$, and thanks to \eqref{opt_IRS_1}-\eqref{opt_IRS_4}, the R-E region of \eqref{MI_FS}-\eqref{MI_2_FS} is expanded as
\begin{align}%\label{SWIPT_opt_problem}
\sup_{\mathbf{\Theta},\left\{\mu_{nr},\mu_{ni},P_{nr},P_{ni}\right\}_{n=0}^{N-1}} \hspace{0.3cm}& \sum_{n=0}^{N-1}I(x_n;y_n) \label{MI_FS_IRS}\\
\mathrm{subject\,\,to} \hspace{0.3cm} &\Tr\left(\mathbf{Q}\right)\leq P,\\
& P_{\dc}^r(x_0,...,x_{N-1},\mathbf{\Theta})\geq \bar{E},\label{MI_2_FS_IRS}\\
&\mathbf{\Theta}=\mathrm{diag}(\mathbf{\Theta}_1, \mathbf{\Theta}_2, \ldots, \mathbf{\Theta}_G),\\
&\mathbf{\Theta}_g^H\mathbf{\Theta}_g=\mathbf{I}_{L_G}, \forall g,\\
&\mathbf{\Theta}_g=\mathbf{\Theta}_g^T, \forall g.\label{MI_6_FS_IRS}
\end{align}
This problem is addressed in \cite{Zhao:2020} under the simpler case of single connected IRS (i.e.\ diagonal $\mathbf{\Theta}$) and symmetric non-zero mean Gaussian inputs.

\section{WPT and WIPT Signal and System Design Methodologies}\label{methodology_section}

We now reflect on the methodology to design WPT and WIPT. Two approaches are discussed, namely the ``model and optimize'' approach and the ``learning'' approach.

\subsection{The ``Model and Optimize'' Approach}

\par All the above communications and signal design/processing techniques for WPT and WIPT were developed following a systematic ``model and optimize'' approach, namely an analytical (physics-based) nonlinear model of the EH (Section \ref{EH_section}) was first derived and tools from signal processing, optimization, communication and information theories were then developed to design those techniques. As demonstrated in the previous sections, the ``model and optimize'' approach that accounts for the EH nonlinearity provides significant potential and gains towards efficient WPT/WIPT design. Importantly, all experimental results in \cite{Kim:2018,Kim:2020,Shen:2021} so far have validated the theory and the designs developed in \cite{Clerckx:2016b,Clerckx:2017,Huang:2017,Huang:2018,Clerckx:2018b,Clerckx:2018c,Clerckx:2019,Varasteh:2017,Varasteh:2017b,Varasteh:2018a,Varasteh:2018,Zeng:2017,Shen:2020a,Shen:2020c}.  
\par The EH nonlinearity has now appeared through this approach to play a crucial role and have a profound impact on WPT and WIPT signal and system designs \cite{Clerckx:2019}. Nonlinearity leads to a WIPT design quite different from that of conventional wireless communication, and changes the basic characteristics of the PHY and MAC layers such as input distribution and modulation, waveform, RF spectrum use, beamforming and multi-antenna, resource allocation as well as the transmitter and receiver architecture. Moreover, ignoring the nonlinearity leads to inefficient designs of WPT and WIPT \cite{Clerckx:2016b,Clerckx:2019,Kim:2018,Shen:2020a,Shen:2020c}.

\par The benefits of the ``model and optimize'' approach are its ability 1) to identify optimal solution and reliable analytic performance guarantees on the accuracy of the solution and 2) to interpret the results and get insights into the signal and system design. Indeed, the approach relies on optimization and communication/information theories to derive optimal input signal for WPT/WIPT accounting for nonlinearity, which led to further insights into implementable approaches. Nevertheless, the ``model and optimize'' approach also comes with two severe \textit{limitations}. 
\par \textit{First}, although nonlinearity is incorporated in the harvester model of Section \ref{EH_section} and can capture the effect of input signal power and shape, several simplifying assumptions are made on perfect impedance matching, ideal low pass filter, diode modeling, parasitics, etc \cite{Clerckx:2016b}. Departing from those assumptions makes the model analytically not tractable, preventing to formulate the optimization of WPT/WIPT. This is illustrated by two examples on WPT waveform design, namely 1) in the presence of imperfect matching: the formulation leads to a chicken-and-egg problem where the waveform design is a function of the impedance mismatch and the impedance mismatch is itself a function of the waveform design due to nonlinearity; 2) in the presence of a non-ideal low pass filter: the choice of the load impacts WPT signal design and one may or may not benefit from using multisine/modulated signals over CW depending on the load and input power, though pinpointing the best set of waveform, input power and load is currently challenging \cite{Bolos:2016,Ouda:2018,Pan:2018}.  
\par \textit{Second}, even if mathematical modelling of the EH nonlinearity is feasible using approximations, finding the fundamental limits of WIPT is mathematically extremely challenging and computationally intensive, as evidenced by works \cite{Varasteh:2017b,Varasteh:2018a}. Generally speaking, identifying optimal inputs of nonlinear channels is known in information theory to be a complicated problem \cite{Fahs:2012}. This often leads to situations where no efficient algorithms are available to solve the problem at hand, and the typical and straightforward approaches to tackle such problems are either considering linearized models or obtaining approximations and lower bounds on capacity \cite{Lomnitz:2011}.   

\subsection{The ``Learning'' Approach}

\par Despite advances in communication and optimization theories, many systems subject to nonlinear responses are unknown in terms of their optimal behaviour (e.g. capacity) and signal/system design. WPT/WIPT are instances of such systems. Their designs face problems that cannot be mathematically formulated and/or for which no efficient solutions exist. The lack of tractable mathematical models of the rectenna, and more generally of the entire WPT chain (transmitter-receiver, including HPA), as well as algorithms to solve WPT/WIPT signal/system optimizations in general settings is a bottleneck towards efficient WPT/WIPT designs. In this regard, machine learning (ML) is instrumental since they can be used to circumvent these modelling and algorithmic challenges. This calls for a ``learning'' approach.
 
\par \textit{First}, let us consider WPT. By collecting data from circuit simulations and prototypes, we can investigate two different WPT design strategies. The first strategy consists in fixing the transmit waveform hypothesis set and learn the rectenna parameters (e.g. input power level, load). By considering load and input power levels as inputs to a deep neural network (NN), the output would be the best waveform taken in a predefined hypothesis class. For instance, consider two possible waveforms $\mathrm{WF}_1$ and $\mathrm{WF}_2$, and a training set consisting of inputs (power level, load) and output (the best waveform among $\mathrm{WF}_1$ and $\mathrm{WF}_2$). Supervised learning (logistic regression) enables to classify under which input power level and load, $\mathrm{WF}_1$ (resp. $\mathrm{WF}_2$) is preferred. This can be generalized to larger hypothesis classes and rectenna parameters, and tackles the limitations of the current approaches used in the RF and microwave literature \cite{Bolos:2016,Ouda:2018,Pan:2018}. The second strategy consists in fixing the rectenna parameters and learn the best transmit waveform. For a given load and input power, reinforcement learning can then be used to design waveform based on sequential feedback (measurements) from the environment (circuit and prototype), despite the presence of nonlinearity, impedance mismatch, etc. This would tackle the limitations of the ``model and optimize'' approach used in \cite{Clerckx:2016b,Clerckx:2017,Huang:2017,Zeng:2017,Huang:2018}.

\par \textit{Second}, let us consider WIPT. By drawing data samples from the model or from measurements, learning can be used to obtain well performing achievable strategies. For instance, a combination of supervised learning and reinforcement learning could be used to learn efficient input distributions for general nonlinear WIPT. This would tackle the limitations of the ``model and optimize'' approach used in \cite{Clerckx:2018b,Clerckx:2019,Varasteh:2017,Varasteh:2017b,Varasteh:2018a,Varasteh:2018,Boshkovska:2015}.

\par A ``learning'' approach towards WPT and WIPT will provide different perspectives on signal and system design, fundamental limits, channel estimation and feedback, low-complexity and efficient strategies.
Learning communications systems \cite{OShea:2016,OShea:2017} seeks to optimize transmitter and receiver jointly without any artificially introduced block structure, and over any type of channel without the need for prior mathematical modeling and analysis. Using ML to address communications-related problems has been studied in the past \cite{Ibnkahla:2000} and has received renewed interests due to the development of deep learning (DL) software libraries \cite{AlRfou:2016,Abadi:2016,Chen:2015b}, and specialized hardware. Several groups have in the past few years investigated DL applications in communications, e.g. interpreting a communication system as an autoencoder (when the system can be properly modeled) \cite{OShea:2016,OShea:2017}, channel decoding \cite{Nachmani:2016,Nachmani:2017,Gruber:2017}, compression \cite{OShea:2016b}, CSI feedback \cite{Wen:2018}, and optical communications \cite{Karanov:2018,Jones:2018}. NNs can also be trained for RF design to learn the behaviour of passive/active components and circuits \cite{Zhang:2003}.

\par In view of the above and that large training datasets can be created from  circuit simulations and experiments, learning appears well suited for WPT/WIPT. Nevertheless, learning is not a one-size-fits-all solution. Indeed, despite the potential to tackle the aforementioned limitations of WPT/WIPT, the drawback of the learning approach, in contrast with the model-and-optimize approach, is the lack of performance guarantees on the accuracy of the solutions and the lack of human-interpretability. This also calls for developing an interplay between model-and-optimize and learning approaches to get the best of both worlds.

\par A learning approach towards WPT and WIPT signal and system design, and an \emph{integrated} approach that leverages the complementarity and synergy of the learning and the model-and-optimize approaches, remain largely uncharted research territories, but physics-based learning may provide suitable directions (see Sec.~\ref{sec:physics-based-learn}). Some early and promising results on this learning approach for WPT and WIPT design have appeared in \cite{Varasteh:2019_ICASSP,Varasteh:2019_SPAWC,Varasteh:2020_TCOM,Kang:2018b,Kang:2020}. In the sequel, we further discuss some learning techniques for WPT and WIPT.

\subsubsection{Model-based and Data-based End-to-End Learning}\label{learning_SWIPT}

\begin{figure}
\begin{centering}
\includegraphics[scale=0.52]{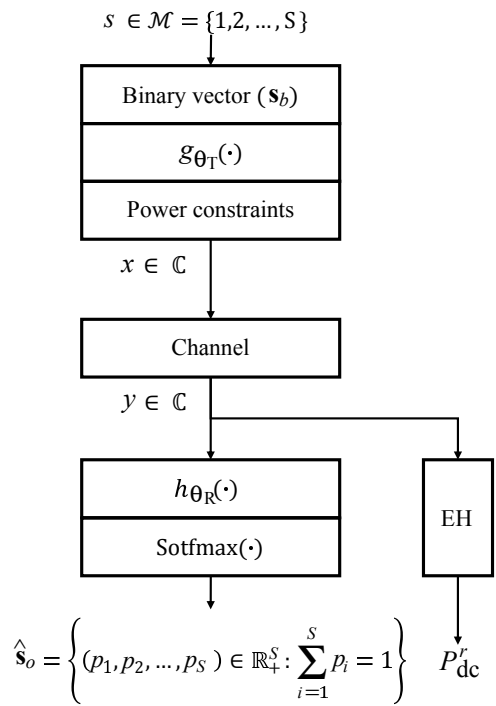}
\caption{NN-based implementation of the SISO single-subband SWIPT \cite{Varasteh:2020_TCOM}.}\label{Fig_nn_pp_structure}
\par\end{centering}
\vspace{-2.1mm}
\end{figure}

\par Considering a SISO single-subband transmission, the goal is to design the channel inputs $x$ such that the receiver demand of information and power is satisfied. Following \cite{Varasteh:2019_ICASSP,Varasteh:2019_SPAWC,Varasteh:2020_TCOM}, we consider the SWIPT system as an NN-based denoising autoencoder, where both the transmitter and receiver are implemented as NNs in order to perform the modulation and demodulation processes, respectively. A general NN-based implementation of the point-to-point system with ideal receiver is illustrated in Fig. \ref{Fig_nn_pp_structure}. At the transmitter, the message $s \in \mathcal{M}$ is converted into a binary vector of length $\lceil\log_{2}(S)\rceil$ denoted by $\mathbf{s}_b$ ($\lceil x \rceil$ returns the smallest integer larger than $x$). The vector $\mathbf{s}_b$ is then processed by the NN and is converted into a codeword $x=g_{\mathrm{\theta_{T}}}(s)$. Accordingly, the set of transmitter design parameters $\mathrm{\theta_{T}}$ are related to the weights and biases across the encoder module. To satisfy the power constraint, a power normalization is performed as the last layer of the transmitter module.

\par The symbol $x$ is corrupted by the channel noise. At the receiver, the estimation is performed by mapping the received noisy observation $y$ to an $S$-dimensional output probability vector denoted by $\mathbf{\hat{s}}_o$ (and estimating the message by returning the index corresponding to the maximum probability). Accordingly, $\mathrm{\theta_{R}}$ refers to the set of receiver parameters in terms the weights and biases across the decoder module.

\par
\begin{figure}
  \centering
    \includegraphics[scale=0.29]{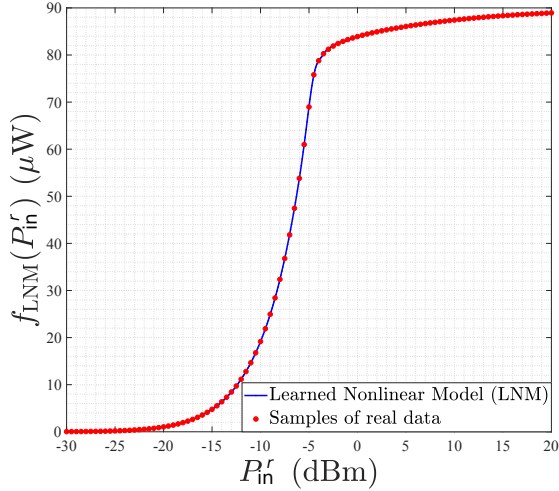}
    \caption{Model of an EH based on applying nonlinear regression over collected real data from rectifier in Fig. \ref{TD_schematic} with a CW input waveform \cite{Varasteh:2020_TCOM}.}\label{Fig_LNM}
\end{figure}

 The delivered DC power at the receiver can be modelled in two different ways, either based on an analytical model e.g. \eqref{vout_def} and \eqref{diode_current_power} \cite{Varasteh:2019_ICASSP,Varasteh:2019_SPAWC}, or by learning its input-signal/output-power relationship from measurement data \cite{Varasteh:2020_TCOM}. The latter is particularly powerful, and the larger the amount and diversity (in terms of a large range of input signal power and shape, load, etc) of the collected data, the higher the accuracy of the learned model. An EH model can obtained by applying nonlinear regression over collected real data \cite{Varasteh:2020_TCOM}. In particular, we study the data collected from the EH
circuit from Fig. \ref{TD_schematic}. The function we consider for modelling the EH is given as
\begin{align}\label{Eq_5}
    \hspace{-0.1cm}f_{\text{LNM}}(P_{\mathrm{in}}^r) = \sigma(W_3\sigma(W_2\sigma(W_1 P_{\mathrm{in}}^r +w_1)+w_2)+w_3),
\end{align}
where $P_{\mathrm{in}}^r$ denotes EH instantaneous input power, i.e., $P_{\mathrm{in}}^r=|y|^2$. $\mathcal{W}=\{W_1^{3\times 1},$ $W_2^{2\times 3},$ $W_3^{1\times 2},$ $w_1^{3\times 1},$ $w_2^{2\times 1},$ $w_3^{1\times 1}\}$ is the set of parameters to be optimized and $\sigma(\cdot)=\tanh(\cdot)$. In order to train the parameters we use Gradient Descent optimization applied over the following objective function
\begin{align}
    L_{\text{EH}}(\mathcal{W}) = \frac{1}{m}\sum_{i=1}^{m} (f_{\text{LNM}}(P_{\mathrm{in}}^r)-P_{\dc}^r)^2,
\end{align}
where $P_{\dc}^r$ is the collected output DC power corresponding to an instantaneous input power $P_{\mathrm{in}}^r$ and $m$ is the number of collected data used for training. In Fig. \ref{Fig_LNM}, the learned model (solid blue line) and the collected data (red dots) are illustrated\footnote{The measurements are obtained using Continuous Wave (CW) signals and assuming that the circuit is operating in steady state (since the effect of the transient state is negligible). Modulation learning using CW measurements is justified because amplitude modulation effectively corresponds to a CW with different power levels.}. 

\par Note that, since for power delivery purposes, the received RF signal is directly fed into the EH, the signal is not processed through the NN. We aim at following a learning approach and training the structure in Fig. \ref{Fig_nn_pp_structure} to minimize the objective function $L(\mathrm{\theta_{T}},\mathrm{\theta_{R}})$ given as
\begin{align} \label{Eq_4}
L(\mathrm{\theta_{T}},\mathrm{\theta_{R}}) = \frac{1}{|\mathcal{B}|}\sum_{l\in \mathcal{B}} \mathcal{L}(s_{o}^{(l)}, \hat{s}_o^{(l)})  +\frac{\lambda}{P_{\dc}^r(s^{(l)})},
\end{align}
where $\mathcal{B}$ is a randomly drawn minibatch of training data, which is assumed to be generated iid with a uniform distribution over the message set $\mathcal{M}$ and $|\mathcal{B}|$ is the cardinality of that batch. $\mathcal{L}(\mathbf{s}_{o}^{(l)}, \hat{\mathbf{s}}_o^{(l)}) = -\sum_{i=1}^S s_{o,i}^{(l)} \log {\hat{s}_{o,i}^{(l)}}$ is the cross entropy function between the one-hot representation of the $l^{th}$ training sample (denoted by $\mathbf{s}_{o}^{(l)}$) and its corresponding output probability vector $\hat{\mathbf{s}}_o^{(l)}$ at the receiver. $s_{o,i}^{(l)}$ and $\hat{s}_{o,i}^{(l)}$ indicate the $i^{\text{th}}$ entry of the vectors $\mathbf{s}_o^{(l)}$ and $\hat{\mathbf{s}}_o^{(l)}$, respectively.

\par The approach can be extended to coded modulation, in which case the output of the mapping function is of higher dimension, i.e.\ a sequence of symbols $x$. Similarly, the approach is extendable to multi-user settings, for instance the broadcast channel, multiple access channel and the interference channel. The end-to-end learning of SWIPT in all those types of channels has been studied in \cite{Varasteh:2020_TCOM}. Further extension to more general multi-carrier settings remains an open area. 
\par In the sequel, we illustrate the benefit of the above approach through two examples of modulation design under EH nonlinearity under linear and nonlinear HPA regime.

%%%%%%%%%%%%%%%%%%%%%%%%%%%%%%%%%%%
\begin{figure*}
\begin{centering}
\includegraphics[scale=0.65]{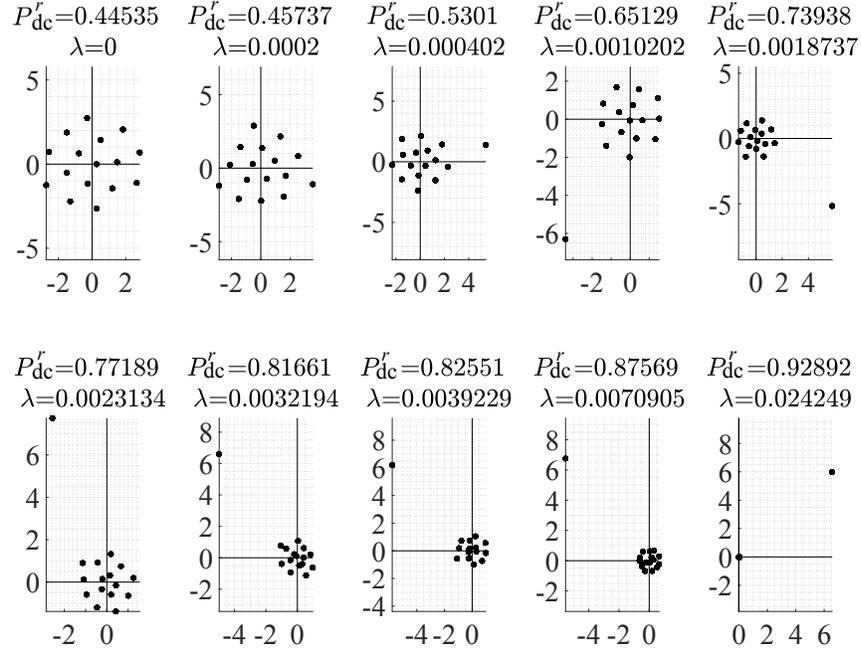}
\caption{Illustration of learned modulations under EH nonlinearity and linear HPA, respectively, with $M=16$, $P_{\rf}^r=10 \mu$W and SNR$=50$ (16.98 dB) for different values of $\lambda$ with corresponding $P_{\dc}^r$.}\label{Fig_absence_HPA}
\par\end{centering}
\end{figure*}

\par \textit{Example 1 - Modulation under Nonlinear EH and Linear HPA}: In Fig. \ref{Fig_absence_HPA}, we illustrate the set of constellation points (with $M=16$) obtained via end-to-end learning considering the EH as shown in Fig. \ref{TD_schematic} and using data-based model \eqref{Eq_5} for $P_{\rf}^r=10 \mu$W. We assume that a linear HPA with $G=1$ and $A_s=\infty$, and therefore do not need to model it. 

\par By increasing $\lambda$, the demand for power at the receiver increases. Accordingly, the modulation loses its symmetry around the origin in a way that some of the transmitted symbols are getting away from the origin. As the receiver power demand increases, the transmit signal modulation is reformed. In the extreme scenario, where the receiver demand for power is at its maximum, the symbols possess only two amplitudes (one away from the origin and the other equal to zero) and becomes an on-off keying signalling. In this example, since only one symbol is shooting away, the probability of the on (high amplitude) signal is 1/16 and the probability of the off (zero amplitude) signal is 15/16. 

\par An interesting observation about the learned modulations in Fig. \ref{Fig_absence_HPA} (in particular focusing on the last sub-figure) is that, the channel input empirical distribution approaches to a distribution with two mass points for the amplitude, one with ``low-probability/high-amplitudes'' and the other with ``high-probability/zero-amplitudes''. This result is inline with \eqref{flash_distr}, where we recall that the information-theoretic optimal channel input distributions of problem \eqref{MI_NL}-\eqref{MI_2_NL} for large power delivery (accounting for EH nonlinearity) follow the same behaviour, i.e., ``low-probability/high-amplitudes'' and ``high-probability/zero-amplitudes''.

\par Though modulation has been computed here using Fig. \ref{Fig_LNM} and illustrated for $P_{\rf}^r=10\mu$W, the same learning approach can be applied to model \eqref{vout_def} as in \cite{Varasteh:2019_ICASSP,Varasteh:2019_SPAWC} and to the curve-fitting model of \cite{Boshkovska:2015} as in \cite{Varasteh:2019_SPAWC}, as well as to other input power levels $P_{\rf}^r$ as in \cite{Varasteh:2020_TCOM}. The shape of the constellation would change depending on the model and the input power level. Interestingly, a major conclusion of \cite{Varasteh:2020_TCOM} is to utilize learning-based results to design non learning-based algorithms, which perform as well. In particular, inspired by the results obtained via learning, an algorithmic approach for coded modulation design has been proposed, which performs very close to its learning counterparts, and is significantly superior due to its high real-time adaptability to new system design parameters.

%%%%%%%%%%%%%%%%%

\begin{figure*}
\begin{centering}
\includegraphics[scale=0.65]{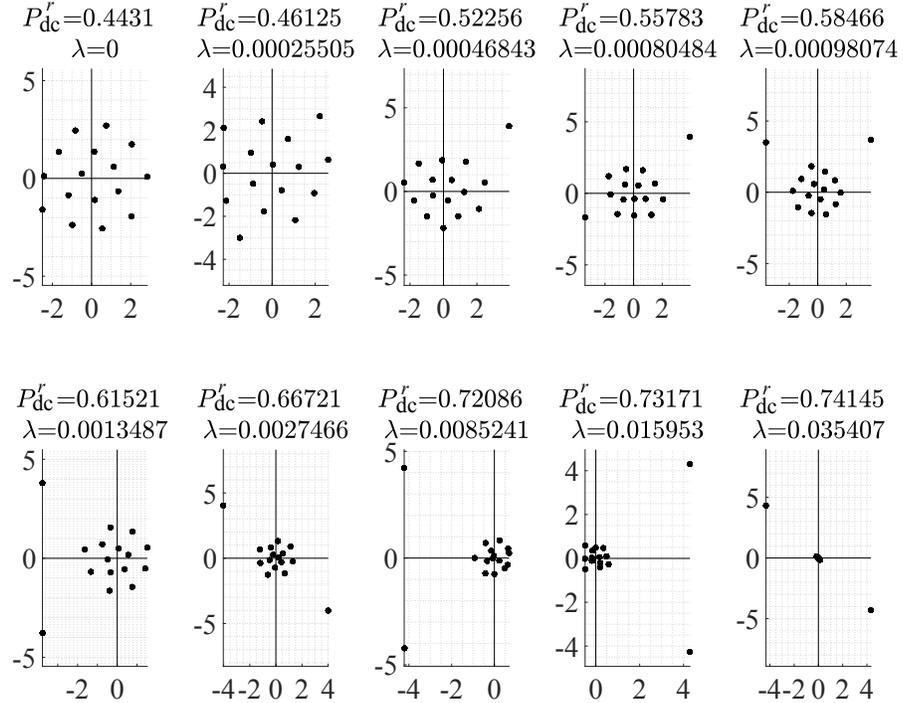}
\caption{Illustration of learned modulations (output constellation of HPA) under EH nonlinearity and HPA nonlinearity, with $M=16$, $P_{\rf}^r=10 \mu$W and SNR$=50$ (16.98 dB) for different values of $\lambda$ with corresponding $P_{\dc}^r$.}\label{Fig_presence_HPA}
\par\end{centering}
\end{figure*}

\par \textit{Example 2 - Modulation under Nonlinear EH and Nonlinear HPA}: In Fig. \ref{Fig_presence_HPA}, we revisit the results of Fig. \ref{Fig_absence_HPA} but accounting for the HPA nonlinearity of \eqref{HPA_NL_model} assuming $\beta=10$ and the output saturation voltage $A_s=4.47mV$. In Fig. \ref{Fig_absence_HPA}, the high amplitude symbol has an amplitude of 12.64mV and was left undistorted by the HPA because its nonlinearity was not modeled. Under HPA nonlinearity, there is no benefit to increase the voltage of one symbol above $A_s$. Instead, we note from Fig. \ref{Fig_presence_HPA} that another symbol is moved away from the origin to increase the EH output power. Comparing Fig. \ref{Fig_absence_HPA} and \ref{Fig_presence_HPA}, we also note that HPA nonlinearity leads to a decrease of $P_{\dc}^r$. This illustrates how deep learning can be used for end-to-end SWIPT design accounting for both transmit and receive nonlinearities. 

\subsubsection{Physics-based Learning}
\label{sec:physics-based-learn}
Physics-based learning has recently emerged as a way to bring together the benefits of learning and model-and-optimize approaches, especially in scientific discovery and engineering system design. Though it has not been used much yet in WPT settings, various manifestations as surveyed in \cite{Willard:2020} hold promise as compelling research directions. These manifestations include physics-guided loss functions for machine learning algorithms, physics-guided initialization of model parameters as a starting state for data-driven training, physics-guided neural network architecture design, residual modeling where the residuals of a physical model are characterized by a data-driven model, and hybrid modeling where physics-based and data-driven models are used together.
 
One example where physical knowledge has been used in WPT is in computing the R-E function for settings where the EH function is only known from simulation/experimental samples \cite{VarshneyS:2019}.  To bound the Shannon capacity in this setting, the known  physical smoothness of energy harvesting circuits is used to reduce the sample complexity needed in learning. A related idea to reduce sample complexity is to constrain deep learning algorithms to yield models that are    
scientifically consistent with known physics
\cite{Karpatne:2017, Jia:2019}. 

Another physics-based learning approach is to learn symbolic expressions (a first stage in model-and-optimize) using data-driven techniques.  The standard technique  for learning such physical laws from data is symbolic regression, which typically yields expressions using basic trigonometric and polynomial functions \cite{UdrescuT:2020, Udrescu:2020}, and may incorporate prior physical knowledge \cite{Hillar:2012}.  Information lattice learning is related to symbolic regression and also performs human-interpretable knowledge discovery from data, but based on physical knowledge of basic symmetries \cite{Yu:2017, Yu:2018}.  Alternatively, one can first train black-box deep learning models and then distill simple scientific laws from them \cite{Cranmer:2020}.  One can imagine learning the nonlinear phenomenology of WPT circuits and systems using these techniques and then optimizing system design.

In machine learning, there is often limited training data available in certain parts of the space.  Moreover, typical black-box machine learning models lack appropriate metacognition and are wildly confident in those regions since they are unconstrained by data.  To address this issue, one can combine data-driven models with models from physical knowledge via \emph{algorithm fusion}, whether using a hard switch or soft Bayesian weights determined by confidence levels estimated from density of training data.  In a way, the physical model acts as a backup to the data-driven model \cite{KshetryV:2019}.  This hybrid approach would allow safe optimization of WPT systems.

\section{Wireless Powered Internet-of-Things}\label{WPT_enabler_section}

The next-generation Internet-of-Things (IoT) is expected to connect tens of billions of edge devices (e.g.,  sensors, smartphones,  and wearable computing devices) to  automate a wide range of services, such as environmental monitoring, transportation, healthcare, traffic monitoring, and public-safety surveillance \cite{Melanie:Sensor:2012}. Among others, a key challenge is the high maintenance cost of  recharging the batteries of an enormous number of sensors and devices. WPT is a promising solution. In this section, we discuss the design of two specific wireless-powered IoT systems: wireless powered mobile edge computing and wireless powered crowd sensing.   

\par As a basic operation of mobile edge computing, \emph{mobile computation offloading} (MCO) refers to offloading  computation-intensive tasks from mobiles  to the cloud, thereby reducing the former's energy consumption and enriching their capabilities and features \cite{kumar2013survey}. On the other hand, realizing MCO involves  the transmission of data and messages across the air interface \cite{kumar:CanOffloadSave2010}. To rein in the incurred transmission-power cost has been driving researchers to  jointly design algorithms for MCO and adaptive transmission under the criterion of  maximum  mobile energy savings (see e.g., the survey in \cite{MEC:Survey} and references therein). On the other hand, active research is also being carrierd out on energy-efficient mobile (local) computing. In the area,  a wide range of  techniques have been proposed to decrease  mobile energy consumption by e.g.,  task  scheduling  \cite{Yao:Scheduling:1995}, dynamic power management \cite{Benini1:DPM:1999}, and control of CPU-cycle frequencies \cite{Pillai:DVS:2001,lorch:DynamicVoltage:2001, yuan:RealCpuScheduling:2003,zhang:MobileMmodel:2013}. In Section~\ref{Section:MEC}, we demonstrate the design of wireless powered mobile edge computing by jointly  designing  three relevant technologies:  1) WPT, 2) MCO, and 3) local computing using the mobile CPU. 
 
\par Traditional solutions for wireless sensor networking are limited in their coverage and scalability, as well as suffer from high maintenance costs~\cite{akyildiz2002wireless}. Recently, leveraging the massive number of sensors on  handheld and wearable equipment has led to the emergence of \emph{mobile crowd sensing} (MCS) \cite{ganti2011mobile}. The key MCS challenges include how to  prolong devices' battery lives and incentivizing  users' involvement. In Section~\ref{Sec:Sensing}, we discuss a MCS design that uses  WPT  as an incentivization mechanism  to  recharge the batteries of  \emph{mobile sensors} (MSs) in return for their participation in MCS.

\par The discussion in the preceding sections targets generic WPT and WIPT systems. Their designs use generic metrics (e.g., communication rates or end-to-end WPT efficiencies) and have not discussed the energy consumption of the receivers. In contrast, the wireless-powered IoT systems discussed in this section target specific applications. As a result, their designs feature more detailed energy consumption models for more elaborate computation and operations (e.g., CPU-cycle frequency control, sensing, and data compression) and application-specific performance metrics (e.g., data utility for crowd sensing). Due to limited space, we consider only two types of wireless-powered IoT systems, MEC and MCS systems. However, the design approach of jointly optimizing WPT,  communication, and computing is general and can be applied to other types of IoT systems. 

One particularly interesting area that is largely uncharted is wireless-powered distributed machine learning such as federated learning, which distributes a large-scale learning task over edge devices. Among others, the design of wireless-powered federated learning should address two issues. First, what is a reasonable model of the energy consumed by a device on updating an AI model (e.g., convolutional NN) using a local dataset? Second, the model convergence rate depends simultaneously on the computation speeds and communication rates of all participating devices. How should we optimally allocate transferred power to devices to accelerate the convergence?
  
\subsection{Wireless-Powered Mobile Edge Computing}\label{Section:MEC}

\begin{figure}[t]
\begin{center}
\subfigure[Wireless powered MEC system]{\includegraphics[width=9cm]{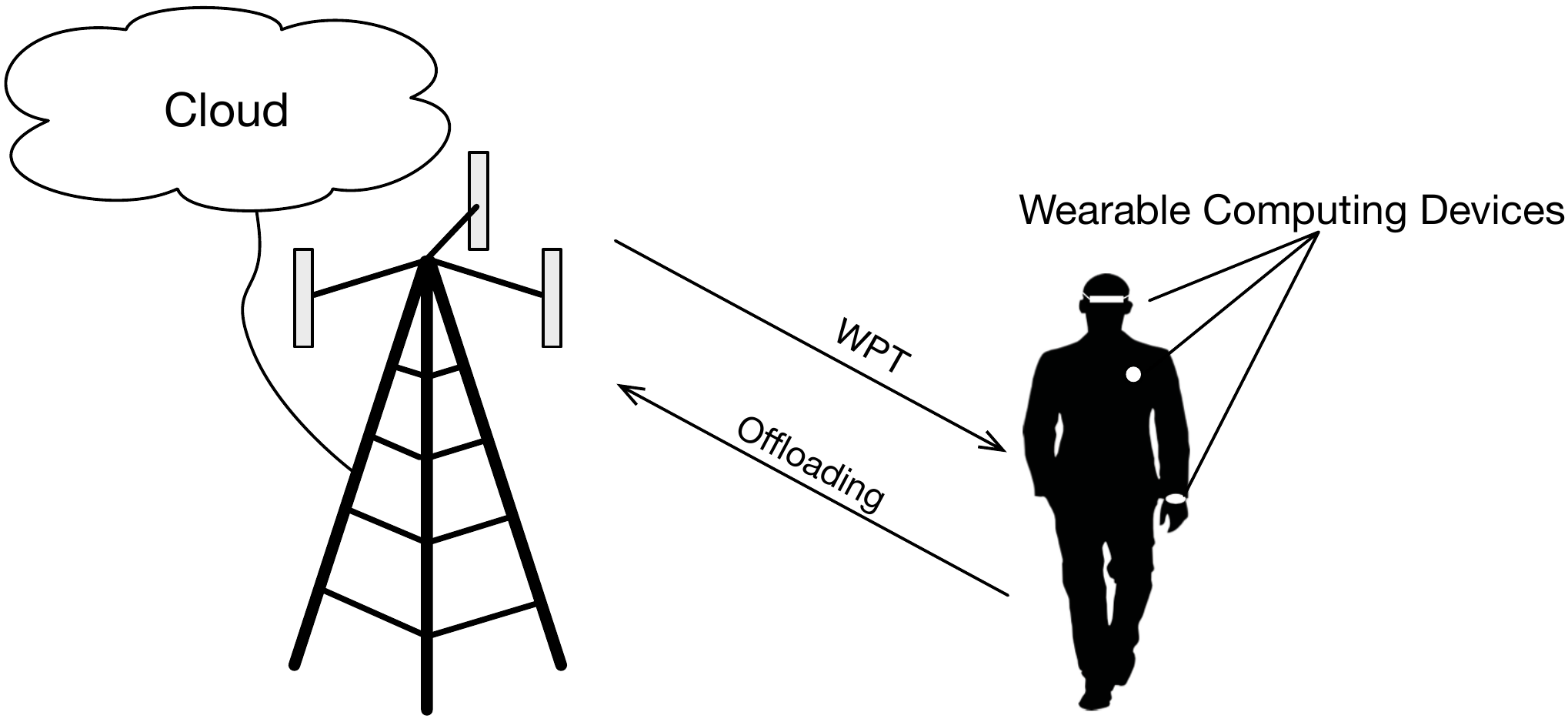}}\\
\subfigure[Mobile operation modes]{\includegraphics[width=9.1cm]{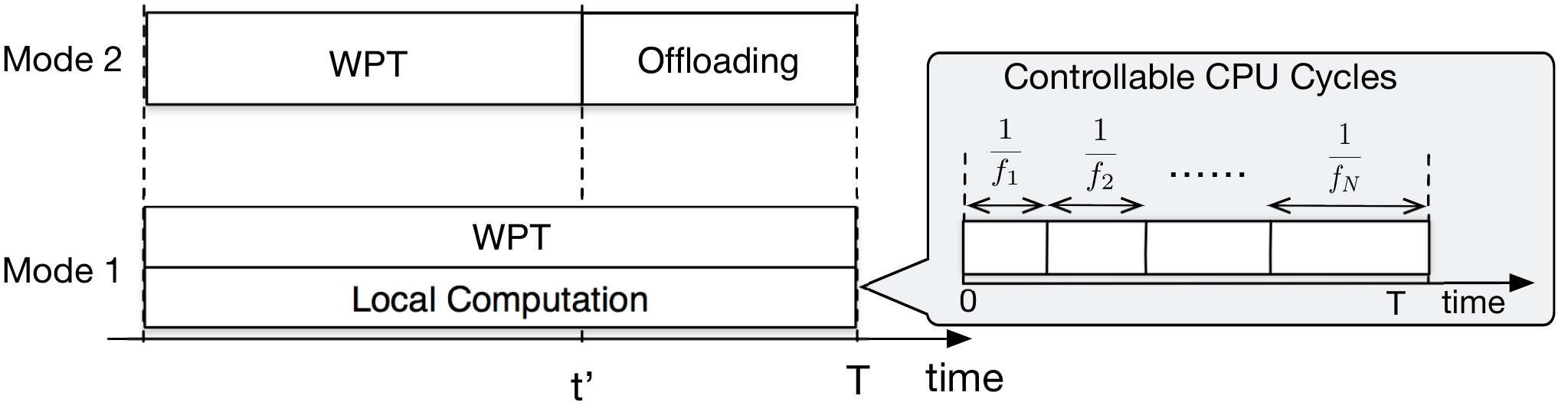}}\\
\caption{(a) Wirelessly powered MEC system and (b) the mobile operation modes.}
\label{system_model}
\end{center}
\end{figure}

A simple  wireless-powered MEC system is illustrated in Fig.~\ref{system_model}(a).  The mobile with milliwatt power consumption  (e.g., mobile sensor and wearable computing device) is equipped with a single   antenna and served by a  multi-antenna BS that is connected to  a cloud. The function of the BS is twofold: 1) performing WPT to the mobile or 2)  offloads a computation task from it. Alternatively, the task can be also executed locally at the mobile. On one hand, local computing and WPT are allowed to coexist in time since only the later need use the antenna. On the other hand, MCO and  WPT requires time sharing of the antenna given  its half-duplex transmission. It is worth mentioning that a more complex design for the case of executing a multi-task program will involve partitioning of the  program into multiple tasks with some offloaded and the rest executed locally. Considering the BS, energy  beamforming is applied for WPT and receive beamforming for receiving the signal during MCO. For simplicity, we assume the existence of channel reciprocity. That allows the   channel power gain to be  represented as a scalar  $g = |h|^2$. Last, the mobile harvests $P_{\dc}^r$  energy per  time unit.

%\vspace{-5pt}
\subsubsection{Edge Device Operations} \label{Section:LC:Model}

\begin{itemize}
\item {\bf Local computing}: The  task of the edge device is to process a fixed number of input data within a duration of  $T$ seconds.  Adopting  an existing  model    \cite{yuan:RealCpuScheduling:2003,  zhang:MobileMmodel:2013}, the required number of CPU cycles, denoted as $X$, can be modeled as a random variable with a given distribution.  To specify the distribution, let $p_k$ denote  the probability that the data processing is not completed  after $k$ CPU cycles and  $N$ the allowed maximum  number of CPU cycles. Mathematically, $p_k = \Pr(X \geq k)$ with $k = 1, 2, \cdots, N$.  It is assumed that mobile has the knowledge of the distribution so as to make a decision on if the computing task should be offloaded or not. 

\item {\bf CPU control}: The CPU-clock frequency can be controlled and its value directly determines the computation energy consumption. A single CPU cycle consumes a certain amount of energy, denoted as $E(f^{\mathsf{clk}})$, which is a function of  the frequency $f^{\mathsf{clk}}$. Based on  the model in \cite{burd:CpuEnergy:1996},  $E(f^{\mathsf{clk}})=\gamma (f^{\mathsf{clk}})^2$ where the given parameter    $\gamma$ depends on the  switched capacitance. The CPU-cycle frequencies for the CPU cycle $1, 2, \dots, N$ are represented by  $f^{\mathsf{clk}}_1, f^{\mathsf{clk}}_2, \dots, f^{\mathsf{clk}}_N$,  respectively. 

\item {\bf Computation offloading}: The device offloads computation by  transmitting  data to  the BS to be  processed  in the cloud. After that,  the computation   result is downloaded to the device.   The capacity of the  uplink channel  (in bit/s) is  denoted as $C$ and can be written as:  
\begin{equation}
 C=W\log\left(1+ \frac{P_{\rf}^t g}{\sigma^2}\right)  \label{rate} 
\end{equation} 
where $\sigma^2$ is the variance of complex white Gaussian channel  noise and $W$ the channel  bandwidth. As the cloud has practically practically infinite computational resources, the time for computing in the cloud   is  much smaller than offloading time. The same is downloading time of computation result because the high BS  transmission power keeps  the downlink transmission delay small. In addition, the computation result usually has  a small size and decoding it  at the mobile consumes negligible energy consumption with respect to that for   offloading and local computing. 
\end{itemize}

 \subsubsection{Wireless Powered Local Computing} 
Consider the case of executing the computation task using the local CPU. The objective  of  energy efficient local computing is  to control the   CPU-cycle frequencies,  $\{f^{\mathsf{clk}}_k\}$, so as to minimize the expected mobile energy consumption, denoted as $\bar{E}_{\textrm{mob}}$. As the computation task is likely to be completed with no more than $N$ CPU cycles, 
\begin{equation}\label{Eq:local:objective}
\bar{E}_{\textrm{mob}}\l(\{f^{\mathsf{clk}}_k\}_{k=1}^X\r) \approx  \sum_{k=1}^{N} \gamma p_k (f^{\mathsf{clk}}_k)^2.
\end{equation}
The frequency control must satisfy two constraints. The first constraint is the computation deadline: $\sum_{k=1}^{N} \frac{1}{f^{\text{clk}}_k} \le T$. The other is the well known  energy harvesting constraint accounting for  the fact that the  consumed energy cannot exceed the harvested energy at any time instant. The constraint can be decomposed into $N$ sub-constraints given as 
 \begin{equation}
\sum_{k=1}^m \gamma (f^{\text{clk}}_k)^2 \leq P_{\dc}^r\sum_{k=1}^m \frac{1}{f^{\text{clk}}_k},\quad   m=1, 2, \cdots, N. 
\end{equation}
At the left side of the above  inequality is the total energy consumed  by the first $m$ CPU cycles while  the total energy harvested by the end of the $m$th cycle is at the right side. Based on above discussion, the design of energy efficient local computing can be formulated as the following optimization problem: 
%\begin{equation}
\begin{align}\label{opt1}
 \min_ {\{f^{\text{clk}}_k\} } \quad  & \sum_{k=1}^{N} \gamma p_k (f^{\text{clk}}_k)^2 \\
\text{s.t.}\quad 
& \sum_{k=1}^m \gamma  (f^{\text{clk}}_k)^2 \leq P_{\dc}^r\sum_{k=1}^m \frac{1}{f^{\text{clk}}_k}, & \forall m,\\ 
& \sum_{k=1}^{N} \frac{1}{f^{\text{clk}}_k}\le T,  \\
&  f^{\text{clk}}_k >0,  & \forall k. \label{opt4}
\end{align}
%\tag{$\textbf{P1}$} 
%\end{equation}
Though the problem is  non-convex, it can be transformed into an equivalent convex  problem by replacing $\{f^{\text{clk}}_k\}$ with a new set of variables $\{y_k\}$ with $y_k= \frac{1}{f^{\text{clk}}_k}$. The equivalent  problem allows the application of Lagrange multiplier theory to shed light on the structure of the optimal policy. To this end, define two positive constants $a$ and $a'$ as 
\begin{equation}\label{Eq:a and a'}
a = \frac{\gamma N^3}{ T^3} \quad \text{and}\quad a' = \frac{\gamma}{T^3}\left( \sum_{k=1}^{N}  p_k^{\frac{1}{3}}\right)^2 \left(\sum_{k=1}^{N}p_k^{-\frac{2}{3}}\right).
\end{equation} Then  the optimal CPU-cycle frequencies $\{f^{\text{clk}\star}_1, f^{\text{clk}\star}_2, \cdots, f^{\text{clk}\star}_N\}$ that solve the optimization problem in \eqref{opt1}-\eqref{opt4} have the following properties \cite{WP:MEC}. 
\begin{enumerate}
\item {\bf Low-power regime}: If the recovered DC power $ P_{\dc}^r < a$, the harvested energy is insufficient for accomplishing the computation task. 

\item {\bf Medium-power regime}: If the recovered DC power $P_{\dc}^r\in [a, a')$,  the optimal CPU-cycle frequencies should be set as
\begin{equation}\label{Eq:CPU:lambda:nonzero}
f^{\text{clk}\star}_k=\l[\frac{1}{T}\sum_{m=1}^{N} (p_m+\lambda)^{\frac{1}{3}} \r] (p_k + \lambda)^{-\frac{1}{3}}, \qquad \forall k 
\end{equation}
where the positive constant $\lambda$ is the Lagrange multiplier associated with  the energy harvesting constraint.

\item {\bf High-power regime}: If the recovered DC power $P_{\dc}^r\geq a'$, $\{f^{\text{clk}\star}_k\}$ are independent of $P_{\dc}^r$: 
\begin{equation}\label{Eq:CPU:lambda:zero}
f^{\text{clk}\star}_k =\l(\frac{1}{T}\sum_{m=1}^{N} p_m^{\frac{1}{3}} \r) p_k^{-\frac{1}{3}}, \qquad \forall k. 
\end{equation}
\end{enumerate}

\subsubsection{Wireless Powered Computation Offloading} 
Consider the case of offloading the computation task to the cloud. The objective of energy efficient computation offloading is to maximize the mobile energy savings, referring to the difference between harvested energy and transmission energy consumption. As shown in Fig.~\ref{system_model}(b),  the time interval $[0, T]$ is divided into two parts: $[0, t']$ and $(t', T]$, corresponding to WPT and offloading, respectively. The  amount of energy harvested over the interval $[0, t']$ can be written as $E_{\textrm{WPT}}(t')=P_{\dc}^r t'$. Consider offloading. The most energy-efficient data transmission policy under a deadline constraint as proved in \cite{PrabBiyi:EenergyEfficientTXLazyScheduling:2001} to be the fixed-rate transmission over  the interval $(t', T]$. Given the result, let  $E_{\textrm{off}}(t')$
represent  the offloading energy consumption and it can be expressed  as $E_{\textrm{off}}(t')=[2^\frac{L}{B(T - t')}-1]\frac{\sigma^2}{h} (T - t')$. It follows that  $ \l[E_{\textrm{WPT}}(t') - E_{\textrm{off}}(t')\r]$ gives  the energy savings. 
As the function need not be  monotone over  $t'$,  its maximum  can be found  by  optimizing the WPT/offloading time partitioning. To simplify notation, define the offloading duration $t = T - t'$. We can write  $E_{\textrm{WPT}}(t)=P_{\dc}^r (T - t)$ and $E_{\textrm{off}}(t)=(2^ \frac{L}{Bt}-1)\frac{\sigma^2}{g} t$. This allows  the objective function  to be simplified as 
\begin{equation}
E_{\textrm{WPT}}(t)-E_{\textrm{off}}(t)\!=\!P_{\dc}^r T + \l(\frac{\sigma^2}{g}\!-\!P_{\dc}^r\r) t- \frac{\sigma^2}{g} t 2^{\frac{L}{Bt}}. \label{Eq:Offloading:Objective}
\end{equation}
It follows that the design of energy efficient offloading can be formulated as the following optimization problem: 
\begin{equation}
\begin{aligned}
 \max_t  \qquad  &E_{\textrm{WPT}}(t)-E_{\textrm{off}}(t)\\
\text{s.t.}\qquad 
 & 0< t < T,\\
 & E_{\textrm{WPT}}(t)-E_{\textrm{off}}(t) \ge 0.\nn
\end{aligned}
\end{equation}
The problem is convex since  the objective function can be easily shown to be  a concave function for $t\in (0, \infty)$.This allows us to obtain the optimal offloading duration, denoted as $t^{\star}$, in closed form. To this end,  define a positive constant $a''$ as
 \begin{eqnarray}
 &a''=\dfrac{\sigma^2 }{e_1}\left\{ 1+  \left[ \frac{L \ln 2}{WT }+\tilde{W}(-e^{-1-\frac{L \ln 2}{WT }})\right] \right.\nn\\ 
&~~~~~~~~~ \left. \times  \exp{\left(\frac{L \ln 2}{WT }+\tilde{W}(-e^{-1-\frac{L \ln 2}{BT } }   )+1\right)} \right\} \label{Eq:Thresh:Off}
\end{eqnarray}
where  $\tilde{W}(x)$ is the Lambert function defined as the solution for $\tilde{W}(x) e^{\tilde{W}(x)}=x$.
Then the optimal offloading duration $t^\star$ has the following properties. 
\begin{enumerate}
\item {\bf Low-power regime}: If the recovered DC power $P_{\dc}^r < a''$, the harvested energy is insufficient for offloading.
\item {\bf Sufficient-power regime}:If $P_{\dc}^r  \ge a''$,  the optimal offloading duration
\begin{equation}
t^\star  =\frac{\ln 2\times L}{W\left[1+ \tilde{W}(\frac{P_{\dc}^r}{\sigma ^2 e} -\frac{1}{e})\right]}.  \label{Eq:rho}
\end{equation}
\end{enumerate}

\subsubsection{Optimal Offloading Decision} \label{Section:ModeSel}
The optimal offloading policy aims at maximizing the  mobile energy savings. Using the results obtained in preceding subsections, we are ready to derive the optimal  policy as follows. 
\begin{itemize}
\item If  either offloading or local computing is feasible but not both, i.e., $P_{\dc}^r \ge a $ for local computing or $P_{\dc}^r\ge a''$ for offloading, then this mode is chosen. 
\item If both  are feasible, we can select the preferred  mode by  comparing  their energy savings corresponding to  the optimal polices derived previously. 
\end{itemize}

\subsection{Wireless Powered Crowd Sensing}\label{Sec:Sensing}

\begin{figure}[!t]
  \centering
  \includegraphics[scale=0.43]{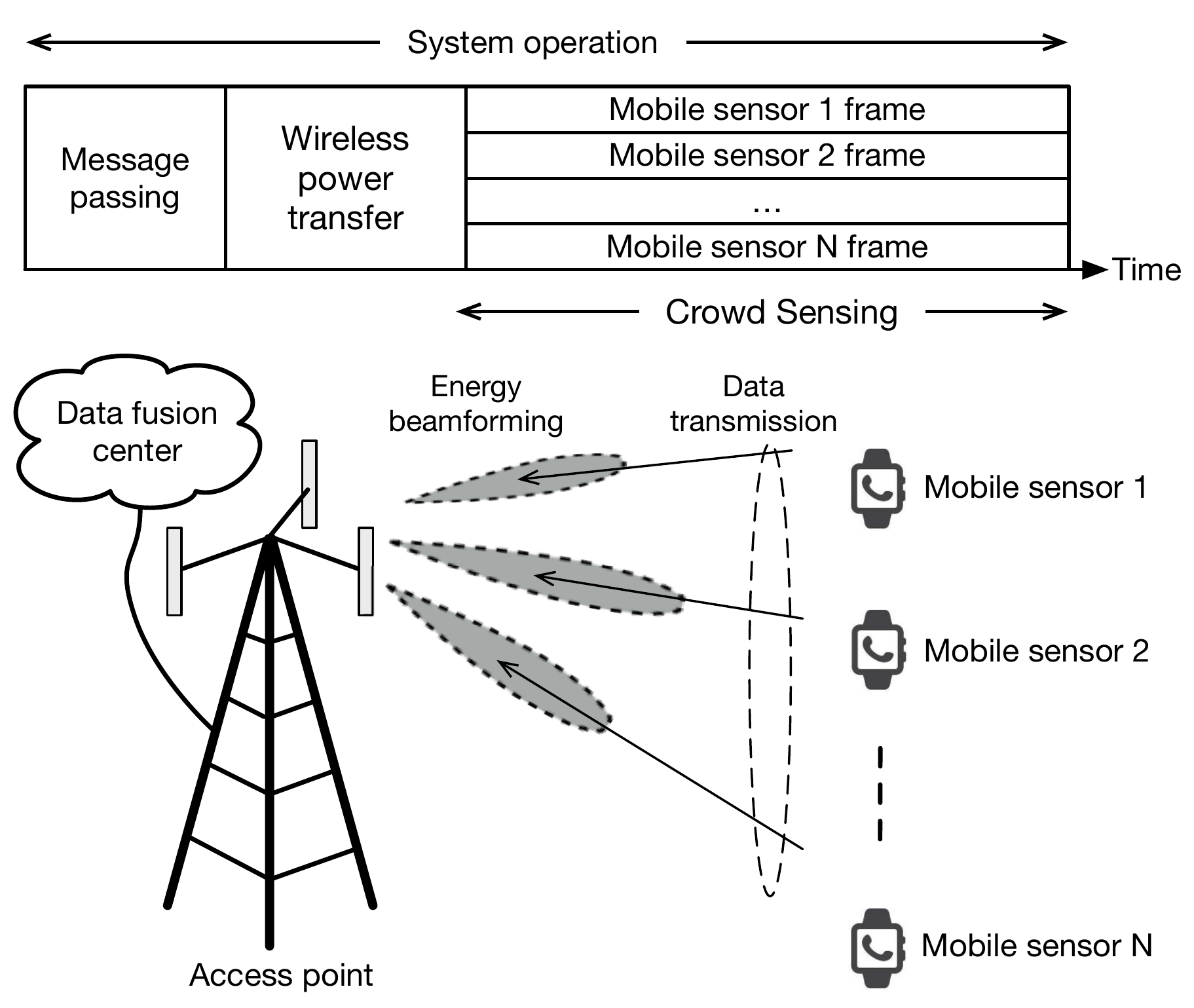}
  \caption{Wireless powered crowd sensing system.}
  \label{FigSys}
\end{figure}

Fig.~\ref{FigSys} illustrates a multi-user WPCS system. There exist in the system multiple single-antenna mobile sensors (MSs) connected to a single  multi-antenna BS. Consider crowd-sensing within some time window of interest. It is divided into  three phases: \emph{message passing}, \emph{WPT}, and \emph{crowd sensing} (see Fig.~\ref{FigSys}), which are desribed as follows.

\begin{itemize}
\item {\bf Message-passing phase}: Via   feedback,  the BS acquires knowledge of the  parameters of each sensor including the channel state, sensing and compression power. Then the  BS applies the knowledge to   jointly control  the  allocation of transferred power to MS's  and their  operations (i.e., data sensing, compression, and transmission). Thereby, the data utility is maximized while the energy consumption is   minimized. After solving the multi-objective optimization problem (elaborated in the sequel), the BS applies the optimal policy to inform  each individual MS to control their  compression ratios, sensing-data sizes, and the time sharing  of its operations.  We assume that  receiving these control parameters with small sizes consumes a sensor negligible energy.

\item {\bf WPT phase}: WPT is adopted by the BS to incentivize MS's to participate in crowd sensing   (see Fig.~\ref{FigSys}). For the  energy transferred by the BS to  each MS, a part is saved as a reward and the rest is used to  execute  the sensing task and transmit sensing  data to the BS.

 \item {\bf crowd-sensing phase} Adopting the settings communicated by the BS, the MS's perform  sensing, data compression, and transmission simultaneously. The data transmission by MS's are over parallel  OFDM sub-channels assigned by the BS. The  crowd-sensing duration, represented by  $T$, is separated into three  slots for sensing, compression, and transmission. Their durations are denoted as $t_n^{(s)}$, $t_n^{(c)}$, and $t_n$,  respectively. This introduces the following time constraint: 
\begin{equation}
\text{(Time constraint)} \quad t_n^{(s)} + t_n^{(c)} + t_n \leq T. \label{Eq:Time:Const}
\end{equation}

\end{itemize}

\begin{figure}[t]
  \centering
  \includegraphics[scale=0.3]{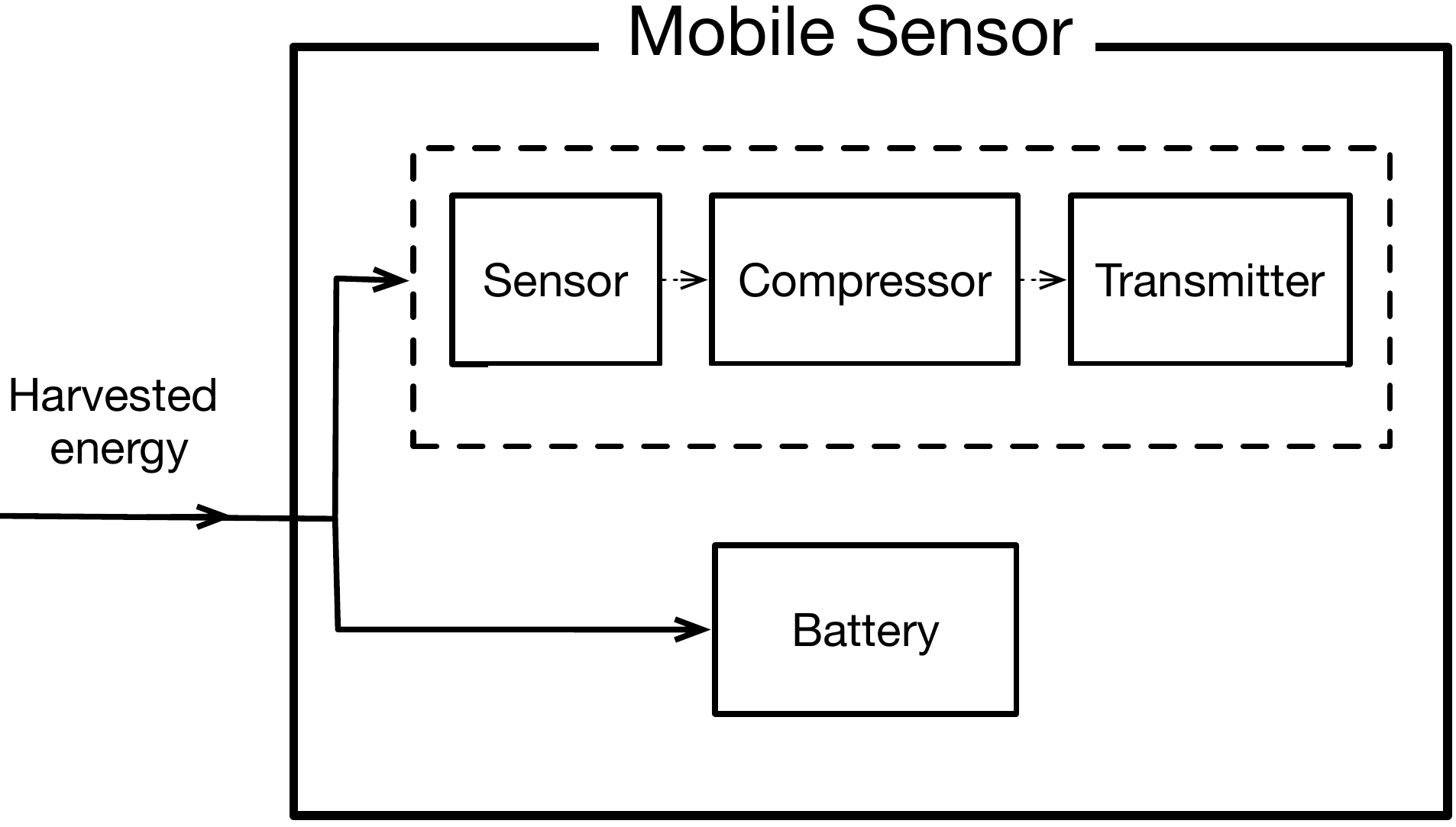}
  \caption{Architecture and  operations of a mobile sensor.}
  \label{FigMob}
\end{figure}

Consider the WPT phase lasting  $T_0$ (seconds).  The BS points $K$ sharp beams to perform simultaneous WPT to $K$ MSs. The transmit power of the $n$-th beam is denoted as $P_{\rf, n}^t$. The total  transmit power of the BS should not exceed  $P$. Mathematically, 
\begin{equation} \label{Eq:Power:Const}
\text{(Power constraint)} \quad \sum_{n=1}^K P_{\rf, n}^t\leq P.
\end{equation}
The  channel between the BS and MS $n$ has a power gain denoted as  $g_n$. Then the amount of energy harvested by  MS $n$ can be written as   $E_n^{(h)}(P_{\rf, n}^t)=e_3 g_n P_{\rf, n}^t T_0= P_{\dc, n}^r T_0$.

\subsubsection{Mobile Sensor Operations}
\par The operations of an arbitrary sensor, say MS $n$, are illustrated  in Fig.~\ref{FigMob} and described as follows. 
\begin{itemize}

\item {\bf Data sensing}: Based on experiments,  the total energy consumption for sensing, denoted as  $E_n^{(s)}(\ell_n^{(s)})$, is approximately proportional to the sensing-data size. Then size, denoted as $\ell_n^{(s)}$, can written as as  $\ell_n^{(s)} =s_n t_n^{(s)}$, where $s_n$ denoting  the output data rate and $t_n^{(s)}$ the sensing duration. Then we can model the sensing energy as  $E_n^{(s)}=q_n^{(s)} \ell_n^{(s)} = q_n^{(s)} s_n t_n^{(s)}$, where $q_n^{(s)}$ denotes the sensing energy consumption per bit.

\item {\bf Data compression}: Sensing data is compressed using lossless method. Some typical ones include Huffman, run-length , or Lempel-Ziv encoding. To simplify exposition,  all MSs are assumed to adopt  an identical   compression method. It has a  maximum compression ratio  $R_{\rm{max}}$. The MS's choose their own compression ratios and the choice of the $n$-th MS is represented by $R_n\in[1, R_{\rm{max}}]$. The gives the size of compressed data as  $\ell_n=\ell_n^{(s)}/R_n$. Some measurement results based on  popular compression techniques, i.e., XZ compression, Zlib, and Zstandard \cite{Arjancompression} suggest that the number of  CPU cycles needed to compress  $1$-bit of data can be fitted to  an exponential function of the compression ratio $R_n$: 
\begin{equation} \label{Eq:Comp:Complexity}
 C(R_n,\epsilon)=e^{\epsilon R_n}-e^{\epsilon},
\end{equation}
where the  positive constant $\epsilon$ depends on  the specific  compression algorithm. For a sanity check,  $C(R_n,\epsilon)=0$ for $R_n=1$ in the special case of   no compression. The  CPU-cycle frequency at MS $n$ is fixed to be  $f^{\text{clk}}_n$. Then we can obtain the compression time duration as $t_n^{(c)}=(\ell_n^{(s)} C(R_n,\epsilon))/f^{\text{clk}}_n$. As in the preceding sub-section, each CPU cycle consumes the energy of $q_n^{(c)}=\gamma (f^{\text{clk}}_n)^2$. The energy consumption for data compression, denoted by $E_n^{(c)}(\ell_n^{(s)}, R_n)$, is given as $E_n^{(c)}(\ell_n^{(s)}, R_n)=q_n^{(c)} \ell_n^{(s)} C(R_n,\epsilon)$ with $C(R_n,\epsilon)$ in \eqref{Eq:Comp:Complexity}. As a result, 
\begin{equation} \label{Eq:Comp:Energy}
E_n^{(c)}(\ell_n^{(s)}, R_n)\!=\!q_n^{(c)}\ell_n^{(s)}(e^{\epsilon R_n}\!-\!e^{\epsilon}).\!
\end{equation}
A similar model can be developed for lossy compression  e.g.,  data truncation after  discrete cosine transform. 

\item {\bf Data transmission}: After compression, each scheduled  MS uploads sensing  data to the BS. For the $n$-th MS, the transmission power and time duration are represented by  $\tilde{P}_{\rf, n}^t$  and $t_n$, respectively. Assuming channel reciprocity, the achievable transmission rate (in bits/s), denoted by $v_n$, can be given as $v_n=\ell_n/t_n=W\log_2\left(1+\frac{g_n \tilde{P}_{\rf, n}^t}{\sigma^2}\right)$.It follows that the transmission energy consumption is $E_n^{(t)}(\ell_n)=\tilde{P}_{\rf, n}^t t_n=\dfrac{t_n}{g_n}z(\ell_n/t_n)$, where we define the function $z(x)$ as $z(x)=\sigma^2(2^{\frac{x}{W}}-1)$.
\end{itemize}

\subsubsection{System Optimization}
Both MSs and (system) operator receive rewards. The reward requested by  a MS is in terms of  energy transferred from the BS minus that for supporting the MS operations. Let $E_n^{(r)}$ denote the energy reward requested by MS $n$.  It is proportional to the size of sensing data denoted as $\ell_n^{(s)}$, namely, $E_n^{(r)}=q_n^{(r)} \ell_n^{(s)}$ with $q_n^{(r)}$ being a fixed scaling factor. On the other hand,  the net operator's reward is the utility of the data collected from MSs minus the energy cost. A commonly  model of data utility is as follows (see e.g., \cite{yang2012crowdsourcing}). We can measure   the utility of $\ell_n$-bit data provided  by MS $n$ using  the logarithmic function $a_n \log(1+b_n \ell_n^{(s)})$, where $b_n$ is  the  loss of information caused by  compression and the weight factor $a_n$ depends on  the  data type. The monotonicity of the  function captures  the fact that more information-bearing data gives a  higher level of data utility, e.g., higher  machine-learning accuracy \cite{nasrabadi2007pattern} or better image/video resolution \cite{russ2016image}. On the other hand, the logarithmic function has the property of reflecting  the diminishing return as the data size increases, resulting in more  repeated and redundant data. With the model,  we write the the sum data utility for the operator as
\begin{equation} \label{Eq:Utility}
U(\boldsymbol{\ell}^{(s)})=\sum_{n=1}^{K}{a_n}\log(1+b_n \ell_n^{(s)}).
\end{equation}
The operator's reward, denoted by $R(\boldsymbol{\ell}^{(s)},\boldsymbol{P})$, can be modelled as
\begin{align} \label{Eq:Reward}
R(\boldsymbol{\ell}^{(s)},\boldsymbol{P})=\sum_{n=1}^{K} a_n \log(1+b_n \ell_n^{(s)})-c\sum_{n=1}^K P_{\rf, n}^t T_0,
\end{align}
where $c$ denotes the price of unit energy as measured against  unit data utility. 

The objective of system optimization is to maximize the operator's reward under two constraints. One is that the totally energy harvested by a MS should exceed its spent energy: 
\begin{equation}\label{Eq:Energy:Const}
E_n^{(r)} + E_n^{(s)} + E_n^{(c)} + E_n^{(t)} \leq P_{\dc, n}^r T_0, \qquad \forall n.
\end{equation}
The other is the time constraint per round: 
\begin{align}
\frac{\ell_n^{(s)}}{s_n}+\frac{\ell_n^{(s)} C(R_n,\epsilon)}{f_n^{\text{clk}}}+t_n\le T.\label{Eq:Time:Const:a}
\end{align}
We can make the observation that if the sensing rate $s_n$ and the CPU-cycle frequency $f_n^{\text{clk}}$ are fixed, the partitioning of crowd-sensing time of sensor $n$ for sensing, compression, and transmission can be determined by the \emph{sensing-data size $\ell_n$}, \emph{compression ratio $R_n$}, and \emph{transmission time $t_n$}. Then the  system operations can be optimized over all or a subset of these variables. 

For example, the problem of jointly optimizing joint  power allocation, sensing, and transmission can be formulated as 
\begin{equation*}
\begin{aligned}
\max_{\substack{P_{\rf, n}^t\ge0, \ell_n^{(s)} \ge0,\\ t_n\ge0}} \quad
&\sum_{n=1}^Ka_n\log(1+\ell_n^{(s)})-c\sum_{n=1}^K P_{\rf, n}^t T_0\\
\text{s.t.} \qquad
&\sum_{n=1}^K P_{\rf, n}^t \le P,\\
&\beta_n \ell_n^{(s)}+t_n\le T,\qquad \qquad \qquad \qquad \qquad \forall~n,\\
&\xi_n \ell_n^{(s)}+\dfrac{t_n}{g_n} f\l(\frac{\ell_n^{(s)}}{R_n t_n}\r)\le P_{\dc, n}^r T_0, ~~~~~ \forall~n,
\end{aligned}
\end{equation*}
where $\xi_n=q_n^{(r)}+q_n^{(s)} +q_n^{(c)} C(R_n,\epsilon)$ and $\beta_n=\frac{1}{s_n}+\frac{C(R_n,\epsilon)}{f_n^{\text{clk}}}$. This can be shown to be a convex program and thus can be solved efficiently using an existing algorithm. Based on the well-known KKT conditions, some light can be shed on the optimal policy \cite{WP:Sensing}. In particular, the optimal sensor-transmission duration ,
\begin{equation}\label{Eq:LosslessAppro}
t_n^\star\propto \frac{T}{W R_n \beta_n}.
\end{equation}
On the other hand, the optimal wireless-power allocation policy has a \emph{threshold}-based structure: 
\begin{equation}\label{Eq:OptPower}
P_{\rf, n}^{t\star}\!=\!
\begin{cases}
\dfrac{1}{e_3 g_n T_0}\l[\dfrac{t_n^\star}{g_n}f\l(\dfrac{T\!-\!t_n^\star}{R_n \beta_n t_n^\star}\r)\!+\!\dfrac{\xi_n (T\!-\!t_n^\star)}{\beta_n}\r],&\phi_n \geq \lambda^\star,\\
0,&\phi_n < \lambda^\star,
\end{cases}\nonumber
\end{equation}
where $\lambda^\star$ is a Lagrange multiplier and  $\phi_n$ represents the \emph{MS-scheduling priority function} given as 
\begin{equation}\label{Eq:LosslessPrio}
\phi_n= \frac{a_n e_3 g_n}{q_n^{(r)}+q_n^{(s)}+q_n^{(c)} C(R_n,\epsilon)+\frac{N_0\ln2}{g_n B R_n}}-c.
\end{equation}
The result suggests that  only the MSs with their priority functional values above  the threshold $\lambda^\star$ should  be scheduled  for participating in the crowd sensing operation and being rewarded with energy in return. Finally, the optimal sensing-data sizes $\{(\ell_n^{(s)})^\star\}$ are proportional to the corresponding sensing durations: 
\begin{equation}
(\ell_n^{(s)})^\star=
\begin{cases}
\dfrac{T-t_n^\star}{\beta_n},&\phi_n \geq \lambda^\star,\\
0,&\phi_n < \lambda^\star.
\end{cases}
\end{equation}

The problem of optimizing all system variables is non-convex. One low-complexity algorithm for finding a local optimal point is to iterate between solving convex sub-problems, each of which is over a subset of variables. For instance, one sub-problem can be that discussed above and the other being the joint optimization of compression and transmission, both of which are convex \cite{WP:Sensing}.

\section{Conclusions}\label{conclusions}

This article has first provided a tutorial overview of various signal processing techniques for WPT and WIPT, then discussed the benefits of two different design methodologies based on \textit{model and optimize} and \textit{learning} approaches, and finally showed how WPT, computing, sensing, and communication need to be jointly designed in future wireless powered applications. One first conclusion of the paper is to highlight that signal processing and machine learning techniques have an important role to play in WPT for future networks, but the techniques need to be developed in light of the physics and hardware constraints of WPT. This calls for abandoning naive and oversimplified linear models and accounting for nonlinearities and non-idealities at the transmitter and receiver ends. A second conclusion is that WPT will act an important enabler for future networks and opens the door to new challenges and opportunities where communications, computing, and sensing have to be jointly designed together with WPT. It is hoped that the signal processing, machine learning, computing, and sensing techniques presented here will help inspire future research in this promising area and pave the way for designing and implementing efficient WPT, WIPT, and wireless-powered systems and networks in the future.

\ifCLASSOPTIONcaptionsoff
  \newpage
\fi

\begin{IEEEbiography} 
{Bruno Clerckx} is a (Full) Professor, the Head of the Wireless Communications and Signal Processing Lab, and the Deputy Head of the Communications and Signal Processing Group, within the Electrical and Electronic Engineering Department, Imperial College London, London, U.K. He received the M.S. and Ph.D. degrees in applied science from the Université Catholique de Louvain, Louvain-la-Neuve, Belgium, in 2000 and 2005, respectively. From 2006 to 2011, he was with Samsung Electronics, Suwon, South Korea, where he actively contributed to 4G (3GPP LTE/LTE-A and IEEE 802.16m) and acted as the Rapporteur for the 3GPP Coordinated Multi-Point (CoMP) Study Item. Since 2011, he has been with Imperial College London, first as a Lecturer from 2011 to 2015, Senior Lecturer from 2015 to 2017, Reader from 2017 to 2020, and now as a Full Professor. From 2014 to 2016, he also was an Associate Professor with Korea University, Seoul, South Korea. He also held various long or short-term visiting research appointments at Stanford University, EURECOM, National University of Singapore, The University of Hong Kong, Princeton University, The University of Edinburgh, The University of New South Wales, and Tsinghua University. 
He has authored two books, 200 peer-reviewed international research papers, and 150 standards contributions, and is the inventor of 80 issued or pending patents among which 15 have been adopted in the specifications of 4G standards and are used by billions of devices worldwide. His research area is communication theory and signal processing for wireless networks. He has been a TPC member, a symposium chair, or a TPC chair of many symposia on communication theory, signal processing for communication and wireless communication for several leading international IEEE conferences. He was an Elected Member of the IEEE Signal Processing Society SPCOM Technical Committee. He served as an Editor for the IEEE TRANSACTIONS ON COMMUNICATIONS, the IEEE TRANSACTIONS ON WIRELESS COMMUNICATIONS, and the IEEE TRANSACTIONS ON SIGNAL PROCESSING. He has also been a (lead) guest editor for special issues of the EURASIP Journal on Wireless Communications and Networking, IEEE ACCESS, the IEEE JOURNAL ON SELECTED AREAS IN COMMUNICATIONS and the IEEE JOURNAL OF SELECTED TOPICS IN SIGNAL PROCESSING. He was an Editor for the 3GPP LTE-Advanced Standard Technical Report on CoMP. He is an IEEE ComSoc Distinguished Lecturer 2021-2022.
\end{IEEEbiography}

\begin{IEEEbiography}
{Kaibin Huang} (Fellow, IEEE) received the B.Eng. and M.Eng. degrees in electrical engineering from the National University of Singapore and the Ph.D. degree in electrical engineering from the University of Texas at Austin. He is currently an Associate Professor with the Department of Electrical and Electronic Engineering, University of Hong Kong, Hong Kong. He received the IEEE Communication Society’s 2019 Best Tutorial Paper Award, the 2015 Asia–Pacific Best Paper Award, the 2019 Asia–Pacific Outstanding Paper Award, the Outstanding Teaching Award from Yonsei University in South Korea in 2011, as well as Best Paper Awards at IEEE GLOBECOM 2006 and IEEE/CIC ICCC 2018. He has served as the Lead Chair for the Wireless Communication Symposium of IEEE Globecom 2017 and the Communication Theory Symposium of IEEE GLOBECOM 2014 and the TPC Co-Chair for IEEE PIMRC 2017 and IEEE CTW 2013. He is an Associate Editor of IEEE Transactions on Wireless Communications and IEEE Journal on Selected Areas in Communications, and an Area Editor of IEEE Transactions on Green Communications and Networking. He has also served on the editorial boards of IEEE Wireless Communications Letters. He has guest edited special issues for IEEE Journal on Selected Areas in Communications, IEEE Journal on Selected Topics on Signal Processing, and IEEE Communications Magazine. He is an IEEE Distinguished Lecturer of the IEEE Communications Society and Vehicular Technology Society. He has been named a Highly Cited Researcher by Clarivate Analytics in 2019 and 2020.
\end{IEEEbiography}

\begin{IEEEbiography}
{Lav R. Varshney} (Senior Member, IEEE) received the B.S. degree (magna cum laude) with honors in electrical and computer engineering from Cornell University, Ithaca, NY, USA, in 2004, and the S.M., E.E., and Ph.D. degrees in electrical engineering and computer science from the Massachusetts Institute of Technology, Cambridge, in 2006, 2008, and 2010, respectively.
He is currently an Associate Professor with the Department of Electrical and Computer Engineering and the Coordinated Science Laboratory, University of Illinois at Urbana–Champaign. From 2019 to 2020, he was on leave as a Principal Research Scientist with Salesforce Research, Palo Alto, CA, USA. From 2010 to 2013, he was a Research Staff Member with the IBM Thomas J. Watson Research Center, Yorktown Heights, NY, USA. His research interests include information and coding theory, statistical signal processing, neuroscience, and artificial intelligence.
Dr. Varshney is a member of Eta Kappa Nu, Tau Beta Pi, and Sigma Xi. He was a recipient of the IBM Faculty Award in 2014 and a Finalist for the Bell Labs Prize in 2014 and 2016. He and his students have received several best paper awards. He currently serves on the advisory board of the AI XPRIZE.
\end{IEEEbiography}

\begin{IEEEbiography}
{Sennur Ulukus} (Fellow, IEEE) received the B.S. and M.S. degrees in electrical and electronics engineering from Bilkent University and the Ph.D. degree in electrical and computer engineering from the Wireless Information Network Laboratory (WINLAB), Rutgers University.,She is currently the Anthony Ephremides Professor of information sciences and systems with the Department of Electrical and Computer Engineering, University of Maryland at College Park, where she also holds a joint appointment with the Institute for Systems Research (ISR). Prior to joining UMD, she was a Senior Technical Staff Member with AT\&T Labs-Research. Her research interests are in information theory, wireless communications, machine learning, signal processing and networks, with recent focus on private information retrieval, age of information, distributed coded computation, energy harvesting communications, physical layer security, and wireless energy and information transfer. She is a Distinguished Scholar-Teacher of the University of Maryland. She received the 2003 IEEE Marconi Prize Paper Award in Wireless Communications, the 2019 IEEE Communications Society Best Tutorial Paper Award, the 2005 NSF CAREER Award, the 2010–2011 ISR Outstanding Systems Engineering Faculty Award, and the 2012 ECE George Corcoran Outstanding Teaching Award. She was a Distinguished Lecturer of the IEEE Information Theory Society from 2018 to 2019. She has been an Area Editor for IEEE Transactions on Green Communications and Networking since 2016 and IEEE Transactions on Wireless Communications since 2019. She was an Editor of IEEE Journal on Selected Areas in Communications—Series on Green Communications and Networking from 2015 to 2016, IEEE Transactions on Information Theory from 2007 to 2010, and IEEE Transactions on Communications from 2003 to 2007. She was a Guest Editor of IEEE Journal on Selected Areas in Communications in 2015 and 2008, Journal of Communications and Networks in 2012, and IEEE Transactions on Information Theory in 2011. She was a TPC Co-Chair of 2019 IEEE ITW, 2017 IEEE ISIT, 2016 IEEE Globecom, 2014 IEEE PIMRC, and 2011 IEEE CTW.
\end{IEEEbiography}

\begin{IEEEbiography}
{Mohamed-Slim Alouini} (Fellow, IEEE) was born in Tunis, Tunisia. He received the Ph.D. degree in electrical engineering from the California Institute of Technology (Caltech), Pasadena, CA, USA, in 1998. He served as a Faculty Member for the University of Minnesota, Minneapolis, MN, USA, and also for Texas A\&M University at Qatar, Education City, Doha, Qatar. In 2009, he joined as a Professor of electrical engineering with the King Abdullah University of Science and Technology (KAUST), Thuwal, Makkah Province, Saudi Arabia. His current research interest includes the modeling, design, and performance analysis of wireless communication systems.
\end{IEEEbiography}

\end{document}